# Strain Functionals: A Complete and Symmetry-adapted Set of Descriptors to Characterize Atomistic Configurations

Edward M. Kober,*[,a] Jacob P. Tavenner,*[,2] Colin M. Adams,*[,3] Nithin Mathew*[,1,a]

*Group T-1, Theoretical Division, Los Alamos National Laboratory, Los Alamos, NM 87544

[1]Group XCP-5, X Computational Physics Division, Los Alamos National Laboratory, Los Alamos, NM 87544

[2]Computational Materials Science and Design Group, Baylor University, Waco, TX 76706

[3]Systems and Technology Research, Woburn, MA 01801

[a] Corresponding authors: emk@lanl.gov, mathewni@lanl.gov

## Abstract

Extracting relevant information from atomistic simulations relies on a complete and accurate characterization of atomistic configurations. We present a framework for characterizing atomistic configurations in terms of a complete and symmetry-adapted basis, referred to as strain functionals. In this approach a Gaussian kernel is used to map discrete atomic quantities, such as number density, velocities, and forces, to continuous fields. The local atomic configurations are then characterized using $n^{\text{th}}$ order central moments of the local number density. The initial Cartesian moments are recast unitarily into a Solid Harmonic Polynomial basis using SO(3) decompositions. Rotationally invariant metrics, referred to as Strain Functional Descriptors (SFDs), are constructed from the terms in the SO(3) decomposition using Clebsch-Gordan coupling. A key distinction compared to related methods is that a minimal but complete set of descriptors is identified. These descriptors characterize the local geometries numerically in terms of shape, density, and orientation descriptors that recognize $n$-fold symmetry axes and net shapes such as trigonal, cubic, hexagonal, etc. They can easily distinguish between most different crystal symmetries using $n = 4$, identify defects (such as dislocations and stacking faults), measure local deformation, and can be used in conjunction with machine learning techniques for *in situ* analysis of finite temperature atomistic simulation data and quantification of defect dynamics.

## 1. Introduction

Macroscopic properties of materials are governed by generation and propagation of defects at the atomic scale and their interaction with the microstructure. Atomistic simulation methods, using *ab initio* or empirical interatomic potentials have enabled us to study these processes with atomic scale resolution. The extraction of information from these atomistic simulation techniques often relies on identifying and categorizing the local configurations/geometries around atomic centers. For example, in simulations of crystalline materials, deformation processes give rise to a variety of defect structures (vacancies and interstitials, point defect clusters, dislocation loops, stacking faults, twins etc.) that can be identified by the distinct local geometries of the atoms associated with the defects. A wide variety of techniques have been developed to achieve these goals and these have been reviewed by Stukowski who has pointed out the relative merits and limitations of the various techniques.[1,2] Many of these methods, such as the common neighbor analysis,[3,4] are specific to particular crystal systems (*e.g. FCC*, *BCC*, *HCP*). Some, such as centro-symmetry analysis,[5] Steinhardt order parameters,[6] bond-angle analysis,[7] diamond structure analysis,[8] and Voronoi-type analyses[4,9] rely on a limited set of symmetry functions/heuristic rules applied to a specific listing of neighbors. These are not necessarily robust under thermal fluctuations. To overcome these limitations, recent methods[10,11] have relied on 'shape/pattern/template matching' techniques using a diverse set of shape and orientation descriptors. Methods using geometric constructions such as the Burger's circuit construction[12,13], Nye tensor calculations,[14] or perturbations of the Voronoi topology[15] have also been applied with good success. Recently, there has been a lot of interest in developing physics-inspired descriptors of atomic environments[16,17-23] which is partly motivated by the application of data-driven approaches for materials modeling. A comprehensive review of these methods and their relationships has recently been published.[24] While these approaches have their own particular advantages, the physical meaning of these descriptors, in terms of symmetry elements of atomic environments, remains unclear. Moreover, the large number of descriptors used in these approaches to describe atomic environments suggests either redundancies in the underlying basis or a strong dependence on thermal fluctuations, especially for the higher order terms.

We present an approach for characterizing atomic geometries and identifying symmetry elements based on the technique of moment analyses. Moment analyses have been used for several decades as a means to characterize three-dimensional (3D) structures, especially for image/pattern analysis.[25-27] The parallels with identifying atomic geometries using such an approach are obvious, albeit the irregular distributions of atoms due to localized strains/defects and thermal fluctuations add computational complexity compared to working with regular arrays of pixels and voxels. We also note that there are inherent parallels between the moment analyses and gradients of the distributions. While the $n^{\text{th}}$ order moments are easy to calculate in a particular reference frame (often expressed in terms of spherical harmonic or harmonic polynomial functions), the comparison between similar objects (geometries) in different reference frames requires using rotationally invariant formulations, and these aspects have not been fully developed. An approach for building rotational invariance into 3D moment analysis was developed by Lo & Don[28] using techniques of angular momentum (or Clebsch-Gordan) coupling.[29] Lo & Don had demonstrated this approach for second and third order moments and identified several different invariant descriptors for those orders. However, they did not present a means by which a unique set of descriptors could be developed, nor did they develop a geometric understanding of what each of the terms implied. A mathematically-related approach based on irreducible tensors has also been presented along with its applicability to physical properties that can be described by third and fourth order tensors.[30] Again, this approach identifies some of the invariants, but is incomplete in their description. In related work on Magnetic Resonance Imaging (MRI), Kindlmann and co-workers[31,32] have formulated an orthonormal set of rotationally invariant fourth order descriptors that is highly relevant to the work here, though their feature space is more limited because of the higher symmetry of the diffusion tensor. They do attach physical significance to their descriptors and point out that there are different frameworks in which these can be cast depending on the desired physical interpretations.

The goal here is to consolidate these types of approaches through fourth order in an extensible framework appropriate for the interpretation of deformations. Fourth is the minimal order required to distinguish most crystal symmetries and deformations, where sixth order would be required to distinguish variations of hexagonal crystal system. We present a complete, general, and symmetry-adapted method that utilizes symmetry elements derived from local neighborhood moments to develop descriptors for describing these atomistic geometries. As the descriptors cannot depend upon the relative orientations of the experimental and observation frames, they can be defined within the 3D rotation group, SO(3), and all features of the deformation must be describable using the rotationally invariant basis functions of that group. These basis functions can be expressed in terms of functions that carry angular momentum (e.g. spherical harmonic or harmonic polynomial functions) that are combined and contracted using the Clebsch-Gordan coefficients. Such a framework has been previously utilized in the development of the Steinhardt parameters,[6] the SOAP descriptors[18] and the related GAP method,[17] the SNAP potentials[33,34], the Moment Tensor Potentials[22,35], and Atomic Cluster Expansion (ACE).[23] A somewhat similar logic was used in the earlier development of the EAM and MEAM potentials.[36-38] While the rotational invariance of the these approaches is mathematically correct, they use a variety of methods for defining a local neighborhood (e.g., simple radial cutoffs, splined functions) that result in complex basis functions that are not readily interpretable. Here, we use a simple Gaussian kernel to define the neighborhood, which allows us to develop a complete, minimal, and orthogonal set of descriptors with clear physical/geometric meaning in terms of shape, density, and orientation of Local Atomic Environments (LAEs).

Beyond the immediate goal of describing the local deformations in atomic neighborhoods, the method of moment expansions has other physical significance. Foremost, there should be a strong relationship to similar tensorial expansions of physical properties.[30] When applied to the local atomic density, the moments are a method for characterizing the nature of the local strain field.[39] That is, the moments measure the gradients in local number density which in turn reflects the existence of local strain and its gradients. We note here that local atomic strain, in the strict continuum mechanics sense, has also been defined in multiple previous works using deformation gradient tensor calculated utilizing different radial kernels.[39-44] Such analyses have provided a good means for characterizing macroscopic deformation processes. Again, by using a simple Gaussian kernel, we can formalize these analyses in terms of a Taylor's series expansion and develop a better understanding of the deformation processes.

## 2. Method

### *2.0. General approach*

All mathematical manipulations and visualizations were performed using the *Mathematica* software package (13.2) available from Wolfram. The phase conventions for the spherical harmonics and Clebsch-Gordan factors are the same as those used by Edmonds.[29]

### *2.1. Continuous fields and their derivatives from discrete quantities*

For this initial development, we will consider a system where all the atoms are of the same type. The standard difficulty in defining strains (and other continuum physical quantities) of atomistic systems is that the atomic positions are treated as discrete points, and some sort of smoothing function or kernel must be applied to map the discrete quantities to a continuum function. As will become apparent below, these functions then define the characteristics of the "neighborhood" of each atom. Various techniques, such as cubic splines or gradated step functions,[43] hard-sphere cut-off with unit weighting,[39] and cosine switching function[17,45] have been employed for performing this smoothing. The problem with the approaches using a cut-off function is that being based on discrete neighbor lists, the atomic "neighborhoods" change discontinuously during dynamic simulations, and this frustrates defining continuum properties with desirable properties such as continuity and curl-free fields. Employing smoothing functions at cutoff distances can alleviate, but not eliminate, these issues, but they do introduce complexities in describing the radial dependence of the resulting descriptors. Here, we will use Gaussian kernels as our smoothing functions which have the useful property of being infinitely differentiable and smooth. These have been used extensively for image processing applications because of these useful mathematical properties.[46] These functions also diminish rapidly with distance such that a cut-off distance based on the amplitude becoming comparable with computational noise is practical. We also note that the Gaussian function arises from a Maximum Entropy argument as the least-biased kernel used for interpolation in mesh-free continuum mechanics simulations.[47,48] The explicit form of the normalized, isotropic kernel used in this work is given in eq. 1 below.

$$w_b(\mathbf{r}) = \frac{1}{(2\pi)^{3/2}\sigma^3} exp\left(-\frac{|\mathbf{r}_b - \mathbf{r}|^2}{2\sigma^2}\right) \tag{1}$$

This is defined with respect to the position of the $b^{th}$ atom, $\boldsymbol{r}_b$, where $\boldsymbol{r}$ is a general coordinate in space and $\sigma$ is the width of the Gaussian which has units of length. The kernel function itself, $w_b$, has units of 1/volume. The integral of this normalized function over all space simply equals 1. This expression is used to define the continuum number density function $N$ (number of atoms per volume) at an arbitrary point in space as shown in eq. 2.

$$N(\mathbf{r}) = \sum_b w_b(\mathbf{r}) = \frac{1}{V_0}\sum_b exp\left(-\frac{|\mathbf{r}_b - \mathbf{r}|^2}{2\sigma^2}\right) \quad (2a)$$

Here, because all the atoms are the same type, we assume the length scale $\sigma$ to be independent of the atomic identities and factor out a constant scalar volume, $V_0$, defined in eq. 2b.

$$V_0 = (2\pi)^{3/2}\sigma^3 \quad (2b)$$

This same approach can be used to define analogous continuum property fields such as mass density, momentum and stress. We now consider defining the Taylor series expansion of the number density in the neighborhood of atom $a$, as shown in eq. 3 given below:

$$n_a(\boldsymbol{r}) = N(\boldsymbol{r_a} + \Delta\boldsymbol{r})$$
$$\approx N(\boldsymbol{r_a}) + \Delta\boldsymbol{r} \odot \left.\frac{\partial N}{\partial \boldsymbol{r}}\right|_{\boldsymbol{r}_a} + \frac{1}{2}(\Delta\boldsymbol{r} \otimes \Delta\boldsymbol{r}) \odot \left.\frac{\partial^2 N}{\partial \boldsymbol{r}^2}\right|_{\boldsymbol{r}_a} + \cdots. \quad (3)$$

This expression is written in tensor form, where the terms associated with the $n^{th}$ derivative are a rank $n$ tensor of dimension 3. The general three-dimensional coordinate $\boldsymbol{r} = [x, y, z]$ is redefined in terms of a small displacement $\Delta\boldsymbol{r}$ with respect to the chosen central position of $\boldsymbol{r}_a$. The symbol $\otimes$ designates the vector outer product where $\Delta\boldsymbol{r} \otimes \Delta\boldsymbol{r}$ is a rank 2 tensor of dimension 3 (or the 3x3 matrix $T_{ij}$), and the next term in the sequence, $\Delta\boldsymbol{r} \otimes \Delta\boldsymbol{r} \otimes \Delta\boldsymbol{r}$, is a rank 3 tensor of dimension 3 (or the 3x3x3 array $T_{ijk}$). The symbol $\odot$ indicates the scalar product (total contraction) between two rank $n$ tensors.

In general, the $n^{th}$ order derivatives of the atomic number density can then be written as in eq. 4.

$$\left.\frac{\partial^n N}{\partial \boldsymbol{r}^n}\right|_{\boldsymbol{r}_a} = \sum_b \left.\frac{\partial^n w_b}{\partial \boldsymbol{r}^n}\right|_{\boldsymbol{r}_a} = \frac{1}{V_0\sigma^n}\sum_b w_b(\boldsymbol{r_a})\,\mathbf{H}_n\left(\frac{\boldsymbol{r}_{ab}}{\sigma}\right)$$
$$= -(1)^n \frac{1}{V_0\sigma^n}\sum_b w_{ab}\,\mathbf{H}_n\left(\frac{\boldsymbol{r}_{ab}}{\sigma}\right) \quad (4)$$

Here, $\boldsymbol{r}_{ab} = \mathbf{r}_b - \mathbf{r}_a$, $w_{ab} = w_b(\boldsymbol{r_a})$ and $\mathbf{H}_n$ is defined as the $n^{th}$ order tensor formulation of the Hermite polynomial expressions,[49] which result from taking the derivatives of the Gaussian function. These formulations could also be expanded out in terms of polynomial combinations of the individual Cartesian components as outlined in eq. 5, where the terms would comprise elements of the general Cartesian tensors written above (e.g. $\Delta\boldsymbol{r} \otimes \Delta\boldsymbol{r}$).

$$\left.\frac{\partial^n N}{\partial x^p \partial y^q \partial z^s}\right|_{\boldsymbol{r}_a} = \frac{1}{\sigma^n}\sum_b w_{ab}\,(x_b - x_a)^p(y_b - y_a)^q(z_b - z_a)^s \quad (5)$$

Here $p, q, s$ are all non-negative integers with the constraint that $p + q + s = n$. These define the general set of harmonic polynomial tensors. The interrelationship of the harmonic and Hermite polynomials involves factors of $r^{2m}$, with $m = 0, 1, 2, \ldots$ which will become apparent below. This is related to transforming the tensors to a traceless basis.

The individual terms in eq. 3 can then be evaluated by substituting in the definition in eq. 2a. The first term for the local density given by equation 2a, with $\boldsymbol{r} = \boldsymbol{r}_a$. The subsequent terms (derivatives), given through fourth order, take the following forms:

$$\left.\frac{\partial N}{\partial \boldsymbol{r}}\right|_{r_a} = \sum_b \left.\frac{\partial w_b}{\partial \boldsymbol{r}}\right|_{r_a} = -\frac{1}{V_0}\sum_b \frac{\boldsymbol{r}_{ab}}{\sigma^2} exp\left(-\frac{|\boldsymbol{r}_b - \boldsymbol{r}_a|^2}{2\sigma^2}\right) = -\frac{1}{\sigma}\sum_b \frac{\boldsymbol{r}_{ab}}{\sigma} w_{ab} \qquad (6a)$$

$$\left.\frac{\partial^2 N}{\partial \boldsymbol{r}^2}\right|_{r_a} = \sum_b \left.\frac{\partial^2 w_b}{\partial \boldsymbol{r}^2}\right|_{r_a} = \frac{1}{\sigma^4}\sum_b (\boldsymbol{r}_{ab} \otimes \boldsymbol{r}_{ab} - \sigma^2 \boldsymbol{I}_2) w_{ab}$$

$$= \frac{1}{\sigma^4}\sum_b \left(\boldsymbol{r}_{ab} \otimes \boldsymbol{r}_{ab} - \frac{\sigma^2}{r_{ab}^2}[\boldsymbol{r}_{ab} \otimes \boldsymbol{r}_{ab}]_S\right) w_{ab} \qquad (6b)$$

$$\left.\frac{\partial^3 N}{\partial \boldsymbol{r}^3}\right|_{r_a} = \sum_b \left.\frac{\partial^3 w_b}{\partial \boldsymbol{r}^3}\right|_{r_a}$$

$$= -\frac{1}{\sigma^6}\sum_b \left(\boldsymbol{r}_{ab} \otimes \boldsymbol{r}_{ab} \otimes \boldsymbol{r}_{ab} - \frac{\sigma^2}{r_{ab}^2}[\boldsymbol{r}_{ab} \otimes \boldsymbol{r}_{ab} \otimes \boldsymbol{r}_{ab}]_S\right) w_{ab} \qquad (6c)$$

$$\left.\frac{\partial^4 N}{\partial \boldsymbol{r}^4}\right|_{r_a} = \sum_b \left.\frac{\partial^4 w_b}{\partial \boldsymbol{r}^4}\right|_{r_a}$$

$$= \frac{1}{\sigma^8}\sum_b \left(\boldsymbol{r}_{ab}\otimes\boldsymbol{r}_{ab}\otimes\boldsymbol{r}_{ab}\otimes\boldsymbol{r}_{ab} - \frac{\sigma^2}{r_{ab}^2}[\boldsymbol{r}_{ab}\otimes\boldsymbol{r}_{ab}\otimes\boldsymbol{r}_{ab}\otimes\boldsymbol{r}_{ab}]_S\right.$$

$$\left. + \frac{\sigma^4}{2r_{ab}^4}[[\boldsymbol{r}_{ab}\otimes\boldsymbol{r}_{ab}\otimes\boldsymbol{r}_{ab}\otimes\boldsymbol{r}_{ab}]_S]_S\right) w_{ab} \qquad (6d)$$

In these equations, the notation ***[M]s*** denotes the symmetric trace of tensor ***M*** where the trace has been expanded back into the full rank of the initial tensor ***M***. For example, the trace of a 3x3 matrix (e.g. eq 6b) is a simple scalar, and the reconstituted symmetric trace tensor is that scalar times the 3x3 identity matrix. This follows the approach of Coope, et al.[50] and Jerphagnon, et al.[30] of expressing a general rank *n* tensor in terms of sums of the irreducible spherical components (e.g. a complete basis set expansion). Further details of the manipulations are given in the Supplemental Material (SM) S.1, along with the definitions of the symmetric trace tensors. Note that the summations are over all atoms *b*, including *b* = *a*. For the expressions as written, the latter terms would simply be zero and there are no mathematical complications from their inclusion. For some applications, it would be more advantageous to evaluate these expansions about a position $\boldsymbol{r}_A \approx \boldsymbol{r}_a$ where $\boldsymbol{r}_A$ is defined as the center of position or mass of the neighborhood associated with atom *a*. The expressions above directly allow such a transformation.

The terms are directly interpretable. The number density at the position of the $a^{th}$ atom depends upon how many other atoms are in its neighborhood, with a Gaussian weight that depends upon the distances to the other atoms $\boldsymbol{r}_{ab} = \boldsymbol{r}_b - \boldsymbol{r}_a$, eq 2a with $\boldsymbol{r} = \boldsymbol{r}_a$. This zeroth order moment is then a metric of the local density of the system. Properties of the Gaussian become quite useful in defining the subsequent derivatives: the derivative of the Gaussian is the product of the scaled vector contained in the argument times the original Gaussian. This recursive property greatly simplifies how the derivative functions are expressed. The right-most term in eq. 6a emphasizes that, by using the Gaussian kernel, the first derivative can be simply related to the Gaussian-weighted sum of the vector distances to the atoms of the system. Consequently, *the gradient of the number density is equivalent to the first order moment of the spatial distribution of the atomic positions* in this Gaussian-weighted neighborhood. In addition, Gaussian weighting functions enable ready access to the higher order derivatives that cannot be achieved with either discrete or low-order polynomial weighting.

The second derivative is given in eq 6b. Considering just the first term in the parentheses in eq. 6b, this contains the second order moment of the spatial distribution of the atomic positions in this neighborhood. This can be related to the rank 2 strain metric and measures the amounts of uniaxial and shear deformations (see SM S.2). The subtraction of the identity matrix basically modifies the trace of the matrix so that the resulting expression has different effective weights for the deviatoric components, which measure the change in shape of the local neighborhood, versus the trace which measures the level of volumetric strain, i.e. change in size. This difference in weighting depends upon the original choice of the weighting function and is also tied to making the subsequent moment expressions orthogonal to one another.

The expressions for the third and fourth derivatives are given in eqs. 6c and 6d, respectively. Overall, these are symmetric rank 3 and rank 4 tensors, respectively, and their geometric attributes correspond to strain gradients and derivatives thereof, respectively. The forms have a structure similar to eq 6b, where there are separate terms based on the traces of highest order tensor. This suggests a further reduction of transforming all of the tensor expressions to traceless versions, denoted by ***[M]****T*, where eqs 6b-d can be rewritten as eqs 7a-c, respectively. Further details are given in Supplemental Material (S.1).

$$\left.\frac{\partial^2 N}{\partial \boldsymbol{r}^2}\right|_{r_a} = \frac{1}{\sigma^4}\sum_b \left( [\boldsymbol{r}_{ab}\otimes\boldsymbol{r}_{ab}]_T - \left\{\frac{\sigma^2}{r_{ab}^2} - \frac{1}{3}\right\} [\boldsymbol{r}_{ab}\otimes\boldsymbol{r}_{ab}]_S \right) w_{ab} \qquad (7a)$$

$$\left.\frac{\partial^3 N}{\partial \boldsymbol{r}^3}\right|_{r_a} = -\frac{1}{\sigma^6}\sum_b \left( [\boldsymbol{r}_{ab}\otimes\boldsymbol{r}_{ab}\otimes\boldsymbol{r}_{ab}]_T - \left\{\frac{\sigma^2}{r_{ab}^2} - \frac{1}{5}\right\} [\boldsymbol{r}_{ab}\otimes\boldsymbol{r}_{ab}\otimes\boldsymbol{r}_{ab}]_S \right) w_{ab} \qquad (7b)$$

$$\left.\frac{\partial^4 N}{\partial \boldsymbol{r}^4}\right|_{r_a} = \frac{1}{\sigma^8}\sum_b \Bigg( [\boldsymbol{r}_{ab}\otimes\boldsymbol{r}_{ab}\otimes\boldsymbol{r}_{ab}\otimes\boldsymbol{r}_{ab}]_{\boldsymbol{T}} - \left\{\frac{\sigma^2}{r_{ab}^2} - \frac{1}{7}\right\} [[\boldsymbol{r}_{ab}\otimes\boldsymbol{r}_{ab}\otimes\boldsymbol{r}_{ab}\otimes\boldsymbol{r}_{ab}]_{\boldsymbol{S}}]_{\boldsymbol{T}} + \left\{\frac{\sigma^4}{2r_{ab}^4} - \frac{\sigma^2}{3r_{ab}^2} + \frac{1}{30}\right\} [[\boldsymbol{r}_{ab}\otimes\boldsymbol{r}_{ab}\otimes\boldsymbol{r}_{ab}\otimes\boldsymbol{r}_{ab}]_{\boldsymbol{S}}]_{\boldsymbol{S}} \Bigg) w_{ab} \qquad (7c)$$

The final terms of these expressions cannot be reduced further. For the even-powered derivatives, the final term becomes a scalar, which would be $r_{ab}^2$ for eq 7a and $r_{ab}^4$ for eq 7c. For the odd-powered derivatives, the final term depends on a vector of 3 terms, which is $(r_{ab}^2 x_{ab}, r_{ab}^2 y_{ab}, r_{ab}^2 z_{ab})$ for eq 7b. This derivation again follows the approach given by Coope[50-52] and Jerphagnon.[30] As discussed further below, the resulting tensor terms of these expansions are irreducible Spherical Tensors (ST), which directly map onto angular momentum vectors. These correspond to subspace groups of the rotationally invariant space symmetry group SO(3).

Very similar expressions would be derived for any other spherically symmetric weighting function besides a Gaussian, such as a simple exponential decay (which occurs in the definition of atomic orbitals) or 1/r (for electrostatic multipolar expansions). The difference would lie only in the terms in the curly brackets in eqs 7a-c, which define where radial nodes would occur that define the orthogonality for closely related subspaces. If finite $n^{th}$-order polynomials were used as the weighting functions, the moments would vanish for orders > *n*. The use of polynomial spline functions to terminate the radial terms at some defined cut-off value would generate similar issues and much more complicated radial node patterns. The angular dependence, captured in the ST, is independent of the spherically-symmetric radial function. At the top level, this is a restatement that many physics problems that involve spherically symmetric fields can be separated into radial and angular components, where the latter can usually be succinctly expressed in terms of spherical harmonic functions. We wish to emphasize a deeper level here where the radial node pattern supplies an additional orthogonal partitioning into the SO(3) subspaces. A well-known example of this is the definition of radial nodes for atomic orbitals which establishes their general orthogonality, subsequent mathematic utility and use for physical interpretations.[53]

### *2.2. Transformation of Cartesian moments to rotationally invariant descriptors*

While complete, recognizing the shapes or deformations expressed in Cartesian moments is difficult and only works well when the coordinate axes happen to be aligned with any symmetry axes of the system. To simplify this, the Cartesian moments can be recast into expressions based on spherical harmonics, which can then be re-expressed in a rotationally invariant format. Both Jerphagnon, et al.[30] and Lo and Don[28] have shown how to transform the harmonic polynomial terms (eq 5) into spherical harmonic expressions, where their approaches follow the algebra of angular momentum.[29] The first step in this process is to linearly transform the Cartesian moments into the regular solid harmonic polynomials, $R_{l(n)}^{m}$, defined in eq. 8.

$$R_{l(n)}^{m}(\boldsymbol{r}) = \left(\frac{r}{\sigma}\right)^{n-l} \left(\frac{r}{\sigma}\right)^{l} Y_{l}^{m} = \left(\frac{r}{\sigma}\right)^{n} Y_{l}^{m} \tag{8}$$

Here, $r$ is the magnitude of the vector $\boldsymbol{r}$, $\sigma$ is the scale factor from the Gaussian weighting process, $n$ is the overall order of the polynomial, $l$ is the overall order of the spherical harmonic polynomial, and $Y_{l}^{m}$ is the standard spherical harmonic function defined in their complex form. The constraints on the integer indices are that $n \geq 0$; $l = n, n\text{-}2, \ldots, \geq 0$; $m = -l, -l+1, \ldots, l-1, l$. We emphasize that while the $Y_{l}^{m}$ functions have no dependence on distance, but only on direction, the solid harmonic $R_{l(n)}^{m}$ functions do have an explicit radial dependence of $\left(\frac{r}{\sigma}\right)^{n}$, in addition to the angular dependence from the $Y_{l}^{m}$ contribution. Thus, this approach supplants the need to build-in further radial dependence or the use of hyper-spherical harmonics.[17,27]

The complex moments about atom $a$ are then defined in eq. 9.

$$v_{l(n)}^{m}(a) = \sum_{b} R_{l(n)}^{m}(\boldsymbol{r}_{ab})\, w_{ab} \tag{9}$$

These $v_{l(n)}^{m}$ terms, which are complex, have the same indices as the solid harmonic polynomials and have the same behavior as angular momentum functions. These terms contain the information of the moment tensors expressed in this solid harmonic basis and are called Spherical representations of the irreducible moment tensors[27] or simply as Spherical Tensors (ST).[30] Each of the irreducible tensors in eqs 7a-c can then be directly mapped onto the ST as shown in Table I. This basis facilitates defining both rotational invariant descriptors and orientation functions.

Table I. Mapping between the irreducible moment tensors (IMT) and spherical tensors (ST). The final column gives the degrees of freedom (DOF) for the tensors.

| IMT | ST | DOF |
|---|---|---|
| $\boldsymbol{r}_{ab}$ | $\boldsymbol{v}_{1(1)}$ | 3 |
| $[\boldsymbol{r}_{ab}\otimes\boldsymbol{r}_{ab}]_{\boldsymbol{T}}$ | $\boldsymbol{v}_{2(2)}$ | 5 |
| $[\boldsymbol{r}_{ab}\otimes\boldsymbol{r}_{ab}]_{\boldsymbol{S}}$ | $\boldsymbol{v}_{0(2)}$ | 1 |
| $[\boldsymbol{r}_{ab}\otimes\boldsymbol{r}_{ab}\otimes\boldsymbol{r}_{ab}]_{\boldsymbol{T}}$ | $\boldsymbol{v}_{3(3)}$ | 7 |
| $[\boldsymbol{r}_{ab}\otimes\boldsymbol{r}_{ab}\otimes\boldsymbol{r}_{ab}]_{\boldsymbol{S}}$ | $\boldsymbol{v}_{1(3)}$ | 3 |
| $[\boldsymbol{r}_{ab}\otimes\boldsymbol{r}_{ab}\otimes\boldsymbol{r}_{ab}\otimes\boldsymbol{r}_{ab}]_{\boldsymbol{T}}$ | $\boldsymbol{v}_{4(4)}$ | 9 |
| $[[\boldsymbol{r}_{ab}\otimes\boldsymbol{r}_{ab}\otimes\boldsymbol{r}_{ab}\otimes\boldsymbol{r}_{ab}]_{\boldsymbol{S}}]_{\boldsymbol{T}}$ | $\boldsymbol{v}_{2(4)}$ | 5 |
| $[[\boldsymbol{r}_{ab}\otimes\boldsymbol{r}_{ab}\otimes\boldsymbol{r}_{ab}\otimes\boldsymbol{r}_{ab}]_{\boldsymbol{S}}]_{\boldsymbol{S}}$ | $\boldsymbol{v}_{0(4)}$ | 1 |

There are *2l+1* components for each order *l* of the $\boldsymbol{\nu}_{l(n)}$ vectors, and they can be generally written as a complex vector as shown in eq 10.

$$\boldsymbol{\nu}_{l(n)} = \left\{\nu_{l(n)}^{l} e^{-il\omega_l}, \nu_{l(n)}^{l-1} e^{-i(l-1)\omega_{l-1}}, \dots, \nu_{l(n)}^{l-1} e^{+i(l-1)\omega_{l-1}}, \nu_{l(n)}^{l} e^{+il\omega_l}\right\} \qquad (10)$$

Here the *l-1* $\nu_{l(n)}^{m}$ coefficients *(1≤m≤l)* are defined as real non-negative quantities which define the magnitude of the *m^th^* component, and the *l* $\omega_m$ angles *(1≤m≤l)* are the phase angles that define the relative rotations of the complex terms. The one exception is that the $\omega_0$ angle is zero by definition because that term has no dependence on the rotation about the *z* axis. The $\nu_{l(n)}^{0}$ coefficient can then be either positive or negative, the significance of which will become apparent below. This defines the *2l+1* number of degrees of freedom that uniquely define these vectors / ST in a fixed arbitrary frame. An alternative method for expressing this vector is to resolve all the complex components into real expressions by appropriate addition / subtraction of conjugate components as below. Such formulations are generally referred to as Real Spherical Harmonics, and the tilde is added to signify that the magnitude has been modified by $\sqrt{2}$.[54]

$$\tilde{\nu}_{l(n)}^{mc} = \frac{1}{i^m\sqrt{2}}\left(\nu_{l(n)}^{m} e^{-im\omega_m} - \nu_{l(n)}^{m} e^{+im\omega_m}\right) = \tilde{\nu}_{l(n)}^{m} cos(m\omega_m) \qquad (11a)$$

$$\tilde{\nu}_{l(n)}^{ms} = \frac{1}{i^m\sqrt{2}}\left(\nu_{l(n)}^{m} e^{-im\omega_m} + \nu_{l(n)}^{m} e^{+im\omega_m}\right) = \tilde{\nu}_{l(n)}^{m} sin(m\omega_m) \qquad (11b)$$

The real spherical tensors are then generally written as in eq 12.

$$\tilde{\boldsymbol{\nu}}_{l(n)} = \left\{\tilde{\nu}_{l(n)}^{0}, \tilde{\nu}_{l(n)}^{1} cos(\omega_1), \tilde{\nu}_{l(n)}^{1} sin(\omega_1), \dots, \tilde{\nu}_{l(n)}^{l} cos(l\omega_l), \tilde{\nu}_{l(n)}^{l} sin(l\omega_l)\right\} \qquad (12)$$

We illustrate these functions in Figure 1. In the left half of the figure are shown representative solid harmonic functions $R_{l(n)}^{m}$ which have positive (red) and negative (blue) phases. Their appearance is very similar to typical atomic orbital representations (*p, d, f* and *g* orbitals for *l* = 1, 2, 3 and 4, respectively) which would have an exponential decay in density with distance compared to the Gaussian damping used here. These show the density changes that would result for that type of deformation. Also included in the figure are the Schoenflies notation for the point group that describes the symmetry of deformation.[55] In the right half of the figure are shown the density distributions that would result from applying those changes to a spherically symmetric Gaussian density. The red regions show where density has been added and the blue regions show where density has been removed. For example, the case of *l=2* and *m=0* in Fig 1B would correspond to a tetragonal distortion (a = b ≠ c) along the vertical (*z*) axis. Neighboring atoms have been brought in closer along the *z* axis, but pushed further away in the *xy* plane. The *l=2* and *m=1,2* distortions show shear strains where the neighborhoods are compressed along one axis, expanded along a perpendicular axis, but left undistorted along the third perpendicular axis (orthorhombic distortion, a ≠ b ≠ c). Thus, the *m*=0 distortion results in a cylindrically symmetric geometry ($\boldsymbol{D}_{\infty h}$), while the *m=1,2* distortions result in a lower symmetry brick-like or orthorhombic geometry (($\boldsymbol{D}_{2h}$). The deformations represented in the *l=2* terms can be directly mapped to strain tensors defined in mechanics literature. These are described in the supplemental material (S.2).

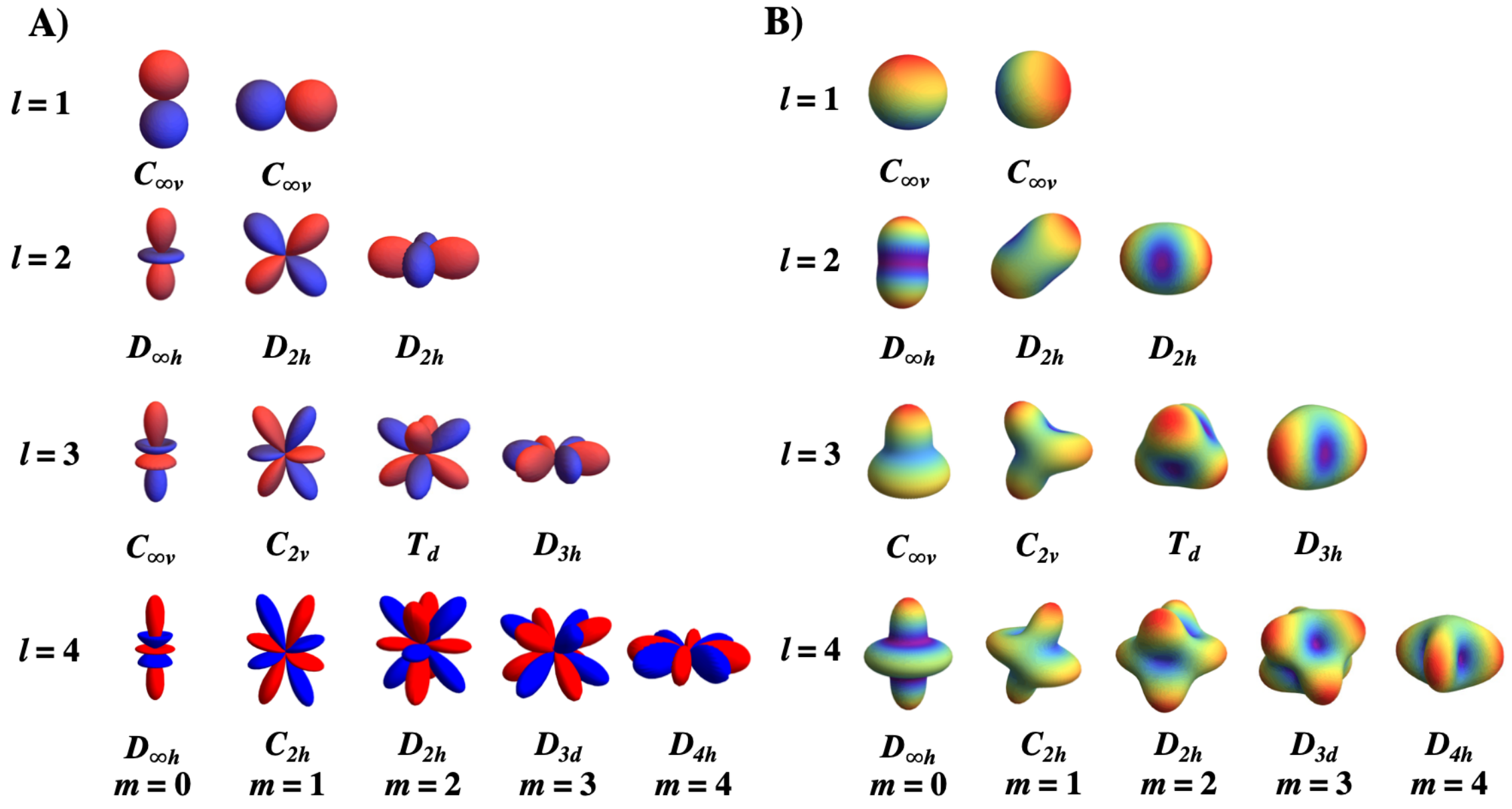

*Figure 1. Comparison plot of the solid harmonic basis functions and the resulting density deformations. A) Illustrative solid harmonic basis functions, $R_{l(n)}^{m}$, for l=n=1-4, and for 0≤m≤l. The surfaces are drawn for an isodensity surface and are colored as red for a positive phase and blue for a negative phase. The vertical axis is the z direction, the x axis points to the right, and with the y axis projecting into the plane of the page. The structures have been slightly rotated and tilted to add perspective. The corresponding functions for -l≤m≤-1 would be obtained by rotating the positive analogs by π/2m radians about the vertical (z) axis. B) Illustrative density deformation plots l=n=1-4 and for 0≤m≤l. These were generated by adding 1/2 of the functions on the left to a spherically symmetric Gaussian distribution: $\left[1+\frac{1}{2}R_{l(n)}^{m}\right]exp\left(-\frac{r^2}{2\sigma^2}\right)$. The surfaces are drawn for an isodensity surface and are colored by how far that surface is from the central point: red for the farthest distance, blue for the closest. Thus, the color simply accentuates the surface plot to highlight where density has been added (red) and depleted (blue). Also included are the Shoenflies notation for the point group of the deformation.*

*Clebsch-Gordan coupling.* The second step in the process is to derive rotationally invariant descriptors from these $v_{l(n)}^{m}$ terms. This is done using the Clebsch-Gordan (*CG*) coupling procedure which defines the addition or coupling process between two angular momentum vectors.[29] While this is normally thought of as a quantum mechanical process for coupling between the spin of an electron and its orbital angular momentum[56], when only integer spins are involved, this describes classical angular momentum processes as well. As shown below, rotational symmetry axes are directly identified by their application.

One can somewhat arbitrarily take the product between any two ST (or between an ST and itself) by the expression in eq. 13, using the complex forms in eq 10.

$$\boldsymbol{N}[\boldsymbol{l_1}(\boldsymbol{n_1}),\boldsymbol{l_2}(\boldsymbol{n_2})]_h^k=\sum_{m=-l}^{m=l}\langle l_1,m,l_2,k-m|l_1,l_2,j,k\rangle v_{l_1(n_1)}^{m}v_{l_2(n_2)}^{k-m} \tag{13}$$

The $N$ terms are the components of the ST $\boldsymbol{N}$ of rank $h$ that results from the $CG$ procedure from coupling two ST that are designated $\boldsymbol{l_1}(\boldsymbol{n_1})$ and $\boldsymbol{l_2}(\boldsymbol{n_2})$ with $|k|\le h$. Here, these are shorthand for the two ST $\boldsymbol{v}_{\boldsymbol{l_1}(\boldsymbol{n_1})}$ and $\boldsymbol{v}_{\boldsymbol{l_2}(\boldsymbol{n_2})}$ that are being operated on, where the $\boldsymbol{v}$ designation has been dropped because it is redundant, and the subscript labels can then be promoted to full text for clarity. They are in bold because they represent vector/tensor quantitites. The terms in the brackets $\langle\,\rangle$ are the $CG$ coefficients which specify how different components of the two ST $\boldsymbol{v}_{\boldsymbol{l_1}(\boldsymbol{n_1})}$ and $\boldsymbol{v}_{\boldsymbol{l_2}(\boldsymbol{n_2})}$ are coupled. There are somewhat restrictive rules as to which terms are allowed to combine. We particularly note the restriction that only values of $h$ = $|l_1\text{-}l_2|$, $|l_1\text{-}l_2+2|$, …, $|l_1+l_2\text{-}2|$, $|l_1+l_2|$ have non-zero coefficients. Notice that there is no explicit dependence of the $CG$ coefficients on the ($n_1$, $n_2$) values in these expressions, only the ($l$, $m$) values. We maintain these to identify the origin of the subspaces. For the cases where $l = n$, we will subsequently drop the $n$ label to simplify the notation.

The resulting vectors $\boldsymbol{N}[\boldsymbol{l_1},\boldsymbol{l_2}]_h$ have the same type of angular momentum properties as the initial vectors, and contain $2h+1$ terms which could be specifically identified by the superscript $k$. These were called "composite" vectors by Lo and Don,[28] where the original definition of the CG coefficients came from the addition or coupling of two angular momentum vectors. Given the uncertainty in the relative directions of those vectors, the CG coefficients define the probability that some desired product vector would be observed from that coupling procedure. This gives rise to a different set of normalization constants compared to other treatments of vector or matrix manipulations. Further additions between these "composite" vectors and the original vectors can then also be formed. As such, there are an unlimited number of such contractions that can be formed. For the case where $h$ = $0$ (and implicitly $k$ = $0$), the result is a scalar which then has no angular dependence. These terms contain the desired rotational invariance properties required to describe the geometries in a coordinate independent frame. That is, these expressions provide direct algebraic links back to the initial vector compositions and thus provide a pathway by which those compositions can be determined.

Unfortunately, there are still an unlimited number of such contractions that can be formed. The approach of Moment Tensor Potentials (MTP)[22] relies on formulating an extensive list of such invariants, and then using Machine Learning methods to formulate a potential function based on this large list.[35] Recently, a similar approach was applied to minimize redundancies in the ACE formulations[57,58] by utilizing singular value decomposition (SVD) as a means to achieve an arbitrary minimal basis set.[59] Such approaches were suggested by earlier expressions for moment expansions which were written in a manner that could suggest that there is an infinite expansion of moments required to capture a particular *l-th* order spherical harmonic term (see eq 3.3 in ref 6).[6] A recent approach uses pattern recognition to produce the minimal set of descriptors.[60] However, as noted above, each *l-th* rank ST contains only *2l+1* degrees of freedom (DOF). As such, it should be possible to find a combination of *2l+1* invariant terms that capture that information, and any additional invariants would represent only redundant information. It is presumed that the simplest invariants that achieve this would be preferable over more complex expressions. That is the critical aspect of this work and we shall now describe an approach for finding these terms. Within the ST framework, it also works out that these terms have rather direct physical interpretations.

The first step in this process is to recognize that the ST are initially expressed in an arbitrary frame. By rotating that frame appropriately, various terms of the vector can be zeroed out.[54] In fact, it is always possible to cause three of the terms to become zero. That is, 3 of the initial *2l+1* DOF of the arbitrary oriented tensor can be recast as the 3 rotation angles that would define the tensor in some more desirable orientation. The remaining *2l-2* DOF then would describe the shape of that rotated tensor in a fewer number of terms. (The exception to this is *l=1* which defines a vector. In that case, only 2 angles are required to describe the orientation and the remaining invariant is the magnitude of the vector.) There are options as to which terms can be zeroed out. Our preferred option, based on the real spherical harmonic forms (eq 12), is to zero out the two $m$=1 terms and the second $m=l$ term. The resulting form of the vectors is shown in eq 14 below, where the rotation matrix to achieve this is discussed in Appendix A. The three DOF that have been eliminated by this process are $\tilde{v}^{1}_{l(n)}$, $\omega_1$, and $\omega_l$. As the $m$=1 terms are the lowest symmetry components of the expansion (see Figure 1), their elimination enables a more ready appreciation of the symmetry of the tensor. We refer to this as the High Symmetry (HS) orientation. This transformation aligns the internal $\bar{z}$ axis with the *l-th* order rotation axis, and the internal $\bar{x}$ axis defines the orientation of the *m=l* component. The resulting magnitudes of the other terms will be altered by this rotation which is noted by the changing the notation from $\tilde{v}$ to $\mu$. As above the $\mu^{0}_{l(n)}$ term can then be either positive or negative. The rotation angles are also altered in the process such that $m\bar{\omega}_m = m\omega_m - l\omega_l$, hence the addition of a bar accent to those terms.

$$\boldsymbol{\mu}_{l(n)} = \left\{\mu^{0}_{l(n)}, 0, 0, \mu^{2}_{l(n)}cos(2\bar{\omega}_2), \mu^{2}_{l(n)}sin(2\bar{\omega}_2), \ldots, \mu^{l}_{l(n)}, 0\right\} \tag{14}$$

*Rank 0 and 1 terms.* First, we consider the rank 1 tensors. These are simply vectors and are somewhat an exception to the rules we will subsequently use. There is one invariant for this class, simply the norm of the vector, which is defined in the CG sense below:

$$\boldsymbol{N}[\mathbf{1},\mathbf{1}]_0 = \frac{1}{\sqrt{3}}\{(\tilde{v}^{0}_{1})^2+(\tilde{v}^{1}_{1})^2\} = \frac{1}{\sqrt{3}}(\mu^{0}_{1})^2 \tag{15}$$

The results are defined in both the initial unrotated frame ($\tilde{v}^{m}_{l}$), and in the HS orientation ($\mu^{m}_{l}$). In the original Cartesian notation, the term in the curly brackets would simply correspond to $\{z^2 + (x^2 + y^2)\}$, where *x, y* and *z* are the Cartesian elements of the vector. In the rotated frame, only $\mu^{0}_{1}$ would be non-zero (and aligned with $\bar{z}$), and two angles would describe the orientation of that axis with respect to the original frame.

We note that the CG normalization factors are different from the normal Cartesian definitions, where these had been defined to achieve a unitary sum of possible combinations of the angular momentum vectors. We retain the CG factors in general because these are generally available in mathematical software. These factors are scalars of order 1, ~O(1), and do not otherwise impact the results, so we do not correct for them. It is appropriate to normalize the invariants by the Gaussian-weighting factors though and convert them into unitless quantities. The first invariant of our overall expansion comes from the zeroth order term which is simply the sum of weights and defines the first SFD:

$$P0I0(a) = V_0 * \sum_b w_{ab} = \sum_b exp\left(-\frac{|\mathbf{r}_b-\mathbf{r}_a|^2}{2\sigma^2}\right) \quad (16a)$$

$$P0I0(a) = \sqrt{\boldsymbol{N}[\mathbf{0},\mathbf{0}]_0} \quad (16b)$$

The label P indicates that this comes from the positions of the atoms, where similar metrics could be defined for other atomic quantities (e.g. velocity, force, etc.). This *P0I0* term is a density metric which has been normalized by the reference volume. We can also define this in terms of the contraction vector $\boldsymbol{N}[\mathbf{0},\mathbf{0}]_0$ to demonstrate the correspondence to the other invariants discussed below. By taking the square root of this term, this effectively defines the term as being a single sum over the atoms in the neighborhood, rather than a square over the sums. The second invariant then comes from the rank 1 tensor which then is normalized by this first term:

$$P1I0(a) = \frac{\sqrt{\boldsymbol{N}[\mathbf{1},\mathbf{1}]_0}}{P0I0(a)} \quad (17)$$

Again, by taking the square root of this invariant, the term is properly normalized by dividing by *P0I0*, rather than its square. And, although the term is unitless, this effectively scales its units as being in terms of $\left(\frac{r}{\sigma}\right)^l$ with *l=1*. This pattern will be followed for the subsequent invariants.

*Rank 2 terms.* Next, we consider the rank 2 tensors, which have previously been considered in a similar context by Kindlmann[31,32] and Lo&Don.[28] Starting from eq 7a, these components correspond to two different subspaces: a rank 2 subspace (the first term on the RHS of eq 7a) and a rank 0 subspace (the second term on the RHS of eq 7a). We consider the rank 2 term first.

The first invariant of the rank 2 tensor is the norm of the $\boldsymbol{v}_2$ vector defined below:

$$\boldsymbol{N}[\mathbf{2},\mathbf{2}]_0 = \frac{1}{\sqrt{5}}\{(\tilde{v}_2^0)^2+(\tilde{v}_2^1)^2+(\tilde{v}_2^2)^2\} = \frac{1}{\sqrt{5}}\{(\mu_2^0)^2+(\mu_2^2)^2\} \quad (18)$$

In the general orientation, this is simply the sum of the magnitudes of the three component terms. In the special HS orientation, this reduces to the sum of the magnitudes of the two possible distortion types, the cylindrically symmetric (tetragonal-like) $\mu_2^0$ and the orthorhombic $\mu_2^2$ distortion (see Fig 1B, *l* = 2). In general matrix theory, this value corresponds to the root mean square (rms) of the difference in the eigenvalues.[31] This measures how asymmetric the matrix is and is analogous to the von Mises stress/strain metrics.

The second invariant of the rank 2 tensor is found by the procedure of Lo & Don:[28] first, the rank 2 outer product of the $v_2^m$ vector with itself is formed, and then this is contracted with the original vector. This is given in eq 19 below.

$$\boldsymbol{N}[\boldsymbol{N}[\mathbf{2},\mathbf{2}]_{\mathbf{2}},\mathbf{2}]_0$$
$$= -\sqrt{\frac{1}{70}}\left\{2(\tilde{v}_2^0)^3 + 3(\tilde{v}_2^0)^2\tilde{v}_2^1 - 6(\tilde{v}_2^2)^2\tilde{v}_2^0 + 3\sqrt{3}(\tilde{v}_2^1)^2\tilde{v}_2^2\cos\big(2(\omega_1-\omega_2)\big)\right\}$$
$$= -\sqrt{\frac{2}{35}}\mu_2^0\{(\mu_2^0)^2 - 3(\mu_2^2)^2\} \tag{19}$$

The expression for this invariant in general coordinates is given on the upper line and is rather complicated and not readily interpreted. However, in the HS orientation given on the second line, the expression simplifies considerably and depends primarily on the difference in magnitudes of the two distortion types. In general matrix theory, this quantity measures the skewness or the cubic difference of the eigenvalues about the mean.[31] This is related to the Tresca stress/strain metrics. These two invariants, along with the three orientation functions to be discussed below, capture the five DOF for the rank 2 ST (see Table I.)

The third invariant of a 3x3 matrix is simply the trace of matrix, which is also the average of the eigenvalues of that matrix. In the current context, this corresponds to a separate rank 0 subspace of a general rank 2 tensor, which can be characterized by the scalar below:

$$v_{0(2)}^0(a) = \sqrt{\frac{2}{3}}\sum_b \left(\frac{r_{ab}^2}{\sigma^2}\right) w_{ab} = \sqrt{\boldsymbol{N}[\mathbf{0(2)},\mathbf{0(2)}]_0} \tag{20}$$

The final term in the expression above casts the term in the composite notation defined above, where the **0(2)** shows that this term arises from the rank 0 subspace of the rank 2 order, and this is distinct from the term in eq 16. The current term nominally characterizes the weighted <$r^2$> value of the neighborhood, whereas the former term was simply the sum of the weights in the neighborhood. This **0(2)** subspace is analogous to the pressure from the stress tensor or volumetric strain from the strain tensor. These elements then have a separable mechanical response compared to the deviatoric elements, which is a direct result of them belonging to a different subspace.

We normalize these invariants in the manner above and define the second order SFDs below:

$$P2I0(a) = \frac{\sqrt{\boldsymbol{N}[\mathbf{2},\mathbf{2}]_0}}{P0I0(a)} \tag{21}$$

$$P2I1(a) = -\frac{\boldsymbol{N}[\boldsymbol{N}[\mathbf{2,2}]_2,2]_0}{\boldsymbol{N}[\mathbf{2,2}]_0 * P0I0(a)} \tag{22}$$

$$P2I2(a) = \frac{\sqrt{\boldsymbol{N}[\mathbf{0(2)},\mathbf{0(2)}]_0}}{P0I0(a)} - \sqrt{\frac{3}{2}} \tag{23}$$

The terms are all defined to effectively have units of $\left(\frac{r}{\sigma}\right)^2$, and that the weighting functions are balanced between the numerator and denominator (e.g. the weighting functions appear as equal powers in both). In the current derivation, the radial function contains a node at $r = \sqrt{3}\sigma$ (see eq 7a), which gives rise to the offset term in eq 23. This nominally makes this element orthogonal to the other totally symmetric subspace $\nu_{0(0)}^0 = \sqrt{\boldsymbol{N}[\mathbf{0},\mathbf{0}]_0}$, which is equivalent to the net density function in eq 2a. For the current application (see below), all of these invariants (along with *P1I0* and *P0I0*) span a range of several tenths in unitless values. These normalization scheme therefore represents a nicely balanced basis set for characterizing LAEs.

The interpretation of the second order terms is illustrated in Figure 2, where this is very similar to a previous illustration by Kindlmann.[31] The *P2I0* metric, corresponding to the norm of the deviatoric components, quantifies the total amount of the second order distortion away from a spherically symmetric object. This quantity is non-negative and there are no constraints on its magnitude. The *P2I1* metric quantifies the skewness of that distribution, where its magnitude is constrained by the value of *P2I0*. For a positive value, the structure is elongated along one particular axis (which corresponds to its internal $\bar{z}$ axis) and contracted about equally in the other two directions. For a negative value, it is contracted along one particular axis and elongated about equally in the other two directions. Both of these limits would correspond to a tetragonal distortion, with a = b < c and a = b > c, respectively. For a value of *P2I1* = 0, there is elongation along one axis, contraction along another, with no distortion in the third direction (e.g. an orthorhombic distortion, a ≠ b ≠ c). From eqs 18 and 19, the following ratio can be defined:

$$\frac{\boldsymbol{N}[\boldsymbol{N}[\mathbf{2,2}]_2,\mathbf{2}]_0}{(\boldsymbol{N}[\mathbf{2,2}]_0)^{3/2}} = -\frac{P2I1(a)}{P2I0(a)} = \sqrt{\frac{2\sqrt{5}}{7}}\cos 3\psi \tag{24}$$

Here, $\psi$ is defined from $\cos\psi = \mu_2^0/\sqrt{(\mu_2^0)^2+(\mu_2^2)^2}$. Thus, one could also consider a ratio between those two moments as a means to characterize the invariants, similar to the approach by Kindlmann.[31] Here, however, we prefer to keep all of these terms defined in the same dimensional units for a more direct comparison between them. In this form, all of the terms can be interpreted as a net physical displacement from some reference geometry, expressed in $\left(\frac{r}{\sigma}\right)^l$ units.

The *P2I2* term is a characterization of the $r^2$ distribution of the neighborhood. In particular, it characterizes whether the weighted values of $r^2$ are larger or smaller than $3\,\sigma^2$. As such, it represents a metric of the size and distribution of the atomic positions in the neighborhood, as suggested in the figure. While this weighting with a radial node function should make this moment orthogonal to the base density metric of *P0I0*, the unevenness of the distribution of the atomic centers (i.e. there are no atoms that are close to the central atom), results in some correlation between those two terms. However, the inclusion of both terms ensures the mathematical completeness of the analysis. These two shape and one size invariants, in combination with three orientation factors to be defined later, cover the original 6 DOF contained in the 3x3 symmetric matrix. As noted by Lo&Don,[28] any higher order invariant contractions can be shown to be algebraic combinations of these three invariants.

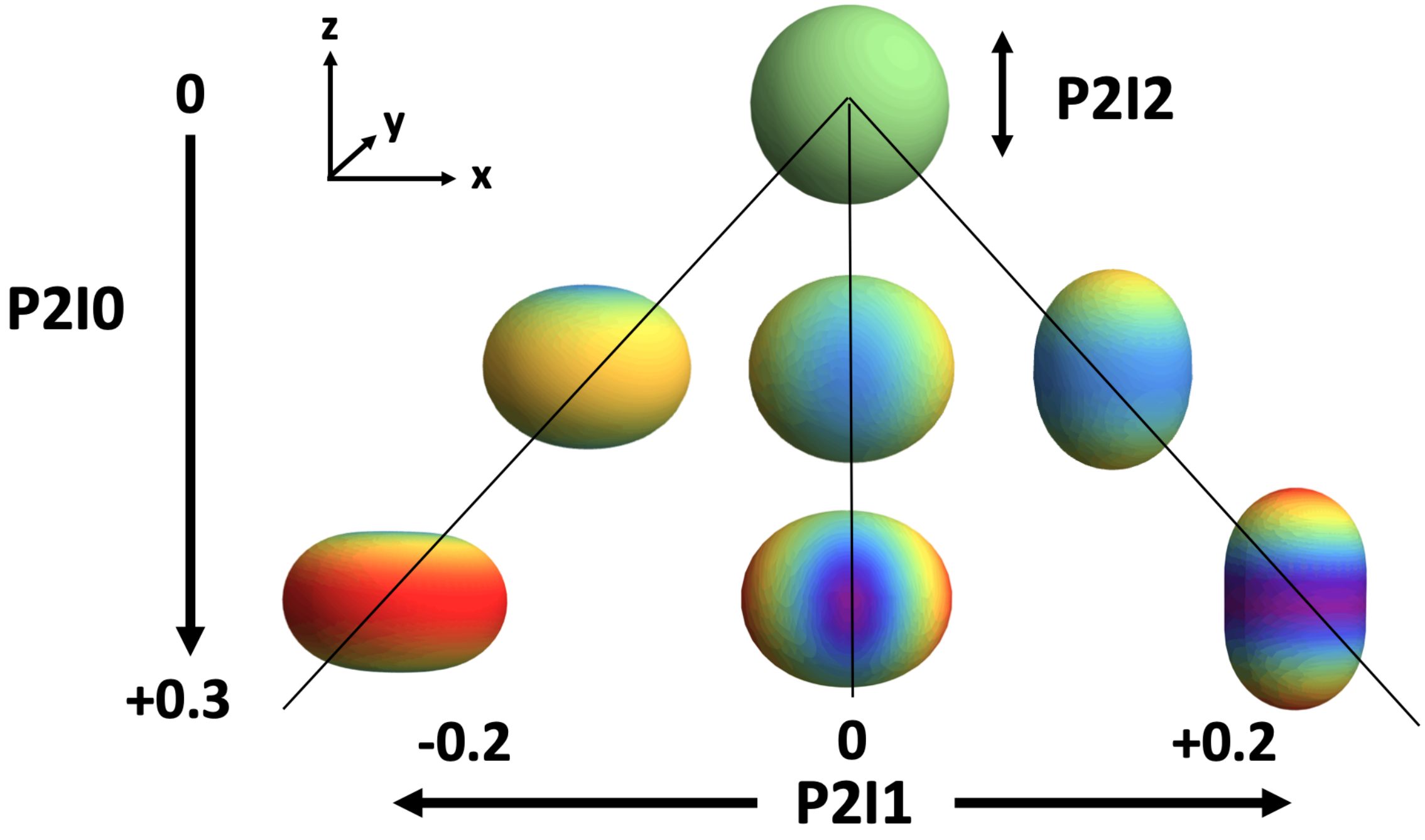

*Figure 2. Illustration of the second order Strain Functional Descriptors. The P2I0 quantity measures the net amount of second order deviatoric distortion. This quantity can have an unlimited positive value, with a value of 0 indicating that there is no second order deviatoric deformation. The P2I1 quantity measures the skewness character of that second order distortion. A positive value corresponds to a prolate distortion and a negative value corresponds to an oblate distortion, with both tending to structures that are cylindrically symmetric about the vertical axis. A value of P2I1 = 0 indicates an asymmetric, orthorhombic distortion. The magnitude of P2I1 is constrained by the value of P2I0, so there is a fan of allowable structures spreading downward from the spherically symmetric structure at the top. The colors and deformations are plotted on the same absolute scale emphasizing the growing displacements as one proceeds downward. The P2I2 quantity is a measure of the radial size of the object or net volumetric strain and measures that separate subspace.*

*Rank 3 terms*. Next, the third rank tensor is considered, building off the approach suggested by Lo&Don.[28] First, the rank 3 subspace will be examined, which is the first tensor term on the RHS of eq 7b. The first invariant is readily defined as the norm of the rank 3 spherical tensor defined in eq 25 below.

$$\boldsymbol{N}[\mathbf{3},\mathbf{3}]_0=\frac{1}{\sqrt{7}}\{(\tilde{v}_3^0)^2+(\tilde{v}_3^1)^2+(\tilde{v}_3^2)^2+(\tilde{v}_3^3)^2\}=\frac{1}{\sqrt{7}}\{(\mu_3^0)^2+(\mu_3^2)^2+(\mu_3^3)^2\} \quad (25)$$

This is the sum of the magnitudes of all of the *l=3* distortions shown in Figure 1. Similar to $\boldsymbol{N}[\mathbf{2},\mathbf{2}]_0$ (eq 18), this measures the net distortion away from spherical symmetry, except now for only the third order terms. For additional terms, the exploration of higher order rank 3 contractions would seem likely. However, these terms contain no additional information where the odd contractions are zero by symmetry (see SM eqs S3.1 and S3.2).

Consequently, only contractions to other even rank orders must then be explored for invariants. The first of these is the norm of the rank 2 contraction. When evaluated in the random orientation, this yields a complicated expression that depends on all seven variables (see SM eq S3.3). However, when rotated to the HS configuration, the dependence on the rotation angles disappears and the term depends only on the magnitudes of the components as shown below.

$$\begin{aligned}&\boldsymbol{N}[\boldsymbol{N}[\mathbf{3},\mathbf{3}]_2,\boldsymbol{N}[\mathbf{3},\mathbf{3}]_2]_0\\&=\frac{1}{84\sqrt{5}}\{16(\mu_3^0)^4+40(\mu_3^0)^2(2(\mu_3^2)^2-(\mu_3^3)^2)+25(\mu_3^3)^2(2(\mu_3^2)^2+(\mu_3^3)^2)\}\end{aligned} \quad (26)$$

We emphasize that the second term above depends on the differences in magnitudes of the $\mu_3^2$ and $\mu_3^3$ terms, demonstrating that this carries distinct information from eq 25. The skewness of this rank 2 contraction can then be calculated similarly to the skewness of the rank 2 tensor above. The result in the HS configuration is somewhat more complicated than the term above and now depends on the three magnitudes and the one rotation angle $\bar{\omega}_2$, as shown below. The presence of that latter term containing $\bar{\omega}_2$ demonstrates that this expression carries distinct information compared to eqs 25 and 26. Note that all three magnitudes must be non-zero for this term to be non-zero.

$$\begin{aligned}&\boldsymbol{N}[\boldsymbol{N}[\boldsymbol{N}[\mathbf{3},\mathbf{3}]_2,\boldsymbol{N}[\mathbf{3},\mathbf{3}]_2]_2,\boldsymbol{N}[\mathbf{3},\mathbf{3}]_2]_0\\&=-\frac{1}{588\sqrt{30}}\{4(\mu_3^0)^2-5(\mu_3^3)^2\}\\&*\{16(\mu_3^0)^4-40(\mu_3^0)^2(6(\mu_3^2)^2+(\mu_3^3)^2)+25(\mu_3^3)^2(3(\mu_3^2)^2+(\mu_3^3)^2)\}\\&+\frac{25}{49\sqrt{2}}\,\mu_3^0(\mu_3^2)^3(\mu_3^3)^2\cos(6\bar{\omega}_2)\end{aligned} \quad (27)$$

Higher order self-contractions of the rank 2 contraction generate terms that depend on the lower order contractions (SM eq S3.4), and so no further information is contained there. The norms of the rank 4 and rank 6 contractions have very similar, though distinct, forms compared to that of the rank 2 (see SM eqs S3.5 and 3.6). It was found that linear combinations of the three terms could be related to the square of the norm of the rank 3 tensor given in eq 25 (see SM eqs S3.7-9). Consequently, only two of these four terms are independent, and any pair could be chosen. Here, we select those in eqs 25 and 26, where these are both the least complicated and have physical significance that becomes apparent below.

From those relationships, it is also apparent that the rank 4 contraction is complementary to that of rank 2 (SM eq S3.7). As both terms are non-negative, where one is large (compared to $\boldsymbol{N}[\mathbf{3},\mathbf{3}]_0$) the other should be small for the same value of $\boldsymbol{N}[\mathbf{3},\mathbf{3}]_0$. As the magnitude of the skewness values are limited by the value of the base functions (either eq 26 or eq S3.5), this implies that the skewness of the rank 4 term must carry different information than that of the rank 2. The value for that term in the HS orientation is given below, with the full expression given in SM eq S3.10.

$$\boldsymbol{N}[\boldsymbol{N}[\boldsymbol{N}[\mathbf{3},\mathbf{3}]_4,\boldsymbol{N}[\mathbf{3},\mathbf{3}]_4]_4,\boldsymbol{N}[\mathbf{3},\mathbf{3}]_4]_0 = \frac{1}{11858\sqrt{13}}\left\{\begin{matrix} -648(\mu_3^0)^6+72(\mu_3^0)^4(37(\mu_3^2)^2+207(\mu_3^3)^2) \\ -12(\mu_3^0)^2(434(\mu_3^2)^4+1263(\mu_3^2)^2(\mu_3^3)^2-621(\mu_3^3)^4)+2744(\mu_3^2)^6 \\ -168(\mu_3^2)^4(\mu_3^3)^2-243(\mu_3^2)^2(\mu_3^3)^4-81(\mu_3^3)^6 \\ +3300\sqrt{15}\,\mu_3^0(\mu_3^2)^3(\mu_3^3)^2\cos(6\bar{\omega}_2) \end{matrix}\right\} \quad (28)$$

The final term in this expression (containing the angular dependence $6\bar{\omega}_2$) is the same in form as the last term in eq 27, but the preceding polynomial terms are distinct. By zeroing out each of the three magnitude terms in turn (see supplemental material SM3.11-13), it becomes apparent that eqs 27 and 28 add a dependence on the difference in magnitudes between $\mu_3^0$ and $\mu_3^2$. This is in addition to the dependence on $\bar{\omega}_2$, so these two invariants constrain the values of those 2 DOF. As there are only 4 DOF for the HS configuration, these four invariants are then sufficient to completely and uniquely describe that shape space. The skewness of the rank 6 contraction produced a similar, but more complex expansion than eq 28, and the higher order contractions that were examined were also demonstrated to be redundant to these lower order expansions.

Those four invariant terms were then normalized similarly to the second order terms, where the terms are all defined to effectively have units of $(r/\sigma)^3$ and that the weighting functions occur to equal order between the numerator and denominator. The two skewness metrics, *P3I2* and *P3I3*, are normalized with respect to their base functions. The resulting functions then have comparable magnitudes of several tenths and are defined below.

$$P3I0(a) = \frac{(\boldsymbol{N}[\mathbf{3},\mathbf{3}]_0)^{1/2}}{P0I0(a)} \quad (29)$$

$$P3I1(a) = \frac{(\boldsymbol{N}[\boldsymbol{N}[\mathbf{3},\mathbf{3}]_2,\boldsymbol{N}[\mathbf{3},\mathbf{3}]_2]_0)^{1/2}}{(\boldsymbol{N}[\mathbf{3},\mathbf{3}]_0)^{1/2} * P0I0(a)} \quad (30)$$

$$P3I2(a) = \frac{\boldsymbol{N}[\boldsymbol{N}[\boldsymbol{N}[\mathbf{3},\mathbf{3}]_2,\boldsymbol{N}[\mathbf{3},\mathbf{3}]_2]_2,\boldsymbol{N}[\mathbf{3},\mathbf{3}]_2]_0}{\boldsymbol{N}[\boldsymbol{N}[\mathbf{3},\mathbf{3}]_2,\boldsymbol{N}[\mathbf{3},\mathbf{3}]_2]_0 * (\boldsymbol{N}[\mathbf{3},\mathbf{3}]_0)^{1/2} * P0I0(a)} \quad (31)$$

$$P3I3(a) = \frac{\boldsymbol{N}[\boldsymbol{N}[\boldsymbol{N}[\mathbf{3},\mathbf{3}]_4,\boldsymbol{N}[\mathbf{3},\mathbf{3}]_4]_4,\boldsymbol{N}[\mathbf{3},\mathbf{3}]_4]_0}{\boldsymbol{N}[\boldsymbol{N}[3,3]_4,\boldsymbol{N}[\mathbf{3},\mathbf{3}]_4]_0 * (\boldsymbol{N}[\mathbf{3},\mathbf{3}]_0)^{1/2} * P0I0(a)} \quad (32)$$

The significance of the first invariant *P3I0* is straightforward. It measures the net distortion away from spherical symmetry in a third order sense; magnitudes of the different possible third order distortions shown in Figure 1B for *l=3*. Similar to *P2I0*, it is a non-negative quantity and could be infinitely large, though it samples a fairly narrow range in the current application. The significance of the other three invariants is illustrated in Figures 3 and 4.

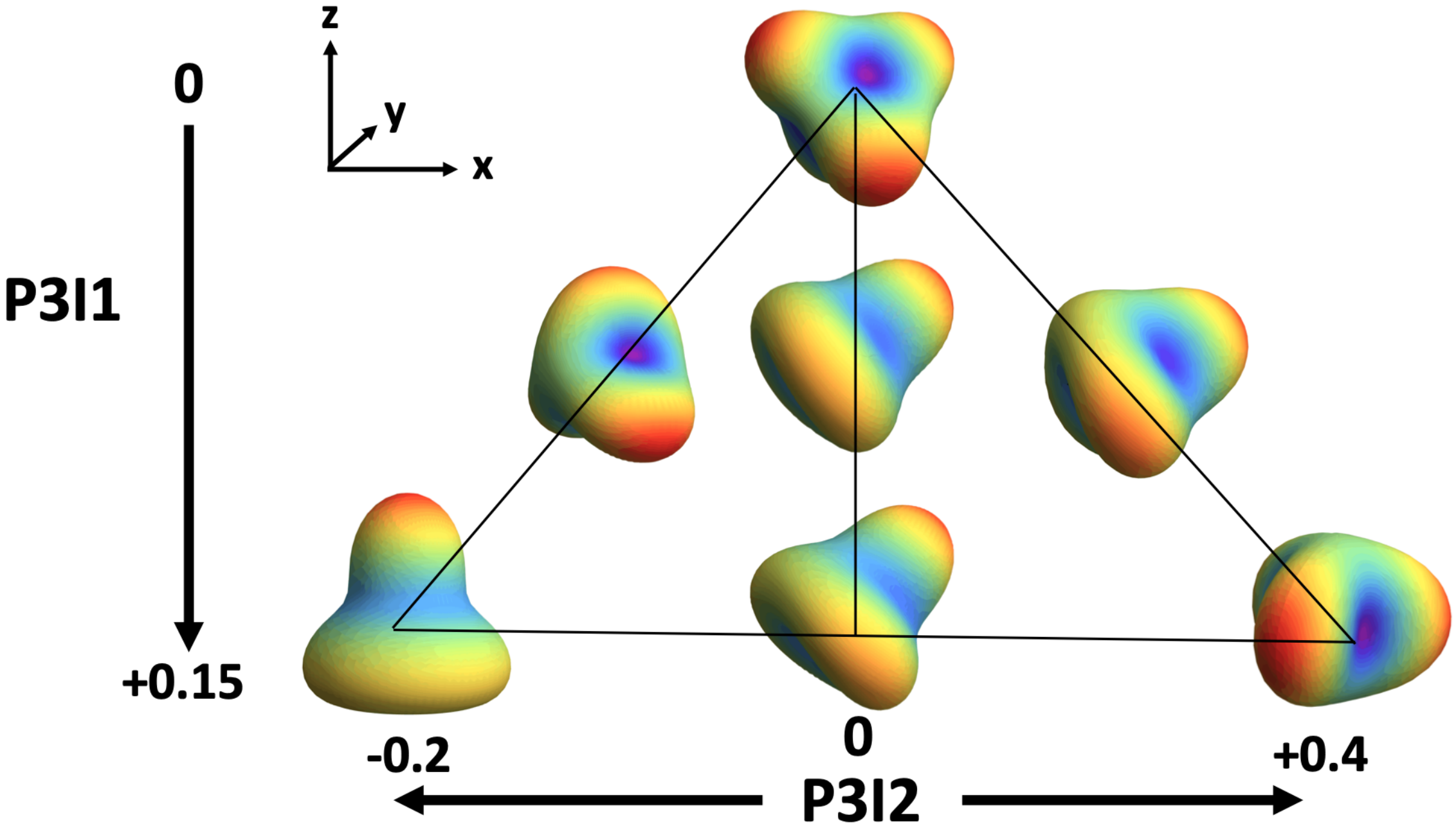

*Figure 3. Illustration of the third order Strain Functional Descriptors P3I1 and P3I2. The magnitudes of the P3I1 and P3I2 terms are constrained by the value of the P3I0 term and so have a finite range. Consequently, this can be viewed as a closed fan structure that originates at the top. The P3I1 quantity measures the net amount of distortion away from the ideal tetrahedral geometry $T_d$ ( $\tilde{v}_3^{\pm 2}$). The P3I2 quantity measures one of the skewness characteristics of that third order distortion. A positive value corresponds to a flattening distortion of the tetrahedron along one of its $C_3$ axes to generate the $D_{3h}$ ($\tilde{v}_3^{\pm 3}$) structure at its maximum limit. A negative value corresponds to a smoothing rotation about one of the $C_3$ axes to generate the bell-shaped $C_{\infty v}$ ($\tilde{v}_3^{0}$) at its minimum limit. These figures are drawn in the HS frame, with that axis frame shown in the upper left. For the P3I2 deformation, the $C_{\infty v}$ and $C_3$ rotation axes coincide with the internal z axis and the xy plane is defined by the threefold outward extensions of the $D_{3h}$ structure. Note the similarity to the P2I1 metric in Figure 2.*

Figure 3 shows the nature of the *P3I1* and *P3I2* invariants. We first note that the magnitudes of these two terms are constrained by the value of the *P3I0*. They have finite ranges limited by that, which has a value of ~0.4 for the illustrative figures shown. The *P3I1* invariant has the notable characteristic that its value is zero for a pure tetrahedral distortion, and its value becomes increasingly positive as the geometry is distorted away from tetrahedral. This can be traced back to the initial definition of the norm of the rank 2 contraction given in equation 26 (and SM eq S3.3), where it can be seen that if only the $\mu_3^2$ term (tetrahedral distortion) is non-zero, then the equation will be zero. Otherwise, this term measures the non-tetrahedral distortions to the geometry. In contrast, the norms of the redundant rank 4 (SM eq S3.5) and rank 6 (SM eq S3.6) terms have finite positive values over their complete ranges and lack such a simplistic and direct interpretation. This is a primary argument for using the rank 2 contraction as our primary metric. We particularly note that the tetrahedron is the highest symmetry object (i.e., the most ideal shape) that can be realized for third order distortions.

The *P3I2* invariant is the primary skewness metric for the *P3I1* distortion which characterizes the nature of the non-tetrahedral distortion. In the negative extreme, the structure becomes the axially symmetric $C_{\infty v}$ $\tilde{v}_3^0$ structure, while in the positive extreme the structure becomes the flatter $D_{3h}$ $\tilde{v}_3^{\pm 3}$ structure. For a value of *P3I2* = 0, the structures have a lower symmetry of $C_s$ or $C_{2v}$. In the Figure, the shapes were drawn in the *HS* configuration. This emphasizes that the $C_{\infty v}$ axis of the $\tilde{v}_3^0$ structure and $C_3$ axis of the $D_{3h}$ $\tilde{v}_3^{\pm 3}$ structure coincide with the internal z axis of the *HS* structure. Further, the three-fold lobes of the $D_{3h}$ $\tilde{v}_3^{\pm 3}$ structure define the xy plane, with one lobe aligned with the x axis. This behavior is very similar to the *P2I1* metric shown in Figure 2, which also derives from a rank 2 skewness metric.

Figure 4 shows the nature of the *P3I2* and *P3I3* invariants. We note again that the magnitudes of these two terms are constrained by the value of the *P3I0* (which is ~0.4 for this illustration). As these are both skewness metrics, they can have positive or negative values whose magnitudes would be further constrained by the norms of the base rank 2 and rank 4 contractions (eq 26 and SM eq S3.5, respectively). At the left of the figure (most negative *P3I3* values), the character of the *P3I2* metric as previously described for Figure 3 is apparent. For the most positive *P3I2* metric (upper left), the structure becomes the $D_{3h}$ $\tilde{v}_3^{\pm 3}$ extreme limit, while for the most negative value of P3I2 (lower left), the structure limits to the $C_{\infty v}$ $\tilde{v}_3^0$ limit. As *P3I3* then transits to positive values, those characteristics remain dominant as the structures morph to tetrahedral structures. They converge to the ideal tetrahedral structure at *P3I3* ≈ +0.4 (not shown). We also note that $C_3$ symmetries are apparent for both the most positive and most negative *P3I2* values. However, for *P3I2* = 0, there is a lack of $C_3$ character, and the structures can be best described as having a symmetry plane ($C_s$) and possibly $C_2$ symmetries. For increasingly positive or negative values of *P3I3*, with *P3I2 = 0*, the structures tend towards $C_{2v}$ structures. For a positive value of *P3I3*, the structure is weighted towards the vertical apex suggesting an acute isosceles triangle, while a negative value is weighted towards the two lower vertices suggesting an obtuse isosceles triangle. For a value of *P3I3* = 0, the structure has only a symmetry plane ($C_s$) with three distinct lobes resembling a scalene triangle. Thus, *P3I3* metric primarily characterizes the bending motion of the structure. In the vertical direction, going towards either positive or negative values of *P3I2*, $C_3$ symmetry axes become more pronounced.

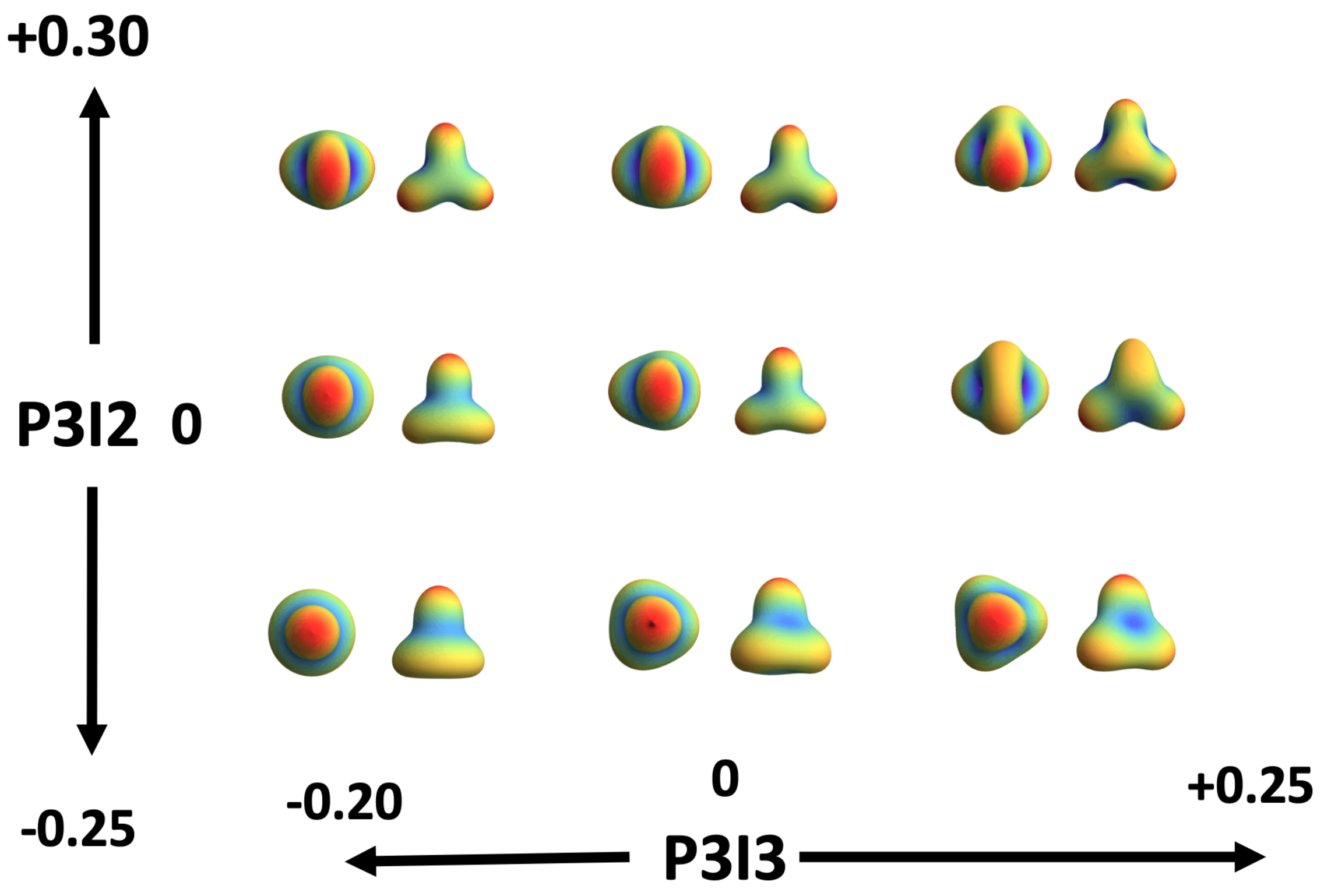

*Figure 4. Illustration of the third order Strain Functional Descriptors P3I2 and P3I3. A total of 9 geometries are shown, with two different orthogonal perspectives being shown to fully illustrate the geometries. The pairs are located at roughly their positions in the P3I2/P3I3 space. The magnitudes of both metrics are constrained by the value of P3I0, which is ~0.4 for this illustration. Both of these are skewness metrics that characterize the nature of the non-tetrahedral distortion and can have positive or negative values. Note that for the most positive or most negative P3I2 values, the structures have identifiable $C_3$ symmetries. For P3I2 = 0 values, the structures have lower symmetries. For P3I3 = 0, the symmetry is lowered down to just a symmetry plane $C_s$ structure. As P3I3 becomes more positive or negative, the structures acquire $C_{2v}$ character. For the most positive P3I3 value (~0.4), the structures all converge to the ideal tetrahedron.*

The other distortion that arises from the 3rd order terms comes from the $\nu_{1(3)}^{m}$ subspace (see Table I), the second tensor term on the RHS of eq 7b. These terms are dipolar in nature (Figure 1, *l*=1), but with $r^3$ weighting and one radial node rather than $r^1$ with no radial nodes. The moments can be described with a vector with three independent values. Those can be transformed into one invariant (the magnitude) and two orientation angles. The magnitude is defined analogously to the previous dipole definition as below

$$\boldsymbol{N}[\mathbf{1(3)},\mathbf{1(3)}]_0 = \frac{1}{\sqrt{3}}\left\{\left(\tilde{\nu}_{1(3)}^{0}\right)^2 + \left(\tilde{\nu}_{1(3)}^{1}\right)^2\right\} = \frac{1}{\sqrt{3}}\left(\mu_{1(3)}^{0}\right)^2 \quad (33)$$

This is then normalized similar to the other invariants as shown in eq 34.

$$P3I4(a) = \frac{(\boldsymbol{N}[\mathbf{1(3)},\mathbf{1(3)}]_0)^{1/2}}{P0I0(a)} \quad (34)$$

The units here are then *(r/σ)*$^3$ with equal weighting factors in both the numerator and denominator. This quantity will have a non-negative value whose value is otherwise unconstrained, though in the current application the values are typically several tenths. We note that this term is not constrained by the value of *P3I0*, as are the other three third order invariants, but is a completely independent descriptor for that distinct subspace.

These third order shape invariant terms are then analogous to the standard strain gradient terms that are defined in solid mechanics.[61,62] This is shown in Figure 5 where we show the comparison between our representation of the Gaussian-weighted deformations and mechanics representations of the strain-gradients.[61]

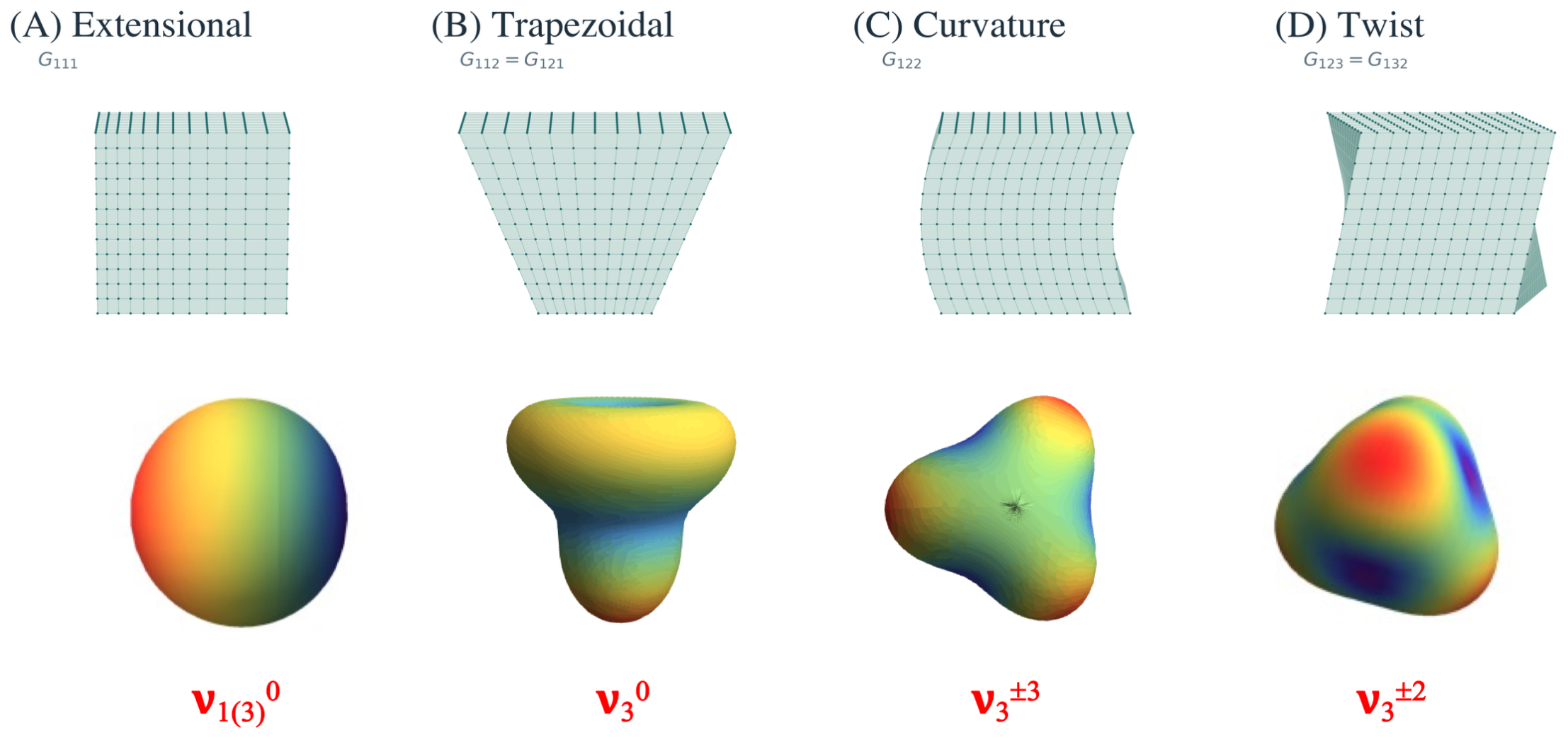

*Figure 5. Comparison of strain-gradient visualizations for a finite element simulation (ref 61) and for the current deformations of spheres. The reddish regions in the lower panels show where the density of the material has been increased, and the bluish regions are where the density has been decreased. Alternatively, this can be interpreted as to regions where material has been added / removed. OpenAI GPT-6.1 Sol was used to visualize the strain gradient deformations.*

There is a fair level of correspondence between these basic deformation patterns. For the extensional gradient (first column, A), there is an increased density at one end of the sample, a decreased density at the other, with no gradients in the transverse direction. This is well reproduced by the gradient term captured in $\nu_{1(3)}^0$. For the trapezoidal distortion (B), there are symmetric transverse gradients which result in an outward expansion of the material at the top and a densification at the bottom similar to the bell-shaped distortion of $\nu_3^0$. For the curvature distortion (C), material is moved forward in the middle of the left, while it is pulled back at the two right corners. This is similar to the trigonal distortion of $\nu_3^{\pm3}$. The twist distortion (D) is often thought of as twisting a cube. However, it can also be thought of as increasing the density or adding material at alternate corners of the cube and decreasing the density at the others. This amounts to a tetrahedral distortion of the cube, captured by the $\nu_3^{\pm 2}$ distortion. The discrepancies partially result from the displacements not being fully orthogonal to one another, and the Gaussian-weighting will foreshorten the displacements as a function of distance from the center. Still there is a clear correspondence between these deformations. We emphasize that in our current analysis, the extensional strain gradient corresponds to a separate subspace than the other third order displacements. Thus, the mechanics of its displacement should be separable in the same manner that volumetric strain and stress (pressure) are separable from their deviatoric components.

*Rank 4 terms.* Next, the fourth rank tensor is considered, building off the approach developed above. First, the rank 4 subspace will be examined (the first tensor term on the RHS of eq 7c; see also Table I), and terms analogous to those derived for the second and third rank tensors will be examined. The first invariant is readily defined as the norm of the rank 4 spherical tensor defined in eq 35 below.

$$\boldsymbol{N}[\mathbf{4},\mathbf{4}]_0 = \frac{1}{3}\{(\tilde{\nu}_4^0)^2+(\tilde{\nu}_4^1)^2+(\tilde{\nu}_4^2)^2+(\tilde{\nu}_4^3)^2+(\tilde{\nu}_4^4)^2\}$$
$$= \frac{1}{3}\{(\mu_4^0)^2+(\mu_4^2)^2+(\mu_4^3)^2+(\mu_4^4)^2\} \tag{35}$$

This is the sum of the magnitudes of all of the *l=4* distortions shown in Figure 1. The expansions of the rank 2 contractions are then explored, similar to the third order terms above. The first of these is the norm of the rank 2 contraction. When evaluated in the random orientation, this yields a complicated expression that depends on all nine variables (SM eq S4.1). However, when rotated to the *HS* configuration, the dependence on the rotation angles simplifies somewhat to a dependence on just $\bar{\omega}_2$ and $\bar{\omega}_2 - 3\bar{\omega}_3$ (SM eq S4.2). For a level of clarity, we just present the functional form in eq 36.

$$\boldsymbol{N}[\boldsymbol{N}[\mathbf{4},\mathbf{4}]_2,\boldsymbol{N}[\mathbf{4},\mathbf{4}]_2]_0$$
$$= F\begin{bmatrix} (\mu_4^0)^4, (\mu_4^2)^4, (\mu_4^3)^4, (\mu_4^4)^4, \\ (\mu_4^0)^2(\mu_4^2)^2, -(\mu_4^0)^2(\mu_4^3)^2, -(\mu_4^0)^2(\mu_4^4)^2, (\mu_4^2)^2(\mu_4^3)^2, -(\mu_4^2)^2(\mu_4^4)^2, (\mu_4^3)^2(\mu_4^4)^2 \\ \mu_4^0(\mu_4^2)^2\mu_4^4\cos(4\bar{\omega}_2), \mu_4^2(\mu_4^3)^2\mu_4^4\cos(2\{\bar{\omega}_2 - 3\bar{\omega}_3\}) \end{bmatrix} \tag{36}$$

Here we only list the polynomial terms (and their sign dependence where appropriate) and omit the numerical coefficients. This makes it obvious that all terms are quartic polynomials, and that only a small number of all possible terms are present here. This is an important aspect for recognizing the completeness of the invariants. The skewness of the rank 2 contraction is also then calculated similar to the skewness of the rank 2 contraction of the third order tensor above. The result in the *HS* configuration is somewhat more complicated than the term above, and the full expression is given in SM eq S4.3. The functional form is given below where this is now a sixth order polynomial, and again it is seen that not all possible terms are present.

$$
\boldsymbol{N}[\boldsymbol{N}[\boldsymbol{N}[\mathbf{4},\mathbf{4}]_2,\boldsymbol{N}[\mathbf{4},\mathbf{4}]_2]_2,\boldsymbol{N}[\mathbf{4},\mathbf{4}]_2]_0
= F\begin{bmatrix}
(\mu_4^0)^6,(\mu_4^2)^6,-(\mu_4^3)^6,-(\mu_4^4)^6,-(\mu_4^0)^4(\mu_4^2)^2,-(\mu_4^0)^4(\mu_4^3)^2,-(\mu_4^0)^4(\mu_4^4)^2,\\
-(\mu_4^0)^2(\mu_4^2)^4,(\mu_4^0)^2(\mu_4^3)^4,(\mu_4^2)^4(\mu_4^3)^2,-(\mu_4^2)^4(\mu_4^4)^2,-(\mu_4^3)^4(\mu_4^2)^2,\\
-(\mu_4^3)^4(\mu_4^4)^2,(\mu_4^2)^2(\mu_4^4)^4,-(\mu_4^3)^2(\mu_4^4)^4,(\mu_4^0)^2(\mu_4^2)^2(\mu_4^3)^2,(\mu_4^0)^2(\mu_4^2)^2(\mu_4^4)^2,\\
(\mu_4^0)^2(\mu_4^3)^2(\mu_4^4)^2,-(\mu_4^2)^2(\mu_4^3)^2(\mu_4^4)^2,\\
[-(\mu_4^0)^2,-(\mu_4^2)^2,-(\mu_4^3)^2,(\mu_4^4)^2](\mu_4^0)(\mu_4^2)^2(\mu_4^4)\mathrm{Cos}[4\bar{\varpi}_2],\\
[(\mu_4^0)^2,-(\mu_4^2)^2,-(\mu_4^3)^2,-(\mu_4^4)^2](\mu_4^2)(\mu_4^3)^2(\mu_4^4)\mathrm{Cos}[2\{\bar{\varpi}_2-3\bar{\varpi}_3\}],\\
-(\mu_4^0)(\mu_4^2)^3(\mu_4^3)^2\mathrm{Cos}[6\{\bar{\varpi}_2-3\bar{\varpi}_3\}],-(\mu_4^0)(\mu_4^2)(\mu_4^3)^2(\mu_4^4)^2\mathrm{Cos}[2\{\bar{\varpi}_2+3\bar{\varpi}_3\}]
\end{bmatrix} \quad (37)
$$

These terms are normalized to their invariant forms as has previously been done for the lower order forms.

$$
P4I0(a) = \frac{(\boldsymbol{N}[\mathbf{4},\mathbf{4}]_0)^{1/2}}{P0I0(a)} \quad (38)
$$

$$
P4I1(a) = \frac{(\boldsymbol{N}[\boldsymbol{N}[\mathbf{4},\mathbf{4}]_2,\boldsymbol{N}[\mathbf{4},\mathbf{4}]_2]_0)^{1/2}}{(\boldsymbol{N}[\mathbf{4},\mathbf{4}]_0)^{1/2} * P0I0(a)} \quad (39)
$$

$$
P4I2(a) = \frac{\boldsymbol{N}[\boldsymbol{N}[\boldsymbol{N}[\mathbf{4},\mathbf{4}]_2,\boldsymbol{N}[\mathbf{4},\mathbf{4}]_2]_2,\boldsymbol{N}[\mathbf{4},4]_2]_0}{\boldsymbol{N}[\mathbf{4},\mathbf{4}]_2,\boldsymbol{N}[\mathbf{4},4]_2]_0 * (\boldsymbol{N}[\mathbf{4},\mathbf{4}]_0)^{1/2} * P0I0(a)} \quad (40)
$$

The units here are now *(r/σ)*$^4$ with equal weighting factors occurring in both the numerator and denominator. Similar to the other orders, the *P4I0* metric captures the net deformation away from spherical symmetry in any of the fourth order deformations. The characteristics of the *P4I1* and *P4I2* metrics are illustrated in Figure 6.

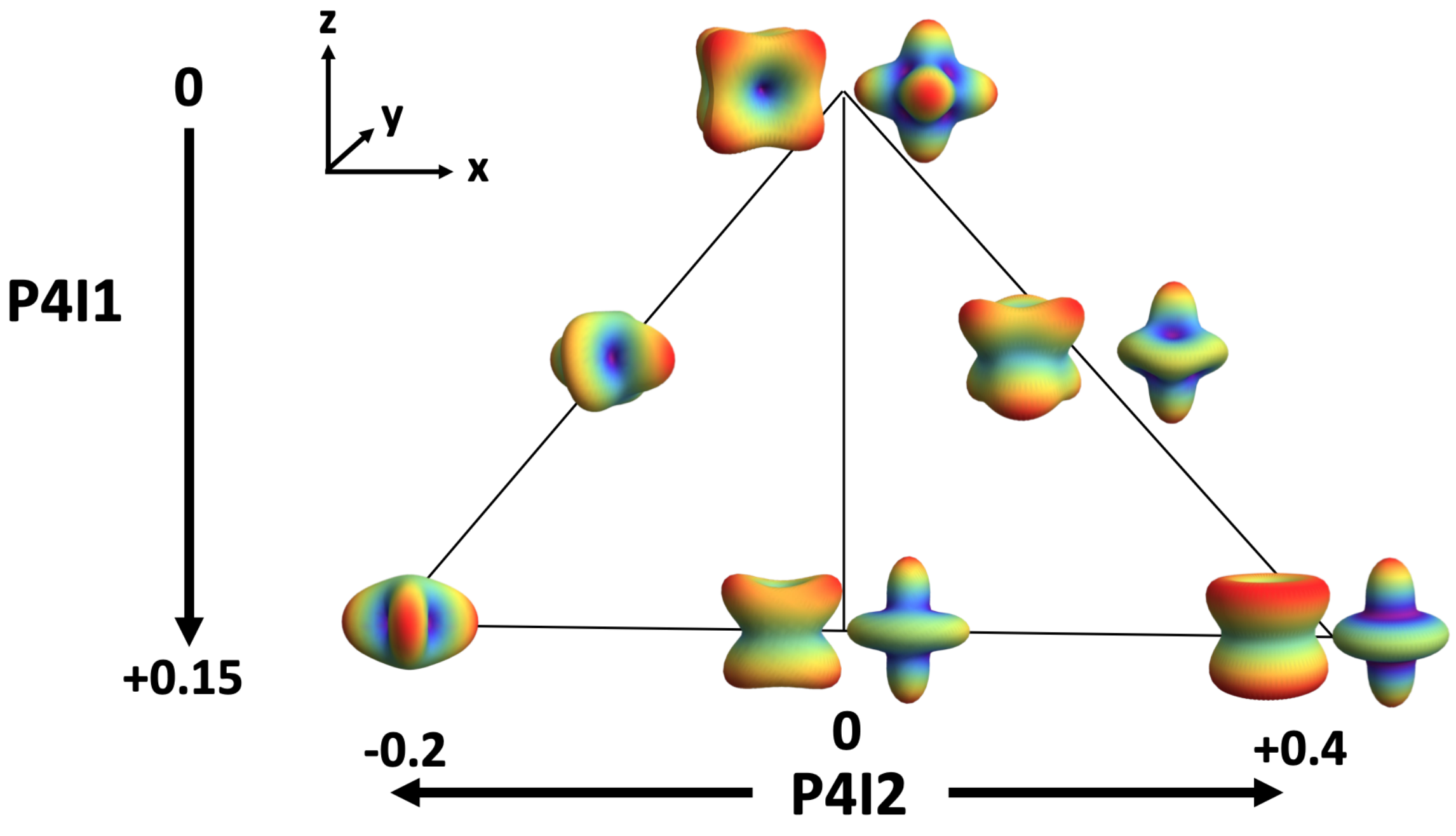

*Figure 6. Illustration of the fourth order Strain Functional Descriptors P4I1 and P4I2. The P4I1 quantity measures the net amount of distortion away from the ideal octahedral geometry of $O_h$ symmetry shown at the top. Here, this can be either a body-centered cubic structure on the left, or the simple cubic structure on the right. The P4I2 quantity measures one of the skewness characteristics of the fourth order distortions. A negative value corresponds to a flattening distortion of the octahedron along one of its three $C_4$ axes to generate the $D_{4h}$ ($\tilde{v}_4^{\pm4}$) structure in the lower left. Basically, one of the $C_4$ distortions is large, and the other two are diminished to zero. A positive value corresponds to diminishing the magnitude of the distortion about one of the $C_4$ axes while maintaining the other two as large to generate the cylindrically-symmetric $D_{\infty h}$ ($\tilde{v}_4^{0}$)* shown in the lower right. *These figures are drawn in the HS frame, with that axis frame shown in the upper left. For the P4I2 deformation, the $D_{\infty h}$ and $C_4$ rotation axes coincide with the internal z axis and the xy plane is defined by the fourfold outward extensions of the $D_{4h}$ structure. Note the similarity to the P2I1 metric in Figure 2 and the P3I1 and P3I2 metrics in Figure 3. The magnitudes of the P4I1 and P4I2 terms will be constrained by the value of the P4I0 term, so this is a closed-fan pattern that originates from the symmetrical structures at the top.*

Figure 6 shows the nature of the *P4I1* and *P4I2* invariants. We first note that the magnitudes of these two terms are constrained by the value of the *P4I0*. They have finite ranges limited by that, which has a value of ~0.4 for the illustrative figures shown. The *P4I1* invariant has the notable characteristic that its value is zero for a pure octahedral distortion, and its value becomes increasingly positive as the geometry is distorted away from octahedral, $O_h$. We particularly note that the octahedron is the highest symmetry object (i.e., the most ideal shape) that can be realized for fourth order distortions. Within this framework, there are two particular $O_h$ variants that cannot yet be distinguished. These are the body-centered cubic (*BCC*) structure shown to the left, and the simple cubic (*SC*) structure shown to the right. Another invariant, to be defined below, is required to distinguish between those structures.

The *P4I2* invariant is the primary skewness metric for the *P4I1* distortion which characterizes the nature of the non-octahedral distortion. In the positive extreme, the structure becomes the axially symmetric $D_{\infty h}$ ($\tilde{v}_4^{0}$) structure, while in the negative extreme the structure becomes the flatter $D_{4h}$ ($\tilde{v}_4^{\pm4}$) structure. For a value of *P4I2* = 0, the structures have a lower symmetry of $D_{2h}$ or $C_{2h}$. In the Figure, the shapes were drawn in the *HS* configuration. This emphasizes that the $D_{\infty h}$ axis of the $\tilde{v}_4^{0}$ structure and $C_4$ axis of the $D_{4h}$ $\tilde{v}_4^{\pm4}$ structure coincide with the internal z axis of the *HS* structure. Further, the four-fold lobes of the $D_{4h}$ $\tilde{v}_4^{\pm4}$ structure define the *xy* plane, with one lobe aligned with the *x* axis. This behavior is very similar to the *P2I1* metric shown in Figure 2, which also derives from a rank 2 skewness metric. For a positive value of *P4I2*, the two variant structures of octahedral geometry persist and are shown. For a negative value of *P4I2*, these structural differences diminish and correspond more to a rotation angle difference.

Similar to what was observed for the moments of the third rank tensor (see SM eqs S3.7-9), it is not difficult to show that the $\boldsymbol{N}[\boldsymbol{N}[\mathbf{4}, \mathbf{4}]_x, \boldsymbol{N}[\mathbf{4}, \mathbf{4}]_x]_0$ moments for *x = 4, 6, 8* can be simply related back the moment for *x = 2* and $\boldsymbol{N}[4,4]_0$ (SM eqs S4.4-6). Consequently, they contain no additional information. Additionally, the terms with $x$ = *odd* are all zero by symmetry. Similar to the third rank tensor, the skewness of the rank 4 contraction is then constructed and the signs of the various terms are given below, with the full expression given in the supplemental material (SM eq S4.7).

$$\boldsymbol{N}[\boldsymbol{N}[\boldsymbol{N}[\mathbf{4},\mathbf{4}]_4,\boldsymbol{N}[\mathbf{4},\mathbf{4}]_4]_4,\boldsymbol{N}[\mathbf{4},\mathbf{4}]_4]_0$$

$$= F\begin{bmatrix} (\mu_4^0)^6, -(\mu_4^2)^6, -(\mu_4^3)^6, (\mu_4^4)^6, -(\mu_4^0)^4(\mu_4^2)^2, -(\mu_4^0)^4(\mu_4^3)^2, (\mu_4^0)^4(\mu_4^4)^2, \\ (\mu_4^0)^2(\mu_4^2)^4, -(\mu_4^2)^4(\mu_4^3)^2, (\mu_4^2)^4(\mu_4^4)^2, (\mu_4^0)^2(\mu_4^3)^4, -(\mu_4^2)^2(\mu_4^3)^4, -(\mu_4^3)^4(\mu_4^4)^2, \\ (\mu_4^0)^2(\mu_4^4)^4, -(\mu_4^2)^2(\mu_4^4)^4, (\mu_4^3)^2(\mu_4^4)^4, (\mu_4^0)^2(\mu_4^2)^2(\mu_4^3)^2, -(\mu_4^0)^2(\mu_4^2)^2(\mu_4^4)^2, \\ -(\mu_4^0)^2(\mu_4^3)^2(\mu_4^4)^2, -(\mu_4^2)^2(\mu_4^3)^2(\mu_4^4)^2, \\ [(\mu_4^0)^2, -(\mu_4^2)^2, -(\mu_4^3)^2, (\mu_4^4)^2](\mu_4^0)(\mu_4^2)^2(\mu_4^4)\mathrm{Cos}[4\varpi_2], (\mu_4^2)^4(\mu_4^4)^2\mathrm{Cos}[8\varpi_2] \\ [-(\mu_4^0)^2, (\mu_4^2)^2, (\mu_4^3)^2, (\mu_4^4)^2](\mu_4^2)(\mu_4^3)^2(\mu_4^4)\mathrm{Cos}[2\{\varpi_2 - 3\varpi_3\}], \\ (\mu_4^0)(\mu_4^2)^3(\mu_4^3)^2\mathrm{Cos}[6\{\varpi_2 - \varpi_3\}], (\mu_4^0)(\mu_4^2)(\mu_4^3)^2(\mu_4^4)^2\mathrm{Cos}[2\{\varpi_2 + 3\varpi_3\})] \end{bmatrix} \quad (41)$$

This expression contains many terms in common with the skewness term given in eq 40 where they are both sixth order polynomial forms. However, there are differences in the signs of some terms as well as some terms that are absent in one or the other. This demonstrates that they capture different attributes. For example, the first four terms (which would describe the shape when there is distortion in only one of the fundamental modes) have different sign patterns: both terms would be positive if a $\mu_4^0$ distortion is present and would be negative if a $\mu_4^3$ distortion is present. However, the two terms would have opposite signs if only a $\mu_4^2$ or a $\mu_4^4$ distortion is present. Also of particular note is the last term in the fifth line: $(\mu_4^2)^4(\mu_4^4)^2\mathrm{Cos}[8\varpi_2]$. This term will be non-zero only when contributions from both $\mu_4^2$ and $\mu_4^4$ are present. And if only those two terms are non-zero, then almost all of the other terms will be zero; only $-(\mu_4^2)^6$, $(\mu_4^2)^4(\mu_4^4)^2$and $-(\mu_4^2)^2(\mu_4^4)^4$ would survive. Consequently, the *P4I3* term should have a strong dependence on $\varpi_2$, the rotation angle of the $\mu_4^2$ contribution relative to $\mu_4^4$ contribution. These characteristics identify this term as the next invariant of the fourth order expansion, the normalized version of which is given in eq 42. It is normalized similarly to the analogous term in eq 35, measuring the skewness of the rank 4 contraction.

$$P4I3(a) = \frac{\boldsymbol{N}[\boldsymbol{N}[\boldsymbol{N}[\mathbf{4},\mathbf{4}]_4,\boldsymbol{N}[\mathbf{4},\mathbf{4}]_4]_4,\boldsymbol{N}[\mathbf{4},\mathbf{4}]_4]_0}{\boldsymbol{N}[\boldsymbol{N}[\mathbf{4},\mathbf{4}]_4,\boldsymbol{N}[\mathbf{4},\mathbf{4}]_4]_0 * (\boldsymbol{N}[\mathbf{4},\mathbf{4}]_0)^{1/2} * P0I0(a)} \quad (42)$$

The characteristic behavior of this metric is illustrated in Figure 7, where it is plotted versus the *P4I1* metric. The base function of *P4I3*, the numerator in eq 42, is rigorously the skewness metric of the function $\boldsymbol{N}[\boldsymbol{N}[\mathbf{4},\mathbf{4}]_4,\boldsymbol{N}[\mathbf{4},\mathbf{4}]_4]_0$. That is a complementary function of $\boldsymbol{N}[\boldsymbol{N}[\mathbf{4},\mathbf{4}]_2,\boldsymbol{N}[\mathbf{4},\mathbf{4}]_2]_0$, which is the base function of *P4I1*. Consequently, *P4I3* spans a range of values and structures for *P4I1* ≈ 0, but limits to a narrow range as *P4I1* increases in value (limited by the value of *P4I0*). This is similar to the behavior of P3I2 and P3I3. The defining characteristic of the $O_h$ structures in the upper left, with an extreme positive value of *P4I3*, is that they have 3 perpendicular and equivalent $C_4$ axes, which then generate 6 equivalent $C_2$ axes and 3 equivalent and perpendicular $C_3$ axes. The defining characteristic of the high symmetry structures in the lower left is that they have *no* $C_4$ axes. The $D_{3d}$ $\tilde{v}_4^3$ structure has one $C_3$ axis with 3 coplanar $C_2$ axes that are perpendicular to the $C_3$ direction, and the $D_{2h}$ $\tilde{v}_4^2$ structure has 3 perpendicular but inequivalent $C_2$ axes. Both structures can be viewed as an overall trigonal distortion of the octahedral geometry. As the value of *P4I1* increases from its minimal value, the structures with a positive value of *P4I3* degrade into $D_{4h}$ structures with only one $C_4$ axis, but with 4 $C_2$ axes and *no* $C_3$ axes. In contrast, the structures with a negative value of *P4I3* maintain the lower symmetry $D_{3d}$ structure with one $C_3$ axis, 3 $C_2$ axes and *no* $C_4$ axes. Along the midline of *P4I3* = 0, the structures have only $C_{2h}$ symmetry (a $C_2$ axis with a perpendicular reflection plane), with no obvious preference for higher $C_4$ or $C_3$ symmetry. Those tendencies develop for further distortions where positive or negative values of *P4I3* develop, respectively.

The effect of the rotation angle $\overline{\omega}_2$ is illustrated by the structures contained in the yellow box in the middle of Figure 7. These structures are all a combination of (2/3) $\mu_4^2$ distortion and (1/3) $\mu_4^4$ distortion, with the only difference being the rotation angle between those two factors. If that phase angle is *0*, the nearly *BCC* $D_{4h}$ structure results (upper right), and if the phase angle is $\pi/4$, the nearly *SC* $D_{4h}$ structure results (upper left). If the phase angle is $\pi/8$, the nearly trigonal $D_{2h}$ structure at the bottom results. For a phase angle of $\pi/16$ or $3\pi/16$, then the $C_{2h}$ structures in the middle result. This further illustrates the points made above.

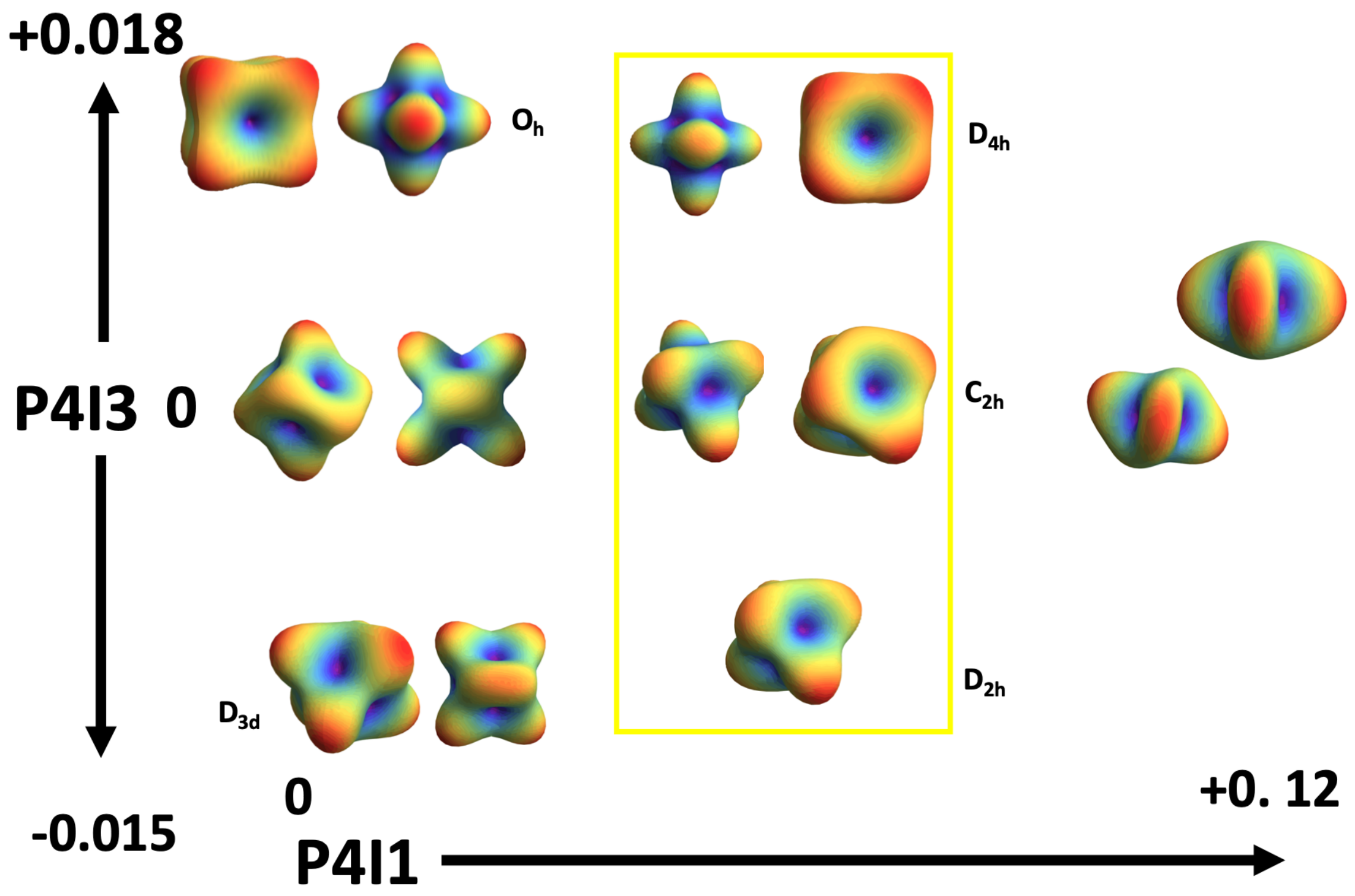

*Figure 7. Illustration of the Strain Functional Descriptor P4I3. Similar to P4I2, this is a skewness metric related to the base P4I1 quantity, though it is more directly related to a complement of that quantity. For a large value of P4I1, the range of P4I3 becomes quite narrow and the limiting structure is a distorted version of the $D_{4h}$ ($\tilde{v}_4^4$) structure (at right). The maximum range for P4I3 occurs when the P4I1 is ~0. There, a positive value limits to*

*the two possible octahedral structures (BCC and SC) in the upper left. Increasing the value of P4I1 while maintaining a positive value of P4I3 maintains the $C_4$ axis symmetry, and a primarily $D_{4h}$ symmetry structure, while degrading the other octahedral symmetries. A negative value of P4I3, corresponds to two possible high symmetry distortions of the octahedron, the trigonally distorted $D_{3d}$ ($\tilde{v}_4^3$) and the orthorhombically distorted $D_{2h}$ ($\tilde{v}_4^2$) structures shown in the lower left. A range of structures with P4I3 = 0 and increasing values of P4I1 is shown along the center line. These structures have only $C_{2h}$ symmetry, and their similarity to an octahedral structure is seen to diminish in progressing to the right. The structures contained in the yellow box are combinations of $\tilde{v}_4^2$ and $\tilde{v}_4^4$ distortions with different phase angles.*

As with the third order expansions, it becomes apparent that the other skewness metrics $\boldsymbol{N[N[N[4,4]_x,N[4,4]_x]_x,N[4,4]_x]_0}$ with $x = 6$ and $8$ can be simply related back the lower order terms with $x = 2$ and $4$. So mixed order expansions and expansions with an odd number of terms were then explored. A fivefold contraction of rank 2 contractions gives the form in eq 43 below, with the full expression given in SM eq S4.8. (This form is equivalent to the mixed fivefold contraction of rank 2 and rank 4 contractions as shown.) One aspect is that the first collection of terms is multiplied by $\mu_4^0$ such that it will only be non-zero if $\mu_4^0 \neq 0$, and that the sign of that term will be reflected by the sign of $\mu_4^0$. This is distinct from all of the previous terms discussed so far. The second aspect is the form of the last two terms in the final line. The last term in particular will be non-zero only if both the $\mu_4^2$ and the $\mu_4^3$ distortions are non-zero. Further if only those two factors are non-zero, then only that term will be non-zero and its sign will be determined by the phase angle difference $\bar{\omega}_2 - \bar{\omega}_3$.

$$\boldsymbol{N}[\boldsymbol{N}[\boldsymbol{N}[\mathbf{4},\mathbf{4}]_2,\mathbf{4}]_2,\boldsymbol{N}[\mathbf{4},\mathbf{4}]_2]_0 = \boldsymbol{N}[\boldsymbol{N}[\boldsymbol{N}[\mathbf{4},\mathbf{4}]_2,\boldsymbol{N}[\mathbf{4},\mathbf{4}]_2]_4,\mathbf{4}]_0 =$$

$$F\begin{bmatrix} \mu_4^0 \begin{bmatrix} (\mu_4^0)^4, -(\mu_4^2)^4, -(\mu_4^2)^2(\mu_4^3)^2, (\mu_4^2)^2(\mu_4^4)^2, (\mu_4^3)^4, -(\mu_4^3)^2(\mu_4^4)^2, (\mu_4^4)^4, \\ -(\mu_4^0)^2(\mu_4^2)^4, -(\mu_4^0)^2(\mu_4^3)^2, -(\mu_4^0)^2(\mu_4^4)^4 \end{bmatrix}, \\ [(\mu_4^2)^2\mu_4^4((\mu_4^0)^2, -(\mu_4^2)^2, (\mu_4^3)^2, (\mu_4^4)^2]\mathrm{Cos}[4\bar{\omega}_2], \\ -\mu_4^0\mu_4^2(\mu_4^3)^2\mu_4^4\mathrm{Cos}[2(\bar{\omega}_2 - 3\bar{\omega}_3)], (\mu_4^2)^3(\mu_4^3)^2\mathrm{Cos}[6(\bar{\omega}_2 - \bar{\omega}_3)] \end{bmatrix} \quad (43)$$

These unique characteristics identify this term as a non-redundant invariant. Since the numerator is formed by a series of rank 2 contractions, it is scaled by the norm of the rank 2 contraction raised to the 3/2 power, as well as the overall norm of the rank 4 term and the sum of the weighting terms. The resulting *P4I4* invariant is defined in eq 44.

$$P4I4(a) = \frac{\boldsymbol{N}[\boldsymbol{N}[\boldsymbol{N}[\mathbf{4},\mathbf{4}]_2,\mathbf{4}]_2,\boldsymbol{N}[\mathbf{4},\mathbf{4}]_2]_0}{(\boldsymbol{N}[\boldsymbol{N}[\mathbf{4},\mathbf{4}]_2,\boldsymbol{N}[\mathbf{4},\mathbf{4}]_2]_0)^{3/2} * (\boldsymbol{N}[\mathbf{4},\mathbf{4}]_0)^{1/2} * P0I0(a)} \quad (44)$$

The significance of this term is illustrated in Figure 8, where the deformations in the *P4I4* term are plotted vs the *P4I1* term. For the extreme values of *P4I1*, the structures limit to values of *P4I4* = 0. These are the equivalent *SC* and *BCC* $O_h$ structures for *P4I1* = 0 at the left, and the $D_{4h}$ $\tilde{v}_4^4$ structure at the right. For intermediate values of *P4I1* however is where the significance of the *P4I4* becomes apparent. As suggested above, the maximum range of *P4I4* occurs when only the $\tilde{v}_4^0$ deformation is present. It is positive for when $\tilde{v}_4^0$ is positive (density present along the z axis and cylindrically distributed in the xy plane), and negative for a negative value of $\tilde{v}_4^0$ (density cylindrically distributed intermediate between the z axis and the xy plane).

The effect of the $\overline{\omega}_2 - \overline{\omega}_3$ phase angle is illustrated by the structures highlighted in the yellow box. These structures are all a combination of ~(1/2) $\mu_4^2$ distortion and ~(1/2) $\mu_4^3$ distortion, with the only difference being the rotation angle between those two factors. If that phase angle is *0*, the nearly $D_{3d}$ distortion of the *SC* structure results (upper inset, $C_{2h}$ actual symmetry). Here the alternate four in-plane apices of the SC structure distort above and below the plane and lose their radial distinction, as they trend towards the $+\tilde{v}_4^0$ structure. If the phase angle is *π/6*, the nearly $D_{3d}$ distortion of the *BCC* structure results (lower inset, $C_{2h}$ actual symmetry). Here, two pairs of diagonally opposite neighbors are coalescing into single nodes and losing their overall radial distinction, as they trend towards the $-\tilde{v}_4^0$ structure. If the phase angle is *π/12* or *π/4*, the low symmetry $C_{2h}$ structure in the middle results, where there are three distinct pairs of vertices apparent. Along the midline of *P4I4* = 0 in general are similar low symmetry structures, though these do limit into high symmetry structures for the extreme values of *P4I1* as noted above.

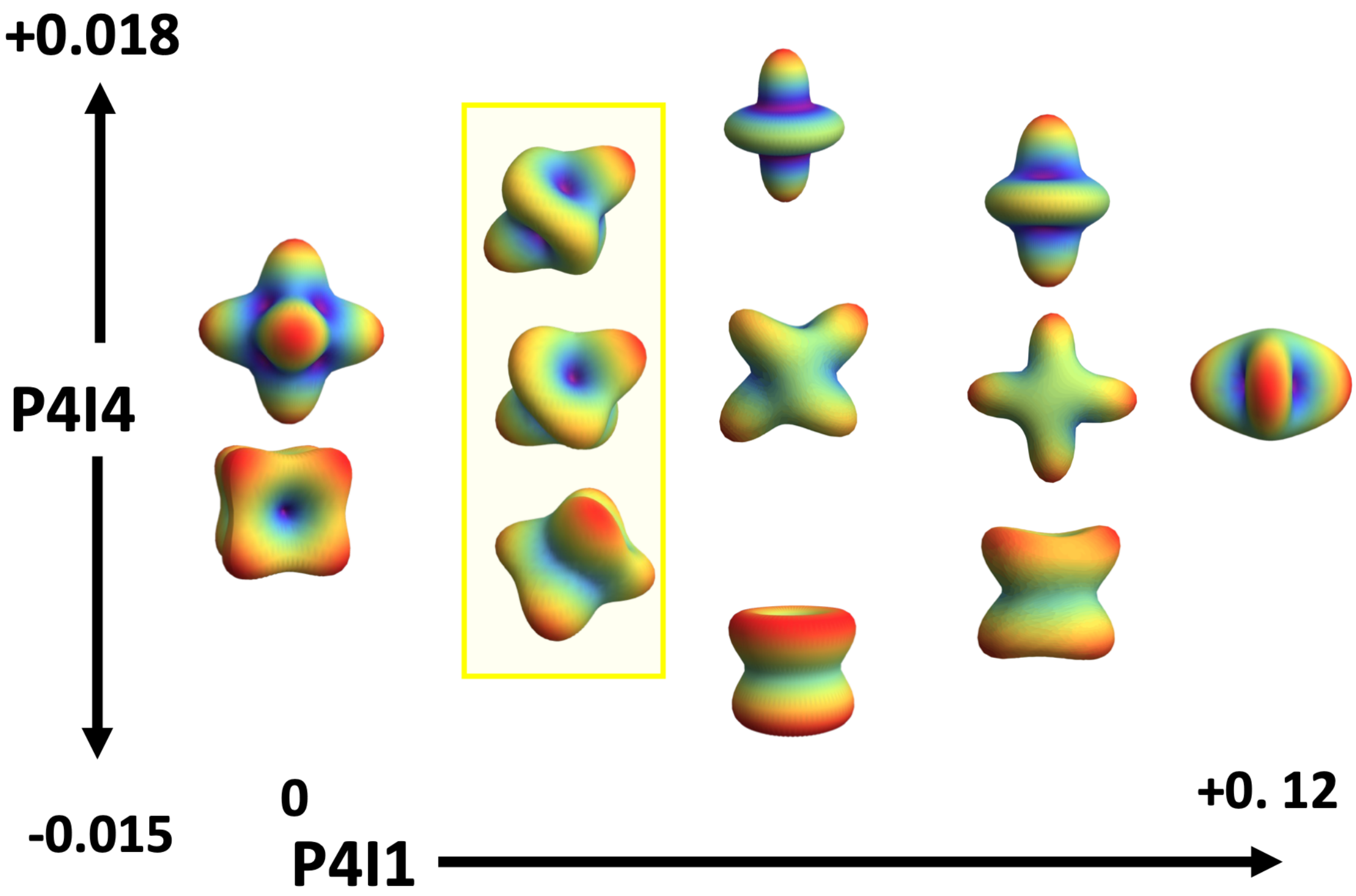

*Figure 8. Illustration of the fifth order Strain Functional Descriptor P4I4. This element has rather distinct behavior plotted vs the P4I1 metric. For a large value of P4I1, the range of P4I4 becomes quite narrow and the limiting structure is the $D_{4h}$ $\tilde{v}_4^4$ structure (at right). For the minimum value of P4I1 = 0, the range of structures also becomes quite narrow and limits to the two possible octahedral structures (BCC and SC). For an intermediate value of P4I1 is where P4I4 has its maximum range. The maximum positive value of P4I4 corresponds to a positive $\tilde{v}_4^0$ distortion, while the maximum negative value corresponds to*

*a negative $\tilde{v}_4^0$ distortion. The structures contained in the yellow box are combinations of $\tilde{v}_4^2$ and $\tilde{v}_4^3$ distortions with different phase angles (see text). A range of structures with P4I4 = 0 and increasing values of P4I1 is shown along the center line. These structures have only $C_{2h}$ symmetry, but limit to high symmetry structures for the extreme values of P4I1.*

A sevenfold contraction of rank 4 contractions generates another invariant with distinct characteristic, where the functional form is given in eq 45, with the full HS form given in SM S4.9. The form here is a contraction between a three-way contraction and a four-way contraction of the base rank 4 vectors, where this form is distinct from all other combinations (e.g. 2 and 5, 1 and 6). The unique term with respect to the other invariants is the final term in that expression which only depends on the value of $\overline{\omega}_3$ itself rather than its value with respect to $\overline{\omega}_2$ (which was discussed previously for eq 43). This term will be non-zero only when $\mu_4^3$ and $\mu_4^4$ are non-zero. And if only those two magnitudes are non-zero, then this is the only non-zero term in the entire expression; the magnitude and sign of eq 45 then depends on those two magnitudes and on $\overline{\omega}_3$, the phase angle between those two terms.

$$\boldsymbol{N}[\boldsymbol{N}[\boldsymbol{N}[\mathbf{4},\mathbf{4}]_4,\mathbf{4}]_4,\boldsymbol{N}[[\mathbf{4},\mathbf{4}]_4,[\mathbf{4},\mathbf{4}]_4]_4]_0 = \tag{45}$$

$$F\begin{bmatrix} (\mu_4^0)^7, -(\mu_4^0)^5(\mu_4^2)^2, (\mu_4^0)^3(\mu_4^2)^4, -(\mu_4^0)(\mu_4^2)^6, (\mu_4^0)^5(\mu_4^3)^2, -(\mu_4^0)^3(\mu_4^3)^4, -(\mu_4^0)(\mu_4^3)^6, \\ -(\mu_4^0)^3(\mu_4^2)^2(\mu_4^3)^2, -(\mu_4^0)(\mu_4^2)^4(\mu_4^3)^2, -(\mu_4^0)(\mu_4^2)^2(\mu_4^4)^4, (\mu_4^0)^5(\mu_4^4)^2, (\mu_4^0)(\mu_4^4)^6, \\ -(\mu_4^0)^3(\mu_4^2)^2(\mu_4^4)^2, -(\mu_4^0)(\mu_4^2)^4(\mu_4^4)^2, -(\mu_4^0)^3(\mu_4^3)^2(\mu_4^4)^2, -(\mu_4^0)(\mu_4^2)^2(\mu_4^3)^2(\mu_4^4)^2, \\ (\mu_4^0)(\mu_4^3)^4(\mu_4^4)^2, (\mu_4^0)^3(\mu_4^4)^4, (\mu_4^0)(\mu_4^2)^2(\mu_4^4)^4, (\mu_4^0)(\mu_4^3)^2(\mu_4^4)^4, \\ \begin{bmatrix} (\mu_4^0)^4, (\mu_4^2)^4, (\mu_4^3)^4, (\mu_4^4)^4, \\ (\mu_4^0)^2\{(\mu_4^2)^2, (\mu_4^3)^2, (\mu_4^4)^2\}, (\mu_4^2)^2\{(\mu_4^3)^2, (\mu_4^4)^2\}, (\mu_4^3)^2(\mu_4^4)^2 \end{bmatrix} (\mu_4^2)^2(\mu_4^4)\mathrm{Cos}[4\overline{\omega}_2], \\ [-(\mu_4^0)^2, (\mu_4^2)^2, (\mu_4^3)^2, -(\mu_4^4)^2](\mu_4^0)(\mu_4^2)(\mu_4^3)^2(\mu_4^4)\mathrm{Cos}[2\{\overline{\omega}_2 - 3\overline{\omega}_3\}], \\ [-(\mu_4^0)^2, (\mu_4^2)^2, (\mu_4^3)^2, -(\mu_4^4)^2](\mu_4^2)^3(\mu_4^3)^2\mathrm{Cos}[6\{\overline{\omega}_2 - \overline{\omega}_3\}], \\ [-(\mu_4^0)^2, -(\mu_4^2)^2, -(\mu_4^3)^2, (\mu_4^4)^2](\mu_4^2)(\mu_4^3)^2(\mu_4^4)^2\mathrm{Cos}[2\{\overline{\omega}_2 + 3\overline{\omega}_3\}], \\ (\mu_4^0)(\mu_4^2)^4(\mu_4^4)^2\mathrm{Cos}[8\overline{\omega}_2], -(\mu_4^3)^4(\mu_4^4)^3\mathrm{Cos}[12\overline{\omega}_3], \end{bmatrix}$$

This distinct characteristic identifies this term as the final non-redundant invariant for this subspace. The other interesting aspect of this invariant is that all of the terms in the first four lines (those that do not include an angular factor) depend on $(\mu_4^0)^x$ where $x$ is odd. Hence, the sign of the overall invariant will be dominated by the sign of $\mu_4^0$.

Since the numerator is formed by a series of rank 4 contractions, it is scaled by the norm of the rank 4 contraction, as well as the overall norm of the rank 4 term and the sum of the weighting terms. The resulting *P4I5* invariant is defined in eq 46.

$$P4I5(a) = \frac{\boldsymbol{N}[\boldsymbol{N}[\boldsymbol{N}[\mathbf{4},\mathbf{4}]_4,\mathbf{4}]_4,\boldsymbol{N}[\boldsymbol{N}[\mathbf{4},\mathbf{4}]_4,\boldsymbol{N}[\mathbf{4},\mathbf{4}]_4]_4]_0}{\boldsymbol{N}[\boldsymbol{N}[\mathbf{4},\mathbf{4}]_4,\boldsymbol{N}[\mathbf{4},\mathbf{4}]_4]_0 * \boldsymbol{N}[\mathbf{4},\mathbf{4}]_0 * P0I0(a)} \tag{46}$$

The significance of this term is illustrated in Figure 9, where the deformations in the *P4I5* term are plotted vs the *P4I1* term. For the extreme value of *P4I1* = 0, the *SC* and *BCC* $O_h$ structures are finally resolved. Those two structures are defined as $\tilde{v}_4^4 \pm \tilde{v}_4^0$, and the significance of the dependence on the sign of $\mu_4^0$ is apparent. For larger values of P4I1 (≈0.5), the extreme values of this metric correspond directly to $\pm\tilde{v}_4^0$ structures. While the right half of this plot is somewhat similar to that of *P4I4* (the $\pm\tilde{v}_4^0$ structures in the middle that asymptote to the $D_{4h}$ $\tilde{v}_4^4$ structure at the right), they have different intermediate structures. This is illustrated by the structures highlighted in the yellow box. These structures are all a combination of ~(1/2) $\mu_4^3$ distortion and ~(1/2) $\mu_4^4$ distortion, with the only difference being the rotation angle $\overline{\omega}_3$ between those two factors. If that phase angle is *0*, the structure shows a distortion towards the *SC* structure, with lobes appearing asymmetrically on the top and bottom of the $D_{4h}$ $\tilde{v}_4^4$ structure. This is shown as the upper highlighted structure. Though not apparent from this perspective, this structure actually has $C_{2h}$ symmetry, where the reflection plane is rotated ~π/4 about the vertical axis with respect to the plane of the page, such that the two equatorial lobes on the left side are mirror images of each other. If the phase angle is *π/12*, the structure shows a distortion towards the *BCC* structure, with all of the four lobes of the $D_{4h}$ $\tilde{v}_4^4$ structure starting to split asymmetrically. This is shown as the lower highlighted structure which also has $C_{2h}$ symmetry (the reflection plane is rotated ~π/2 about the vertical axis with respect to the plane of the page, such that the right and left sides of the structure are mirror images). If the phase angle is *π/24*, the low symmetry $C_i$ structure in the middle results, where the lobe projecting out of the plane of the page does *not* have $C_2$ or mirror symmetry. Along the midline of *P4I5* = 0 in general are similar low symmetry $C_i$ structures.

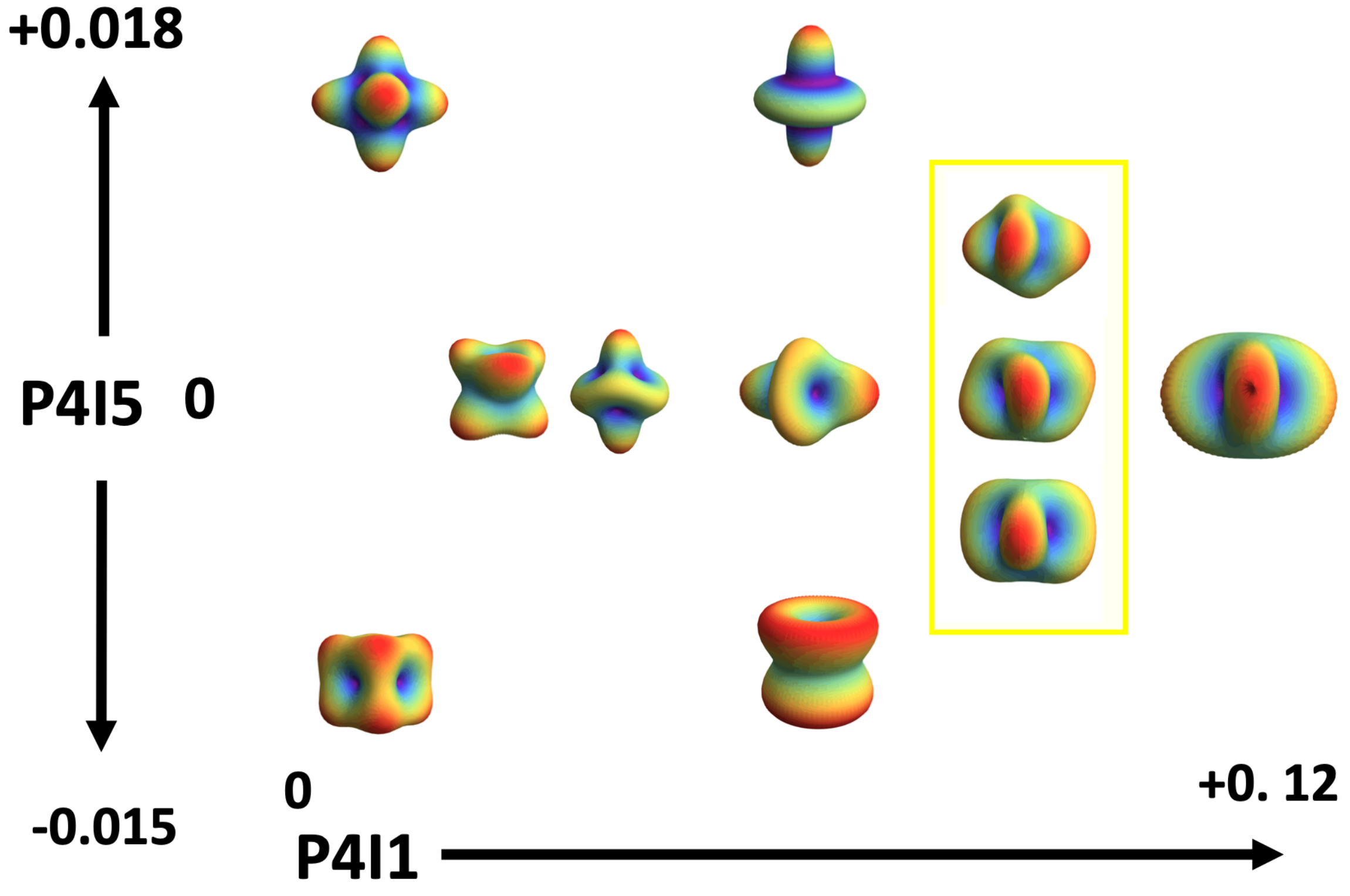

*Figure 9. Illustration of the seventh order Strain Functional Descriptor P4I5. Again plotted vs. the P4I1 metric, this element has the distinct characteristic of distinguishing the two $O_h$ structures at P4I1 = 0: the simple cubic (SC) in the upper left and the BCC structure in the lower left. This also distinguishes the positive and negative $\tilde{v}_4^0$ distortions for intermediate*

*values of P4I1, similar to P4I4. The structures contained in the yellow box are combinations of $\tilde{v}_4^3$ and $\tilde{v}_4^4$ distortions with different phase angles (see text). This is another unique attribute of this descriptor. A range of structures with P4I5 = 0 and increasing values of P4I1 is shown along the center line. These structures generally have only $C_i$ symmetry.*

The 2nd order and 0th order subspaces are then readily defined by analogy to the rank 2 terms (eqs 21-23) and the scaled invariants are given below. For *P4I8*, this term is spherically symmetric (no angular dependence), but it nominally has two radial nodes to make it orthogonal to *P2I2* and *P0I0*. Hence its more complicated structure.

$$P4I6(a) = \frac{\sqrt{\boldsymbol{N}[\mathbf{2(4)},\mathbf{2(4)}]_0}}{P0I0(a)} \tag{47}$$

$$P4I7(a) = -\frac{\boldsymbol{N}[\boldsymbol{N}[\mathbf{2(4)},\mathbf{2(4)}]_2,\mathbf{2(4)}]_0}{\boldsymbol{N}[\mathbf{2(4)},\mathbf{2(4)}]_0 * P0I0(a)} \tag{48}$$

$$P4I8(a) = \frac{\sqrt{\boldsymbol{N}[\mathbf{0(4)},\mathbf{0(4)}]_0}}{P0I0(a)} - \sqrt{5}\left(P2I2(a) + \sqrt{\frac{3}{2}}\right) + \sqrt{\frac{15}{8}} \tag{49}$$

*Orientation functions.* There are several options available for defining the orientation functions which can depend upon the particular application and the information desired. At least one of the ranks must be defined with respect to the external axis frame. The remaining ranks can then either be defined with respect to that chosen rank (defining a completely internally defined structure) or with respect to the external axis frame (defining a response to some external perturbation). For the current application of characterizing MD simulations of metals, we will choose the latter. As discussed further below, the rank 4 tensors carry information about the orientation of the local crystal domain. The lower rank tensors tend to carry information about the deformation of that crystal (shear strains, strain gradients) which tend to be predominantly oriented by an externally applied strain. For the subspaces within each rank, we choose to define their orientation with respect to the highest order subspace. This is because the deformations within those subspaces are likely interrelated.

We will start by considering these internal rotations between the subspaces. As pointed out in the discussions above, the rank 2 contractions of any of the rank *n* tensors identify the internal *xyz* axes of that rank defining the HS coordinate frame. There are then three degrees of freedom (rotations) that define the relative orientations between any of these axis frames. These can be found from the 2- and 3-way rank 2 contractions of those tensors. This is illustrated in Appendix A where it is shown how to extract the Euler angles that define the orientation of a rank 2 ST with respect to an arbitrary coordinate frame. As discussed there, the selection of an absolute direction orientation has subjective aspects to it. However, the invariants which define the relative orientations provide fixed values. For the rank 4 space, we define these invariants as follows:

$$P4I9(a) = \boldsymbol{N}[\boldsymbol{N}[\mathbf{4},\mathbf{4}]_2,\mathbf{2(4)}]_0 \frac{(\boldsymbol{N}[\mathbf{4},\mathbf{4}]_0 * \boldsymbol{N}[\mathbf{2(4)},\mathbf{2(4)}]_0)^{1/4}}{(\boldsymbol{N}[\boldsymbol{N}[\mathbf{4},\mathbf{4}]_2,\boldsymbol{N}[\mathbf{4},\mathbf{4}]_2]_0 * \boldsymbol{N}[\mathbf{2(4)},\mathbf{2(4)}]_0)^{1/2} * P0I0(a)} \tag{50}$$

$$P4I10(a) = \boldsymbol{N}[\boldsymbol{N}[\boldsymbol{N}[\mathbf{4},\mathbf{4}]_2,\mathbf{2}(\mathbf{4})]_2,\boldsymbol{N}[\mathbf{4},\mathbf{4}]_2]_0 \frac{(\boldsymbol{N}[\mathbf{4},\mathbf{4}]_0 * \boldsymbol{N}[\mathbf{2}(\mathbf{4}),\mathbf{2}(\mathbf{4})]_0)^{1/4}}{\boldsymbol{N}[\boldsymbol{N}[\mathbf{4},\mathbf{4}]_2,\boldsymbol{N}[\mathbf{4},\mathbf{4}]_2]_0 * (\boldsymbol{N}[\mathbf{2}(\mathbf{4}),\mathbf{2}(\mathbf{4})]_0)^{1/2} * P0I0(a)}$$
(51)

$$P4I11(a) = \boldsymbol{N}[\boldsymbol{N}[\boldsymbol{N}[\mathbf{4},\mathbf{4}]_2,\mathbf{2}(\mathbf{4})]_2,\mathbf{2}(\mathbf{4})]_0 \frac{(\boldsymbol{N}[\mathbf{4},\mathbf{4}]_0 * \boldsymbol{N}[\mathbf{2}(\mathbf{4}),\mathbf{2}(\mathbf{4})]_0)^{1/4}}{(\boldsymbol{N}[\boldsymbol{N}[\mathbf{4},\mathbf{4}]_2,\boldsymbol{N}[\mathbf{4},\mathbf{4}]_2]_0)^{1/2} * \boldsymbol{N}[\mathbf{2}(\mathbf{4}),\mathbf{2}(\mathbf{4})]_0 * P0I0(a)} \quad (52)$$

These terms are written in a way that emphasizes the three possible rank 2 contractions of two rank 2 ST multiplied by normalization factors. The denominators of the latter are all distinct and scale the contractions by the magnitudes of the component tensors. Multiplied by the leading contractions, this would constrain those products to have values in the range of [-1,1] and could be defined as an inverse cos function of an angle between the component tensors. The numerators of the normalization factors are the same and are the geometric mean of the magnitudes of main rank 4 ST and the rank 2 subspace of that. This gives all three invariants effective units of $(r/\sigma)^4$, balancing their magnitudes with the other P4 invariants.

Similar functions can be defined for the rank 3 space. A distinction here is that the rank 1 subspace is only defined by the direction of a line, so only two angles (and invariants) are required to define its orientation with respect to the rank 3 ST. Those invariants are defined below, where eqs 53 and 54 are analogous to eqs 50 and 51. The analog to eq 52 does not contain any additional information.

$$P3I5(a) = \boldsymbol{N}[\boldsymbol{N}[\mathbf{3},\mathbf{3}]_2,\boldsymbol{N}[\mathbf{1}(\mathbf{3}),\mathbf{1}(\mathbf{3})]_2]_0 * \frac{(\boldsymbol{N}[\mathbf{3},\mathbf{3}]_0 * \boldsymbol{N}[\mathbf{1}(\mathbf{3}),\mathbf{1}(\mathbf{3})]_0)^{1/4}}{(\boldsymbol{N}[\boldsymbol{N}[\mathbf{3},\mathbf{3}]_2,\boldsymbol{N}[\mathbf{3},\mathbf{3}]_2]_0 * \boldsymbol{N}[\boldsymbol{N}[\mathbf{1}(\mathbf{3}),\mathbf{1}(\mathbf{3})]_2,\boldsymbol{N}[\mathbf{1}(\mathbf{3}),\mathbf{1}(\mathbf{3})]_2]_0)^{1/2} * P0I0(a)} \quad (53)$$

$$P3I6(a) = \boldsymbol{N}[\boldsymbol{N}[\boldsymbol{N}[\mathbf{3},\mathbf{3}]_2,\boldsymbol{N}[\mathbf{1}(\mathbf{3}),\mathbf{1}(\mathbf{3})]_2]_2,\boldsymbol{N}[\mathbf{3},\mathbf{3}]_2]_0 * \frac{(\boldsymbol{N}[\mathbf{3},\mathbf{3}]_0 * \boldsymbol{N}[\mathbf{1}(\mathbf{3}),\mathbf{1}(\mathbf{3})]_0)^{1/4}}{\boldsymbol{N}[\boldsymbol{N}[\mathbf{3},\mathbf{3}]_2,\boldsymbol{N}[\mathbf{3},\mathbf{3}]_2]_0 * (\boldsymbol{N}[\boldsymbol{N}[\mathbf{1}(\mathbf{3}),\mathbf{1}(\mathbf{3})]_2,\boldsymbol{N}[\mathbf{1}(\mathbf{3}),\mathbf{1}(\mathbf{3})]_2]_0)^{1/2} * P0I0(a)} \quad (54)$$

Finally, we consider the orientation of the main ST themselves. One obvious choice would be to simply use the orientation framework that defines the HS orientation. While this would be useful for some applications, it is not very useful for the current application of MD simulations of crystalline materials. Considering *BCC* and *FCC* metals in particular, the *x*, *y* and *z* axes can be readily interchanged for the ideal structure and they cannot be distinguished. Given some low level of thermal fluctuations, there will be distortions away from that ideal structure, often dominated by an orthorhombic distortion. The "best" $C_4$ axis of that structure will be defined as the *Z* axis for the HS structure, where thermal fluctuations would likely eliminate ideal $C_4$ symmetry. Given that the initial structure was aligned with the external *xyz* frame, one tends to observe structures that are randomly aligned with one of the *x*, *y* or z axes and these will fluctuate with time. For somewhat higher temperatures, one would also observe trigonal distortions that would align the *Z* axis with the <111> crystallographic direction. For higher temperatures, though still below melting, all sense of directionality could be lost because of the large amount of instantaneous distortion. In such situations, time averaging of the structures would be useful in identifying persistent distortions.

An alternative approach would be to use a template matching approach, as in the Polyhedral Template Matching method.[11] Here, one would define the ideal structure (e.g. FCC or BCC) in terms of its shape invariants and oriented in a preferred reference frame. A comparison with the shape invariants would define how close it is to the ideal structure, and then the rank 2 contractions of the rank 4 ST (eqs 50-52) would define the amount of misorientation. This avoids the explicit definition of the Z axis of the HS configuration and would identify the best match with a minimal rotation. A minor limitation of this is that the ideal geometry must be defined at some reference density with a specific value of σ. The state of compression (e.g. from shock-loading) will then be reflected in the P0I0 values. There then may be issues with identifying highly distorted geometries under these compressed states. One could additionally define a reference compressed structure under those circumstances.

For more general applications, we favor a compromise approach. For each rank *n* tensor $\boldsymbol{M}_n$, we calculate the dot products with the $\boldsymbol{\chi}_n^0$, $\boldsymbol{\chi}_n^n$ and $\boldsymbol{\chi}_n^{-n}$ tensors that define the external reference frame $\boldsymbol{X}_n$. By normalizing these with the magnitude of the rank *n* ST, these orientation functions will have values in the range of [-1, 1]. These are defined below:

$$\begin{aligned} O1I0(a) &= \boldsymbol{N}[\boldsymbol{M}_1, \boldsymbol{\chi}_1^0]_0/(\boldsymbol{N}[\boldsymbol{M}_1, \boldsymbol{M}_1]_0)^{1/2} \\ O1I1(a) &= \boldsymbol{N}[\boldsymbol{M}_1, \boldsymbol{\chi}_1^1]_0/(\boldsymbol{N}[\boldsymbol{M}_1, \boldsymbol{M}_1]_0)^{1/2} \\ O1I2(a) &= \boldsymbol{N}[\boldsymbol{M}_1, \boldsymbol{\chi}_1^{-1}]_0/(\boldsymbol{N}[\boldsymbol{M}_1, \boldsymbol{M}_1]_0)^{1/2} \end{aligned} \tag{55}$$

$$\begin{aligned} O2I0(a) &= \boldsymbol{N}[\boldsymbol{M}_2, \boldsymbol{\chi}_2^0]_0/(\boldsymbol{N}[\boldsymbol{M}_2, \boldsymbol{M}_2]_0)^{1/2} \\ O2I1(a) &= \boldsymbol{N}[\boldsymbol{M}_2, \boldsymbol{\chi}_2^2]_0/(\boldsymbol{N}[\boldsymbol{M}_2, \boldsymbol{M}_2]_0)^{1/2} \\ O2I2(a) &= \boldsymbol{N}[\boldsymbol{M}_2, \boldsymbol{\chi}_2^{-2}]_0/(\boldsymbol{N}[\boldsymbol{M}_2, M_2]_0)^{1/2} \end{aligned} \tag{56}$$

$$\begin{aligned} O3I0(a) &= \boldsymbol{N}[\boldsymbol{M}_3, \boldsymbol{\chi}_3^0]_0/(\boldsymbol{N}[\boldsymbol{M}_3, \boldsymbol{M}_3]_0)^{1/2} \\ O3I1(a) &= \boldsymbol{N}[\boldsymbol{M}_3, \boldsymbol{\chi}_3^3]_0/(\boldsymbol{N}[\boldsymbol{M}_3, \boldsymbol{M}_3]_0)^{1/2} \\ O3I2(a) &= \boldsymbol{N}[\boldsymbol{M}_3, \boldsymbol{\chi}_3^{-3}]_0/(\boldsymbol{N}[\boldsymbol{M}_3, \boldsymbol{M}_3]_0)^{1/2} \end{aligned} \tag{57}$$

$$\begin{aligned} O4I0(a) &= \boldsymbol{N}[\boldsymbol{M}_4, \boldsymbol{\chi}_4^0]_0/(\boldsymbol{N}[\boldsymbol{M}_4, \boldsymbol{M}_4]_0)^{1/2} \\ O4I1(a) &= \boldsymbol{N}[\boldsymbol{M}_4, \boldsymbol{\chi}_4^4]_0/(\boldsymbol{N}[\boldsymbol{M}_4, \boldsymbol{M}_4]_0)^{1/2} \\ O4I2(a) &= \boldsymbol{N}[\boldsymbol{M}_4, \boldsymbol{\chi}_4^{-4}]_0/(\boldsymbol{N}[\boldsymbol{M}_4, \boldsymbol{M}_4]_0)^{1/2} \end{aligned} \tag{58}$$

*Summary.* In Table II, we summarize the different invariant functions that we have defined both in their equation form and our labels. In the first column, we have identified the base class of functions from which they come, including the analogous hydrogen-atom orbital functions. In the second column are listed the base Gaussian vectors with their radial-node functions. The third column are the base formulations of the invariants (not normalized) and the fourth column lists their labels. The functions are grouped according to their subspaces to emphasize how the dimensionality of the vector spaces maps onto the number of invariants. This can include orientation functions with respect to either other subspaces or the external frame.

Table II. Summary of the defined invariant functions. The base class (subspace) is listed in the first column, the base vectors with radial functions are listed in the second column, and the invariant functions and their labels are listed in the third and fourth columns.

| Class | Vector | Invariant | Label |
|---|---|---|---|
| $R_{000}$ (1s) | $\nu_0^0$ | $\boldsymbol{N}[\mathbf{0},\mathbf{0}]_0$ | P0I0 |
| | | | |
| $R_{11m}$ (2p) | $\nu_1^1$ | $\boldsymbol{N}[\mathbf{1},\mathbf{1}]_0$ | P1I0 |
| | $\nu_1^0$ | $\boldsymbol{N}[\mathbf{1},\boldsymbol{\chi}_1^0]_0$ | O1I0 |
| | $\nu_1^{-1}$ | $\boldsymbol{N}[\mathbf{1},\boldsymbol{\chi}_1^1]_0$ | O1I1 |
| | | | |
| $R_{22m}$ (3d) | $\nu_2^2$ | $\boldsymbol{N}[\mathbf{2},\mathbf{2}]_0$ | P2I0 |
| | $\nu_2^1$ | $\boldsymbol{N}[\boldsymbol{N}[\mathbf{2},\mathbf{2}]_2,\mathbf{2}]_0$ | P2I1 |
| | $\nu_2^0$ | $\boldsymbol{N}[\mathbf{2},\boldsymbol{\chi}_2^0]_0$ | O2I0 |
| | $\nu_2^{-1}$ | $\boldsymbol{N}[\mathbf{2},\boldsymbol{\chi}_2^2]_0$ | O2I1 |
| | $\nu_2^{-2}$ | $\boldsymbol{N}\left[\mathbf{2},\boldsymbol{\chi}_2^{-2}\right]_0$ | O2I2 |
| | | | |
| $R_{200}$ (2s) | $\nu_0^0 \cdot (r^2\text{-}a)$ | $\boldsymbol{N}[\mathbf{0}(\mathbf{2}),\mathbf{0}(\mathbf{2})]_0$ | P2I2 |
| | | | |
| $R_{33m}$ (4f) | $\nu_3^3$ | $\boldsymbol{N}[\mathbf{3},\mathbf{3}]_0$ | P3I0 |
| | $\nu_3^2$ | $\boldsymbol{N}[\boldsymbol{N}[\mathbf{3},\mathbf{3}]_2,\boldsymbol{N}[\mathbf{3},\mathbf{3}]_2]_0$ | P3I1 |
| | $\nu_3^1$ | $\boldsymbol{N}[\boldsymbol{N}[\boldsymbol{N}[\mathbf{3},\mathbf{3}]_2,\boldsymbol{N}[\mathbf{3},\mathbf{3}]_2]_2,\boldsymbol{N}[\mathbf{3},\mathbf{3}]_2]_0$ | P3I2 |
| | $\nu_3^0$ | $\boldsymbol{N}[\boldsymbol{N}[\boldsymbol{N}[\mathbf{3},\mathbf{3}]_4,\boldsymbol{N}[\mathbf{3},\mathbf{3}]_4]_4,\boldsymbol{N}[\mathbf{3},\mathbf{3}]_4]_0$ | P3I3 |
| | $\nu_3^{-1}$ | $\boldsymbol{N}[\mathbf{3},\boldsymbol{\chi}_3^0]_0$ | O3I0 |
| | $\nu_3^{-2}$ | $\boldsymbol{N}[\mathbf{3},\boldsymbol{\chi}_3^3]_0$ | O3I1 |
| | $\nu_3^{-3}$ | $\boldsymbol{N}[\mathbf{3},\boldsymbol{\chi}_3^{-3}]_0$ | O3I2 |
| | | | |
| $R3_{1m}$ (3p) | $\nu_1^1 \cdot (r^2\text{-}b)$ | $\boldsymbol{N}[\mathbf{1}(\mathbf{3}),\mathbf{1}(\mathbf{3})]_0$ | P3I4 |
| | $\nu_1^0 \cdot (r^2\text{-}b)$ | $\boldsymbol{N}[\boldsymbol{N}[\mathbf{3},\mathbf{3}]_2,\boldsymbol{N}[\mathbf{1}(\mathbf{3}),\mathbf{1}(\mathbf{3})]_2]_0$ | P3I5 |
| | $\nu_1^{-1} \cdot (r^2\text{-}b)$ | $\boldsymbol{N}[[\boldsymbol{N}[\mathbf{3},\mathbf{3}]_2,\boldsymbol{N}[\mathbf{3},\mathbf{3}]_2]_2,\boldsymbol{N}[\mathbf{1}(\mathbf{3}),\mathbf{1}(\mathbf{3})]_2]_0$ | P3I6 |
| | | | |
| $R_{44m}$ (5g) | $\nu_4^4$ | $\boldsymbol{N}[\mathbf{4},\mathbf{4}]_0$ | P4I0 |
| | $\nu_4^3$ | $\boldsymbol{N}[\boldsymbol{N}[\mathbf{4},\mathbf{4}]_2,\boldsymbol{N}[\mathbf{4},\mathbf{4}]_2]_0$ | P4I1 |
| | $\nu_4^2$ | $\boldsymbol{N}[\boldsymbol{N}[\boldsymbol{N}[\mathbf{4},\mathbf{4}]_2,\boldsymbol{N}[\mathbf{4},\mathbf{4}]_2]_2,\boldsymbol{N}[\mathbf{4},\mathbf{4}]_2]_0$ | P4I2 |
| | $\nu_4^1$ | $\boldsymbol{N}[\boldsymbol{N}[\boldsymbol{N}[\mathbf{4},\mathbf{4}]_4,\boldsymbol{N}[\mathbf{4},\mathbf{4}]_4]_4,\boldsymbol{N}[\mathbf{4},\mathbf{4}]_4]_0$ | P4I3 |
| | $\nu_4^0$ | $\boldsymbol{N}[\boldsymbol{N}[\boldsymbol{N}[\mathbf{4},\mathbf{4}]_2,\mathbf{4}]_2,\boldsymbol{N}[\mathbf{4},\mathbf{4}]_2]_0$ | P4I4 |
| | $\nu_4^{-1}$ | $\boldsymbol{N}[\boldsymbol{N}[\boldsymbol{N}[\mathbf{4},\mathbf{4}]_4,\mathbf{4}]_4,\boldsymbol{N}[[\mathbf{4},\mathbf{4}]_4,[\mathbf{4},\mathbf{4}]_4]_4]_0$ | P4I5 |
| | $\nu_4^{-2}$ | $\boldsymbol{N}[\mathbf{4},\boldsymbol{\chi}_4^0]_0$ | O4I0 |
| | $\nu_4^{-3}$ | $\boldsymbol{N}[\mathbf{4},\boldsymbol{\chi}_4^4]_0$ | O4I1 |
| | $\nu_4^{-4}$ | $\boldsymbol{N}[\mathbf{4},\boldsymbol{\chi}_4^{-\mathbf{4}}]_0$ | O4I2 |
| | | | |
| $R_{42m}$ (4d) | $\nu_2^2 \bullet (r^2\text{-}c)$ | $\boldsymbol{N}[\mathbf{2}(\mathbf{4}),\mathbf{2}(\mathbf{4})]_0$ | P4I6 |
| | $\nu_2^1 \bullet (r^2\text{-}c)$ | $\boldsymbol{N}[\boldsymbol{N}[\mathbf{2}(\mathbf{4}),\mathbf{2}(\mathbf{4})]_2,\mathbf{2}(\mathbf{4})]_0$ | P4I7 |
| | $\nu_2^0 \bullet (r^2\text{-}c)$ | $\boldsymbol{N}[\boldsymbol{N}[\mathbf{4},\mathbf{4}]_2,\mathbf{2}(\mathbf{4})]_0$ | P4I9 |
| | $\nu_2^{-1} \bullet (r^2\text{-}c)$ | $\boldsymbol{N}[\boldsymbol{N}[\boldsymbol{N}[\mathbf{4},\mathbf{4}]_2,\mathbf{2}(\mathbf{4})]_2,\boldsymbol{N}[\mathbf{4},\mathbf{4}]_2]_0$ | P4I10 |
| | $\nu_2^{-2} \bullet (r^2\text{-}c)$ | $\boldsymbol{N}[\boldsymbol{N}[\boldsymbol{N}[\mathbf{4},\mathbf{4}]_2,\mathbf{2}(\mathbf{4})]_2,\mathbf{2}(\mathbf{4})]_\mathbf{0}$ | P4I11 |
| | | | |
| $R_{400}$ (3s) | $\nu_0^0 \bullet (r^4\text{-}dr^2\text{-}e)$ | $\boldsymbol{N}[\mathbf{0}(\mathbf{4}),\mathbf{0}(\mathbf{4})]_0$ | P4I8 |

In Table III, we summarize the physical interpretations of the SFDs. Our definitions are primarily aimed at designing appropriate metrics for interpreting deformations in LAEs, characterized as shape, density, and orientation metrics.

Table III. Physical interpretations of SFDs.

| SFD | Interpretation | Type |
|---|---|---|
| P0I0 | Sum of weights | Density |
| | | |
| P1I0 | Gradient in density | Shape |
| | | |
| P2I0 | Net deviatoric strain | Shape |
| P2I1 | Skewness of deviatoric strain: axial vs. equatorial | Shape |
| P2I2 | Net volumetric strain | Density |
| | | |
| P3I0 | Net deviatoric strain gradient | Shape |
| P3I1 | Distortion from tetrahedral symmetry: ax vs eq | Shape |
| P3I2 | Skewness of non-tetrahedral distortion | Shape |
| P3I3 | Symmetry of curvature/bending | Shape |
| P3I4 | Uniaxial strain gradient | Shape |
| | | |
| P4I0 | Net fourth order deformation | Shape |
| P4I1 | Distortion from octahedral symmetry | Shape |
| P4I2 | Skewness of non-octahedral distortion: ax vs eq | Shape |
| P4I3 | Skewness metric for trigonal distortion | Shape |
| P4I4 | Skewness metric of density along primary axis | Shape |
| P4I5 | Triclinic skewness metric: bcc vs sc/fcc | Shape |
| P4I6 | Net deformation in 2nd rank subspace | Shape |
| P4I7 | Skewness of deformation in 2nd rank subspace | Shape |
| P4I8 | Net $r^4$ term volumetric strain | Density |

## 3. Results

### *3.1. Choosing the size of the Gaussian neighborhood*

The size of the Gaussian neighborhood, i.e., the value of σ, strongly affects the magnitudes and relative values of the SFDs. Figure 10 shows the SFDs, P4I0 and P3I0 (normalized to their maximum values), for 0 K *FCC* and *HCP* (in panel a) and *BCC* and Diamond (abbreviated as DIA, in panel b) as function of the σ normalized by the nearest-neighbor distance ($r_{NN}$). The vertical lines in Fig. 10 shows the choice of σ corresponding to $(2\pi)^{3/2}\sigma^3 = V_a$ (solid line) or $4/3\ \pi\sigma^3 = V_a$ (dotted line), where $V_a$ where is the per-atom volume calculated using the equilibrium lattice constant. The former case corresponds to the volume of the integrated Gaussian kernel, and the latter case corresponds to the volume of a simple sphere. In cases where the underlying lattice structure is not known or does not exist (i.e. a fluid), $V_a$ can be approximated locally using a spherical neighborhood or globally using the volume of the simulation cell. The following observations can be made from Fig. 10: (1) There exists an optimal value for σ which maximizes the SFDs, thus maximizing the amount of information that can be extracted from the neighborhood moments. (2) The optimal value of σ for *FCC*, *HCP* (dense, close packed structures), *BCC* and *DIA* (less dense structures) are close to the solid line where one approximates the volume of atom using a Gaussian i.e., a normalization of $(2\pi)^{3/2}$. (3) The distribution of SFDs tend to get broader for less dense structures and also with increase in value of $n$. (4) Using the normalization of $4/3\ \pi$ for $V_a$, which is equivalent to approximating the atoms as spheres in contact, is a poor choice as the SFDs provide very little information with this choice of $\sigma$.

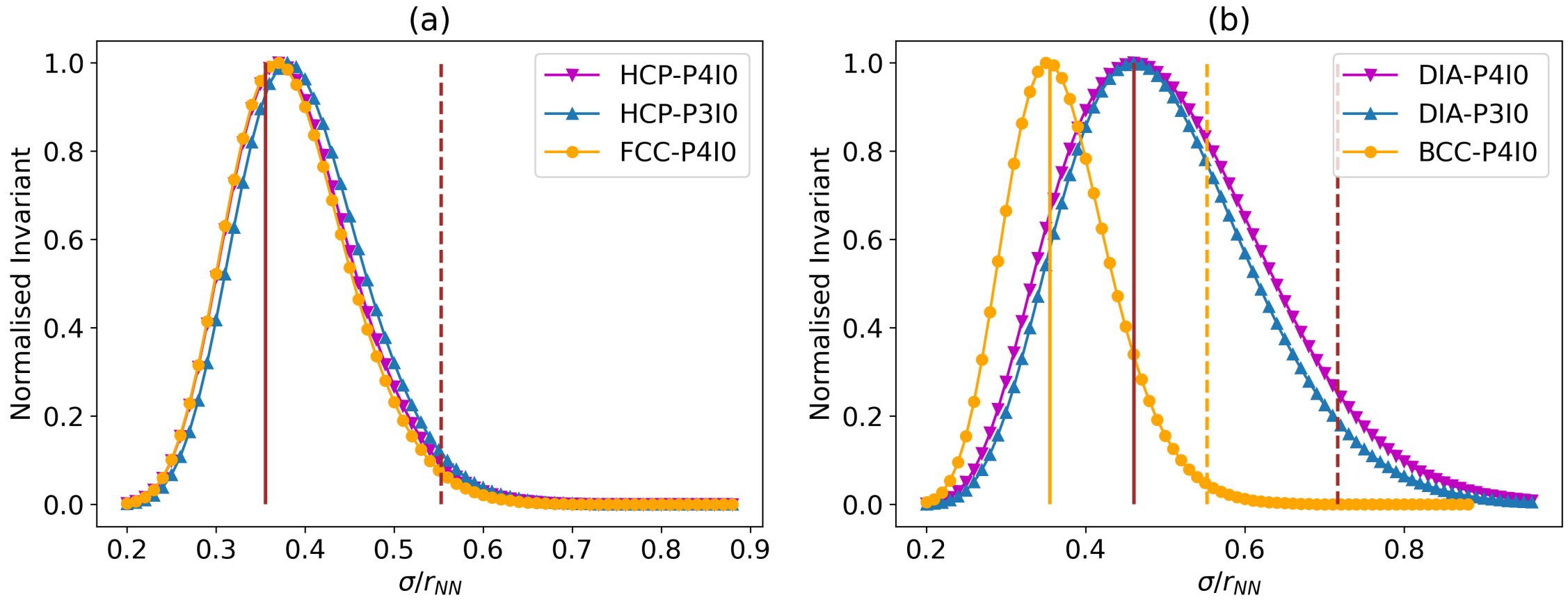

*Figure 10: Normalized SFDs as a function of the size of the Gaussian neighborhood for crystals in their ideal 0 K structure. (a) FCC and HCP (b) BCC and Diamond (DIA). See text for definitions of solid and dotted vertical lines.*

The presence of an optimum σ for calculating the SFDs is easy to rationalize: If the width of the neighborhood is very small, no neighbors are detected and the SFDs quickly go to zero. Conversely, too many neighbors are included if the width is too large *i.e.*, the neighborhood becomes isotropic and the SFDs rapidly go to zero. We subsequently employ the choice of σ corresponding to $(2\pi)^{3/2}\,\sigma^3 = V_a$ for all the results presented in this manuscript. Such a Gaussian neighborhood includes mostly contributions from the first nearest neighbor shells and some from the second nearest neighbors, respectively, but contributions from further atoms fall off rapidly in significance. For an *FCC* structure in equilibrium, the contribution is down to $10^{-6}$ by the $4^{th}$ nearest-neighbor shell. Note that the SOAP/GAP analysis treats σ as an arbitrary value, typically 0.5 Å, for both amorphous[21] and crystalline[19] systems.

In this section, we present some applications of the SFD analysis to MD simulations of defect-free and defective crystals. All MD simulations were performed using the LAMMPS[63] molecular dynamics software. All analysis involving Common Neighbor Analysis, Centro symmetry parameter, and Dislocation Extraction Algorithm (DXA) were performed using the OVITO[64] software, which was also used for visualizing atomic geometries.

### *3.2. Defect-free crystals at finite temperature*

Figure 11 demonstrates how atoms in different defect-free crystal structures at finite temperatures are distinguished by the SFDs. Cu (FCC), W (BCC), Ti (HCP) and Si (DIA) were chosen as the representative systems. Finite temperature, equillibrium, molecular dynamics simulations were performed in the microcanonical ensemble for these systems using standard inter-atomic potentials.[65-68] Each simulation cell contained between 10000 and 14000 atoms and was initially equillibrated to a temperature $T = 0.5 * T_m$, where $T_m$ is the melting point. We then plotted the histogram for the number of atoms (heat maps) in P2I2 and P0I0 in panel (a) and P4I0 and P0I0 in panel (b). The coordinates were averaged over 10 snapshots, each of which was 50 fs apart.

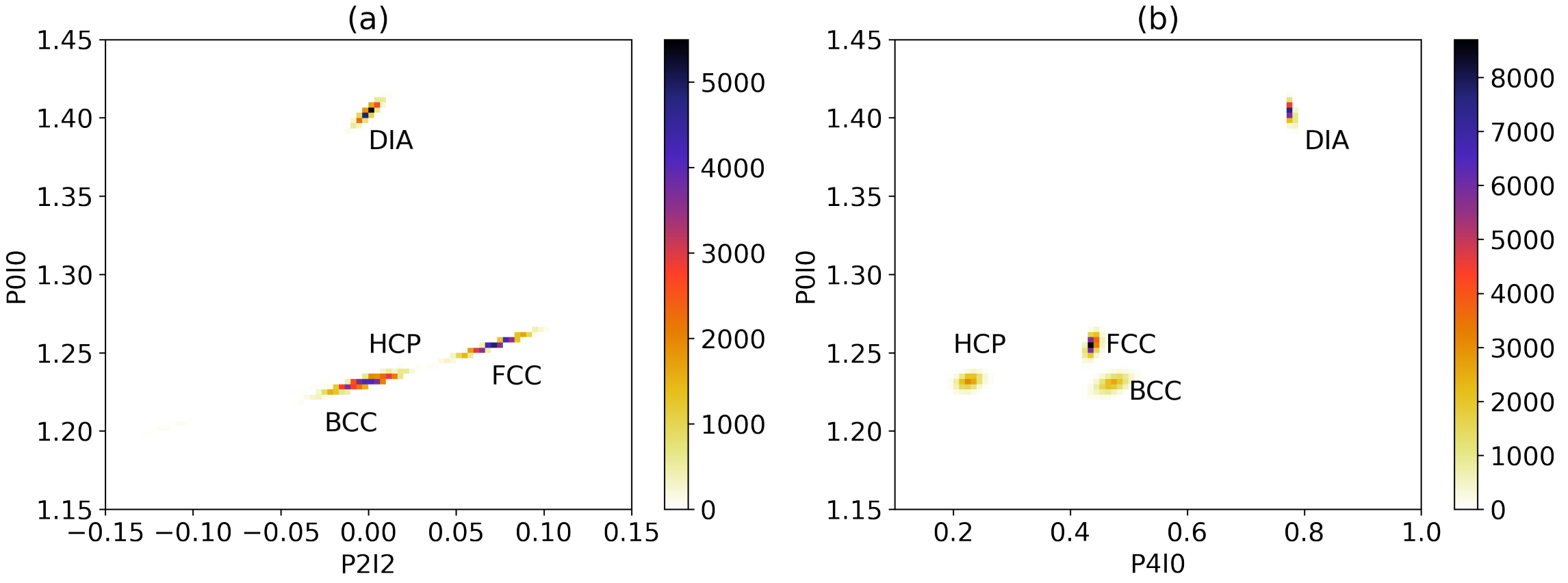

*Figure 11: Distribution of SFDs in FCC, BCC, HCP and DIA crystals thermalized at $0.5*T_m$. (a) Density metrics representing net sum of weights and average $r^2$ over the*

*neighborhood (b) Net sum of weights and net fourth order deformation metric. SFDs were calculated using coordinates averaged over 10 snapshots, each of which was 50 fs apart.*

Figure 11(a) shows the histograms in terms of two density metrics. While this alone is not enough to distinguish the different crystal structures, it is readily apparent from panel (b) that fourth-order SFDs in conjunction with other metrics can clearly do this. This demostrates the power of SFDs to characterize different crystal structures at finite temperatures. Distinguising between different hexagonal symmetries would require the extension of SFDs to sixth order and this will be pursued in future work.

### *3.3. Volume-preserving deformation paths between crystal structures and melting*

The behavior of SFDs during volume-preserving trigonal deformation and tetragonal deformation (along the Bain path) of the FCC structure was investigated. Figure 12 shows the fourth-order shape SFDs as a function of the strain along the deformation axis, *i.e.* <111> and <100>, for the trigonal path (panel (a)) and the Bain path (panel (b)) respectively. Note that these are volume-preserving deformations and therefore, we used a constant value for $\sigma$ throughout the deformation.

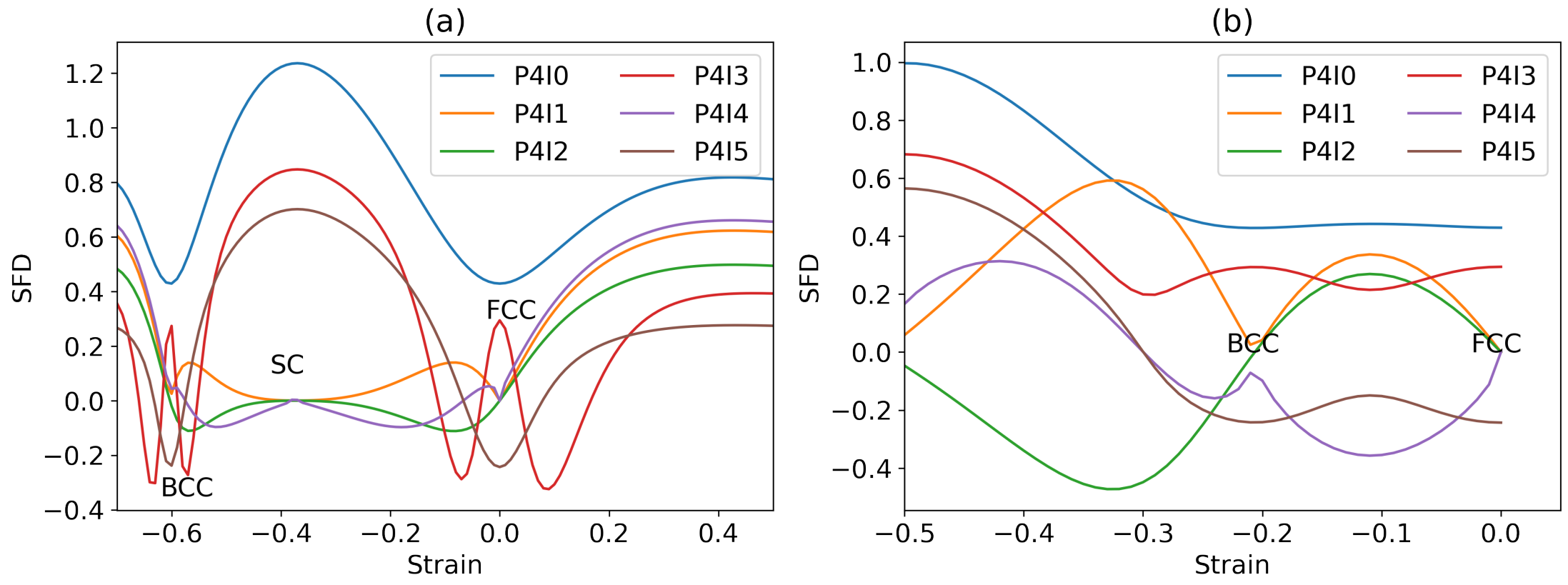

*Figure 12: Fourth-order SFDs as a function of strain for volume-preserving deformations of an FCC lattice. (a) Deformation along the $C_3$ axis or the Trigonal path. (b) Deformation along the $C_4$ axis or the Bain path.*

The SFDs smoothly track the deformation from FCC-SC-BCC in the trigonal path and FCC-BCC in the Bain path, with special structures (as noted in the panels) at the extremums. As demonstrated in Figure 6 and associated discussion, *P4I1* and *P4I2* are zero for the octahedral geometries (SC and BCC) and increase in magnitude for distortions away from those ideal structures. Since *P4I1* is defined to be non-negative, it tends to go to zero for ideal structures, where these can be either cusps (for BCC and FCC) or smooth minimum (for SC). *P4I2*, which can be either positive or negative, generally transits through zero for these ideal structures. Local extrema in other metrics are also often associated with the ideal structures, which reflects their base definition in terms of symmetry metrics. One exception to this is the local minimum in *P4I3*, especially in Fig 12a. Given its ability to identify trigonal distortions (Fig 7), this is highlighting a symmetric transition state structure in the deformation path. In line with this, we note the coincidence of extremum between BCC and FCC structures in Fig 12b, which also identify the likely symmetric transition state. This demonstrates that SFDs can smoothly characterize deformations between different crystal structures, thus providing useful descriptors for phase transformations. We do note that 4th order shape SFDs *by themselves* cannot clearly distinguish between FCC and BCC structures; these need to be used in conjunction with density metrics (*P0I0*, *P2I2* and *P4I8*) or 6th order SFDs to uniquely identify these structures.

We also investigated the robustness of SFDs during melting. As a representative case, we studied homogeneous melting of Cu at $P = 0$ atm. The simulation was performed with 3D periodic boundary conditions using a 12×12×12 supercell with the $C_4$ axes oriented along the Cartesian x, y, and z directions. The simulation cell was initially equilibrated to a temperature of 300 K using a Nose-Hoover thermostat. Following the equilibration, NPT simulations were performed at $P = 0$ atm using a heating rate of $1.0e^{12}$ $Ks^{-1}$. As the volume is not preserved, we used a variable $\sigma$ for the SFD analysis, which was calculated as $\sigma = \left[(\frac{V}{N})/(2\pi)^{3/2}\right]^{\frac{1}{3}}$, where $V$ is the instantaneous volume of the simulation cell and $N$ is the total number of atoms. Figure 13 shows the variation of the third- and fourth-order shape SFDs, shown in panels a and b, respectively, as a function of the simulation time.

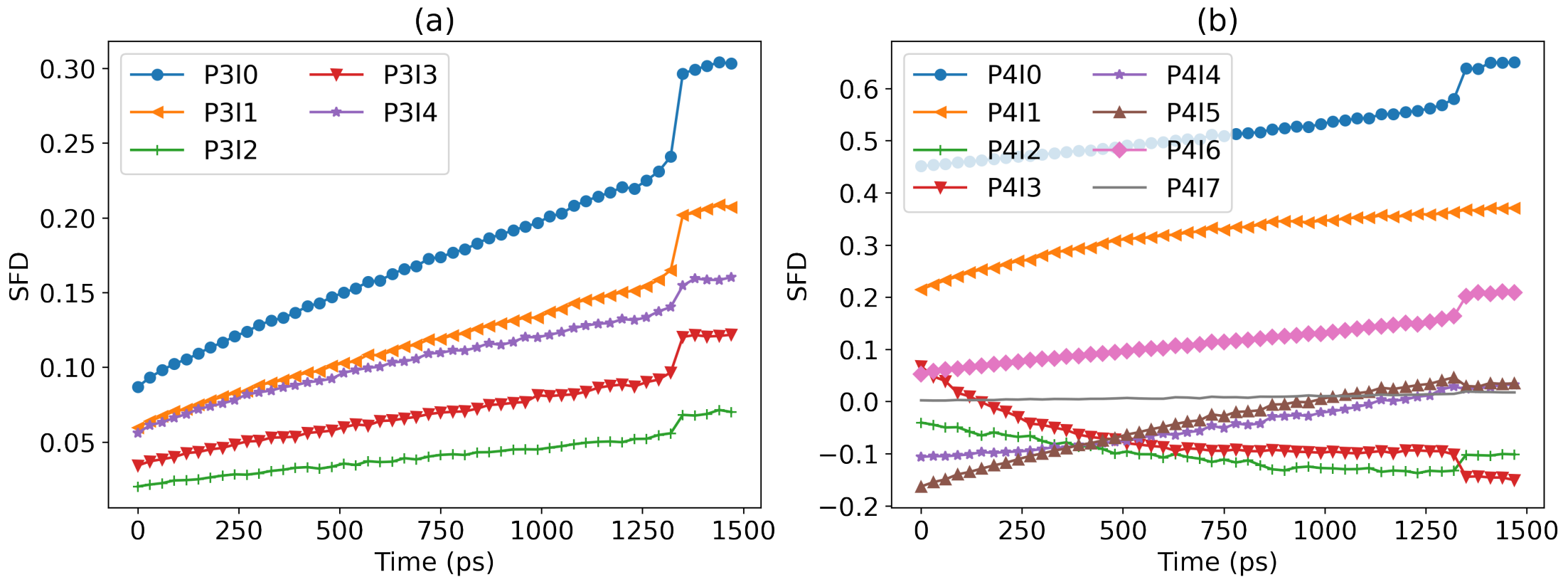

*Figure 13: SFDs as a function of simulation time for homogeneous melting of Cu at P= 0 atm. (a) Third-order SFDs. (b) Fourth-order SFDs.*

The SFDs shown in Figure 13 are averaged over all the atoms in the simulation cell. As is evident from Figure 13, the third- and fourth- order descriptors are smooth and continuous till the point of melting with a steep jump as the system goes through the solid-liquid phase transformation. This demonstrates that the SFDs are robust descriptors for melting.

### *3.4. Dislocations at 0 K*

Plastic deformation in crystalline materials is closely linked to generation, propagation, and interaction of dislocations. Figure 14 shows dislocation loops at $T = 0$ K undergoing expansion under a pure resolved shear stress in Ta (BCC) (panel a) and Cu (FCC) (panel b). The simulations were performed using EAM potentials[65,69] and contained approximately 1.69 million atoms in Ta and 900,000 atoms in Cu.

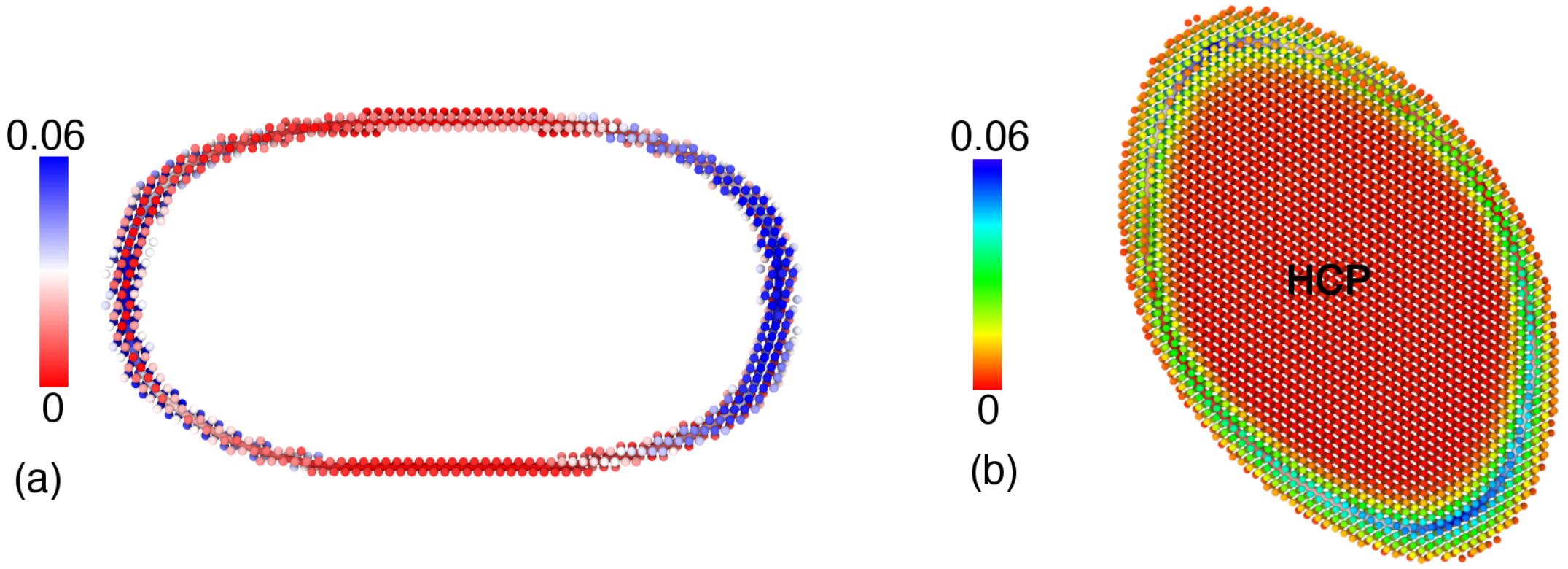

*Figure 14: Expanding dislocation loops identified using third-order SFDs. The atoms are colored according to P3I4. The solid lines are dislocation lines identified by DXA. (a) Ta (BCC) and (b) Cu (FCC). The atoms in the stacking fault are also shown in (b).*

Atoms were filtered using P3I0, which is the net magnitude of the different possible third order distortions. In both cases the atoms are colored according to P3I4, which characterizes the extensional strain gradient (see discussion of Figure 5). Also shown in the figures are dislocation lines identified by the DXA[12,13,70] colored according to screw (red)/ edge (blue) character of the dislocation loop. It is evident that P3I0, the SFD which characterizes net third-order deformation is able to distinguish atoms in the dislocation core and stacking fault from BCC/FCC atoms. It can also be seen that P3I4 is sensitive to the character of the dislocation. Regions of the loop with an edge character have a relatively large value for P3I4 (blue color in both panels (a) and (b)) based on the location of the 'extra half-plane', thus providing a metric to distinguish between screw and edge dislocations. This is easy to rationalize, as edge dislocation cores will have a strong gradient along the direction of the Burgers vector in the local number density. Thus, the third-order SFDs, which essentially measure third-order gradients in the number density, can identify dislocation cores and their edge/screw character. This follows from the fact that third-order SFDs characterize deformations that are represented using strain gradients defined in continuum mechanics, which are in turn directly related to the curvature of the crystal lattice (or equivalently elements of the Nye tensor[71]). Determination of slip planes and Burger's vectors of the dislocations is possible with the use of the orientation SFDs but we defer that aspect to a subsequent publication. Also note that the dislocation core, which represents the boundary between "slipped" and "un-slipped regions" of the crystal is clearly demarcated in panel (b), which shows a smooth transition between HCP stacking fault and FCC regions.

*3.5 Grain boundary characterization*

A particularly useful application of the SFDs is to the analysis of grain boundaries (GB)[72-74] We illustrate this in Figure 15 for the particular case of $\Sigma 25\{100\}$ symmetric tilt in Cu with a misorientation angle of ~74°. The perspective has the tilt axis normal to the page and the GB plane is horizontal. In panel (a), the atoms are colored according to P4I1, which measures how non-octahedral the coordination geometries are (see Figure 6). Away from the GB, these values go to zero (dark blue) as the structure relaxes to the ideal FCC ($O_h$) geometry. The repeating pattern at the GB shows three atoms in particular that are highly distorted (red) in this manner. In panel (b), the atoms are colored according to P3I0, which measures the net third order distortions. As above, these values go to zero away from the GB, with pockets of localized deformation at the GB. In this case, six atoms are found to be highly distorted (orange to dark red), three of which were also found to be non-octahedral. This emphasizes the ability of the SFD to differentiate the types of distortion. In panel (c), the atoms are colored according to P2I2, which is a density metric. As is generally expected, several atoms at the GB are noticeably under dense (bluish), and three of these bluish atoms can be seen to be surrounding a vacancy. Dark red vertical striations passing through the GB are also apparent, showing that there are also regions of mild compression alternating with the under dense regions. In panel (d), the atoms are colored according to O4I2. This measures the rotation angle of the HS frame of the atoms about the y axis of the system, which is normal to the page. For the bulk material, the orientation corresponds to orange in the upper crystal and medium blue in the lower crystal. These orientation features are useful in defining to which crystalline grain (or twin) different atoms belong, as well as highlighted regions of misorientation. Note that there are regions of both greater and lesser rotation as the GB is approached, and that there are pockets of intermediate rotation (greenish) located at the GB.

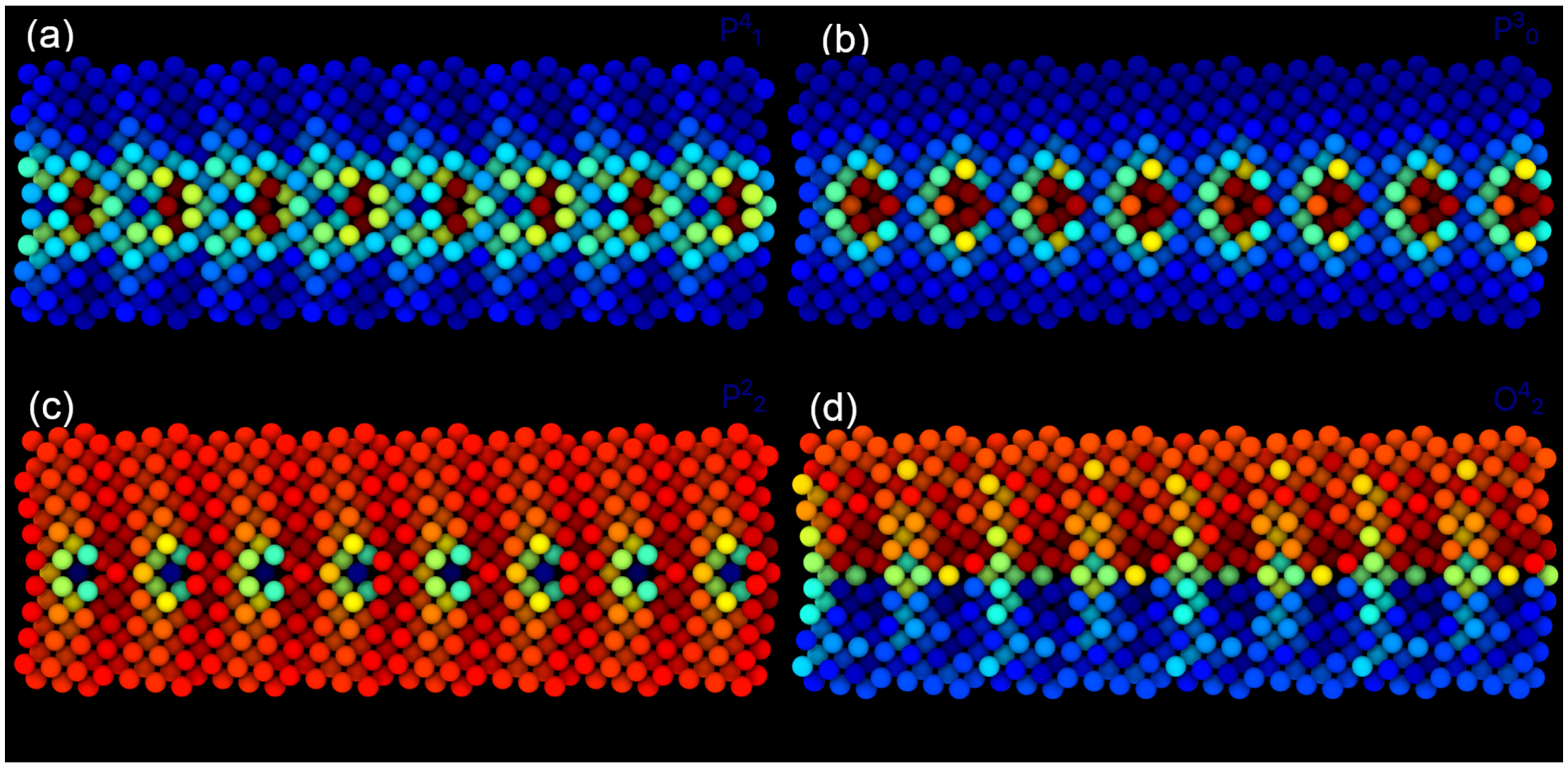

*Figure 15. Application of the SFD to the characterization of a grain boundary. The metrics shown here are (a) P4I1, (b) P3I0, (c) P2I2, and (d) O4I2. The atoms are colored by a jet color scheme for that metric, with red being the largest and blue being the smallest. The range is set by the values in the plot and the units are arbitrary.*

*3.6. Machine learning analysis of finite temperature MD simulations*

Our final test was to apply machine learning classification algorithms (*i.e.*, unsupervised learning) to automate detection of defect structures in a finite temperature MD simulation. The SFD basis comprise an excellent descriptor set for classification algorithms, being minimal, complete and interpretable. As a test case, we performed MD simulations of isothermal compression of Cu along [001], with a barostat applied to [100] and [010] directions to maintain zero normal stresses. Simulations were performed at $T$= 77 K and 300 K, at a constant strain rate of 1e$^8$ s$^{-1}$. The initial simulation cell consisted of a 43×39×41 unit cells with a total of 275028 atoms and was equilibrated in an NPT ensemble at zero stress for 50 picoseconds (ps) before the start of deformation. The simulation cell was deformed to a total strain of 12.5% to generate defective structures. This was followed by a 100 ps long equilibration in the micro-canonical ensemble to achieve the steady state defect structures. Atomic positions were averaged in both 300 K (40 snapshots, 10 fs apart) and 77 K (20 snapshots, 10 fs apart) simulations to reduce the noise from thermal fluctuations.

Among the variety of classification algorithms available, we chose the Gaussian Mixture Model (GMM),[75] as implemented in Scikit-learn,[76] for our analysis. GMMs assume that the underlying data can be represented using a mixture of *m* Gaussians, where *m* is a hyper-parameter specified by the user. The unknown parameters of the *m* Gaussians are estimated from the training data using the well-known expectation-maximization algorithm.[75] For the purposes of this study, we used a 4$^{th}$ order expansion. We did not use the orientation SFDs and therefore our descriptors space consisted of 19 SFDs/atom describing the shape and size of the local atomic environments. We tried *m*= 2-10 and used the Davies-Bouldin (DBI) and the Calinski and Harabasz (CHI) scores as implemented in sklearn to evaluate the quality of the GMM fits. Clusters that are further apart and less dispersed result in a lower DBI score, whereas a higher CHI score defines more well-defined clusters. Figure 16 shows the classification of the final simulation snapshot for the 300 K case.

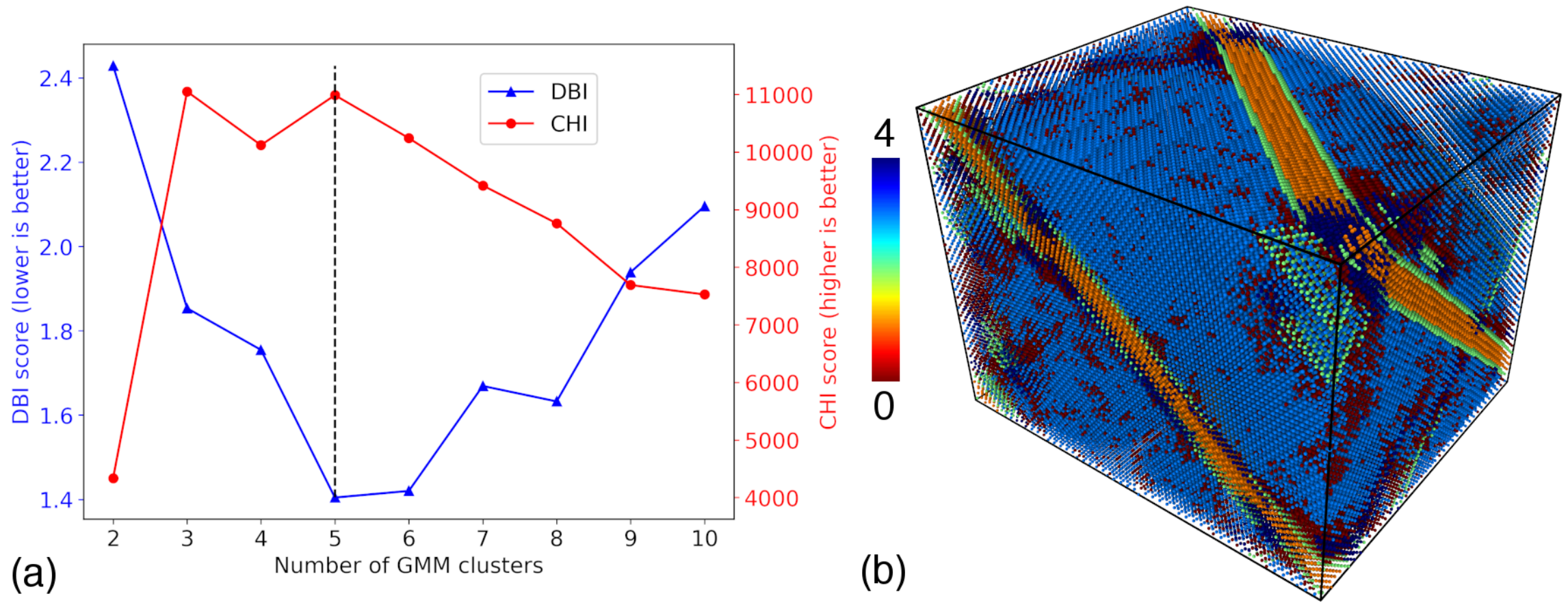

*Figure 16. Application of the SFDs in conjunction with unsupervised machine learning for automated detection of defects in finite-temperature molecular dynamics simulations (a) Clustering-quality scores for different GMM fits. The dotted line denotes the optimal value of clusters, which results in simultaneous local extrema in the scores (b) Atoms in the simulation cell of deformed Cu at 300 K, colored according to their cluster identifiers.*

For the 300 K case, *m*=5 was predicted to be optimal, as both the scores are at local extrema (Figure 16(a)). Figure 16(b) shows the atoms in the simulation cell colored according to their GMM cluster ids. Clear identification of twins (orange), deformed FCC (light blue), stacking faults (green) is readily apparent. Figure 17 shows atomic arrangements of each of these classes in detail in conjunction with the dislocation lines identified by the DXA approach.

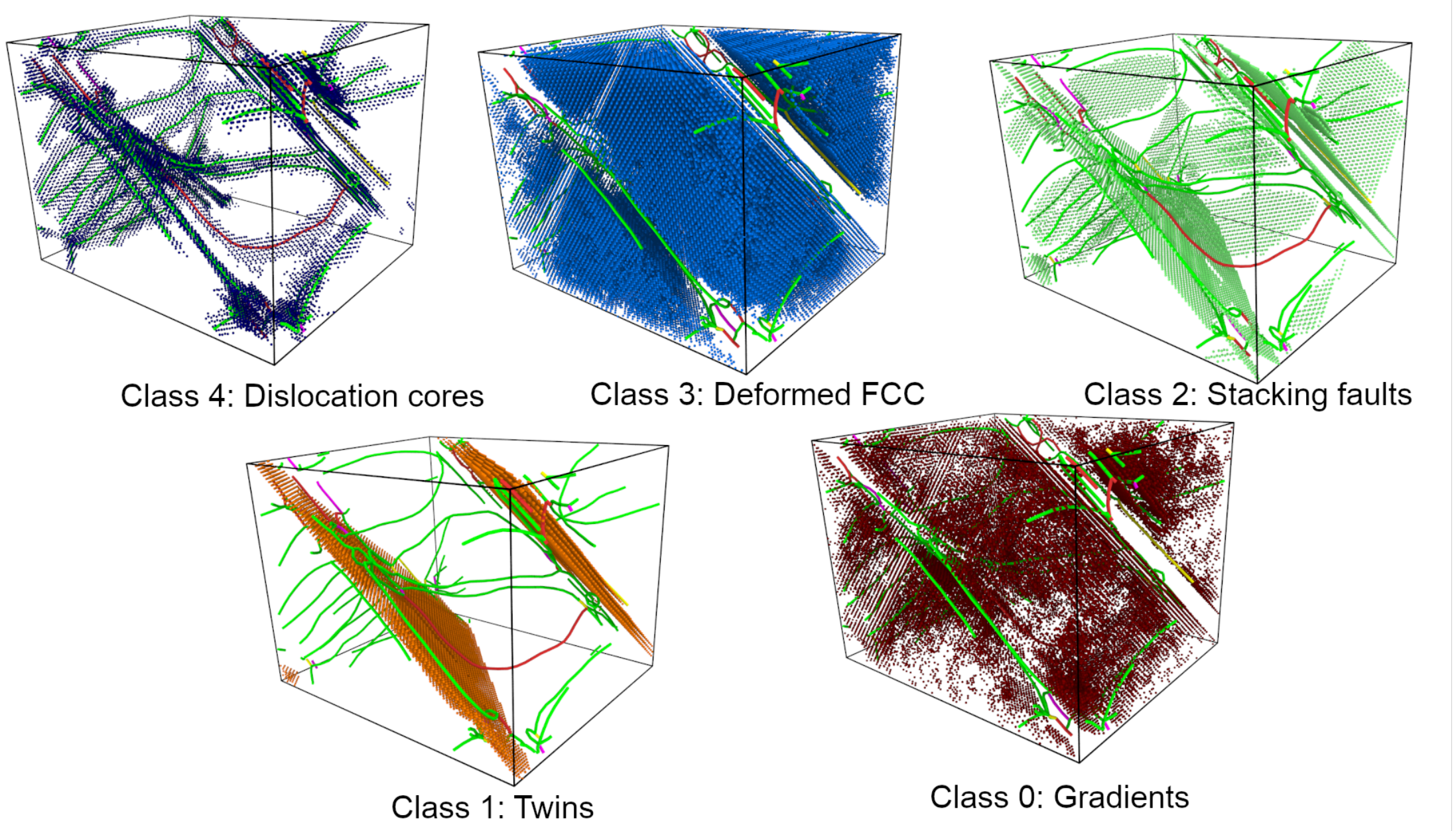

*Figure 17. Clusters identified by GMM from 300 K molecular dynamics simulation of high-strain rate compression of Cu. The GMM classifier clearly identifies dislocation cores, stacking faults, twins, along with deformed FCC and a group characterized as "gradients".*

Additional details are evident in Figure 17. Dislocation cores (dark blue) are clearly identified as class 4, in good agreement with DXA. The HCP stacking faults (class 2) are found at the boundary of the twinned regions and inside the slipped regions of the FCC crystal, in proximity to dislocation lines. The dislocations thread through the twinned regions and our GMM classifier separates these two classes clearly. Class 0 does not clearly map to a defect class but is comprised of atoms that generally have high values of gradients (either P1 or P3 values). These are generally associated with the regions neighboring the defect structures.

We repeated this exercise for a simulation at 77 K. Classification of atomistic structures where qualitatively similar to the 300 K data. In Figure 18(a), we compare the dislocation cores identified by our method to DXA. The atoms are colored by the P3I4 invariant, which measures the extensional strain gradient. The comparison is excellent, as in the case of 300 K data. Moreover, there are additional defective regions that are identified by the GMM classification – these correspond to atoms with high values of P3I4 (colored yellow). Zoom-in of these "defects" are shown in Figure 18(b). Some of them are isolated vacancy defects that gets classified along with the dislocations. There are also groups of these defects which form lines/ triangular loops (prismatic loops) – they form from breaking of dislocation junctions/crossing of dislocation lines during intense plastic deformation. The DXA defect mesh is present at some of these defects due to the disorder in these regions, but many of them are not even identified as defective regions. Nevertheless, we can clearly identify them and quantify the higher order deformations present in these regions.

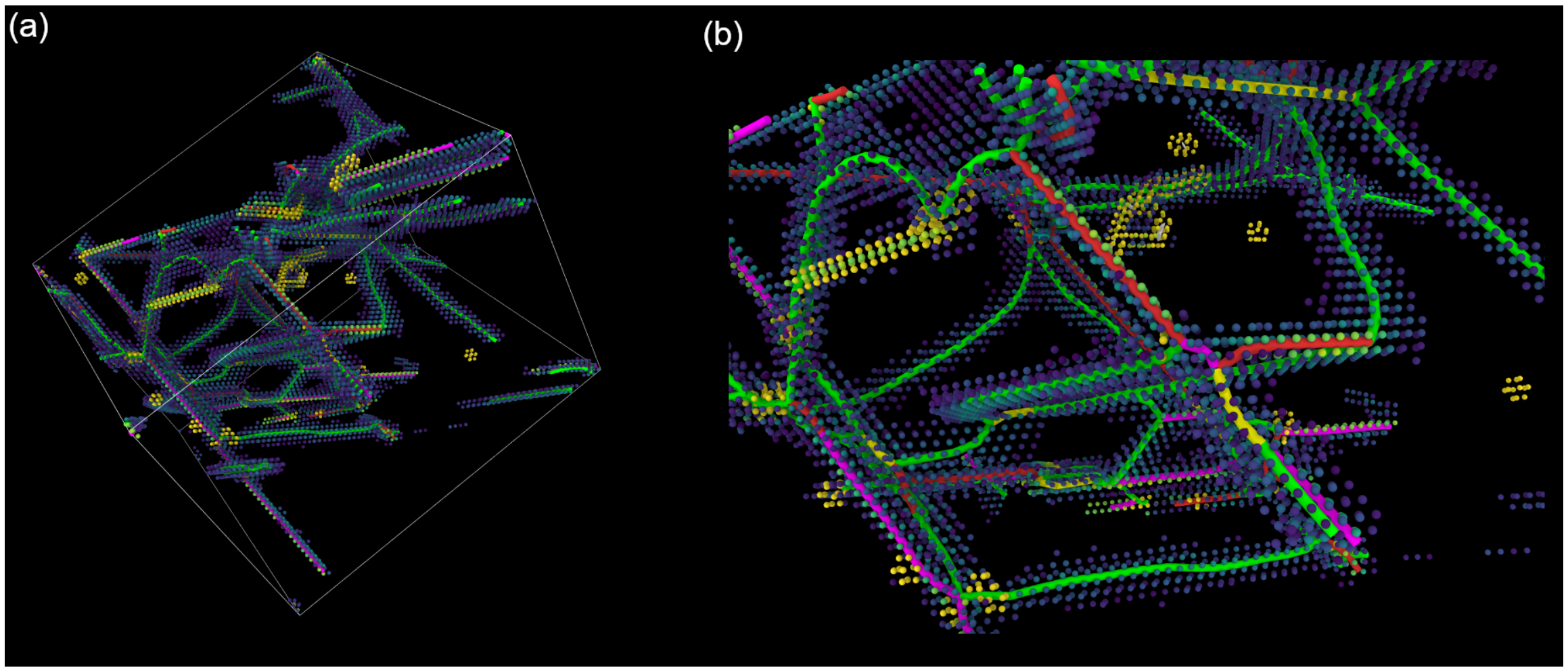

*Figure 18. Dislocation identified by GMM from 77 K molecular dynamics simulation of high-strain rate compression of Cu. (a) the GMM classifier clearly identifies dislocations that are found by DXA (b) Zoom-in showing additional defects not identified by DXA, comprised of regions with high values for the extensional strain gradient (P3I4).*

HCP atoms are present as part of stacking faults bounded by Shockley partials and twin boundaries. In comparison with Adaptive Common Neighbor Analysis, our analysis predicts a lower number of HCP atoms: 6.3% (compared to 8.1%) at 77 K and 6.4% (compared to 8.1%) at 300 K. We have not performed a detailed analysis on the origin of these differences. We note here that classification using GMM performed better compared to other clustering algorithms such as DBSCAN[77] and K-means,[75] possibly due to the fact that the SFD distributions resulting from thermal vibrations are approximately Gaussian distributed.

## 4. Discussion

The major accomplishment of this work is to demonstrate how to derive a 4th-order complete and minimal set of rotational invariants that describe a 3D spatial neighborhood. This was demonstrated here using a neighborhood defined by Gaussian-weighting, where extensions to other neighborhood definitions based on similar continuously differentiable functions (e.g. simple exponential decay) can easily be made. The method could also be extended to splined functions if those are continuously differentiable to the desired order of expansion. This method can also be extended to higher orders (i.e. sixth order expansions would be required to fully differentiate all possible crystal structures), and this is being pursued. In achieving this reduction to a minimal number of invariants, there is the great side benefit of finding that these descriptors relate to simple physical deformations and can be readily interpreted. All previous developments have treated the Clebsch-Gordan coupling procedure as a complex mathematical abstraction, whereas it contains quite critical symmetry information as shown here. Additionally, it has been demonstrated that using this same procedure can also define the orientation of these shaped objects in an arbitrary shape frame so that no information has been lost in this transformation to rotationally invariant descriptors. These are general conclusions that can be applied to similar problems in 3D image processing.[28,32]

We do wish to emphasize that the current selection of a minimal set of invariants is not unique. Given the infinite number of possible invariants, one should also be able to derive an infinite number of minimal and complete sets. For the 4th invariants in particular, we identified two additional low-order invariants that could be used in place of *P4I3* $\propto N[N[N[4,4]_4, N[4,4]_4]_4, N[4,4]_4]_0$ and *P4I4* $\propto N[N[N[4,4]_2, 4]_2, N[4,4]_2]_0$. Those two terms would be based on the invariants $N[N[4,4]_4, 4]_0$ and $N[N[N[4,4]_4, 4]_4, N[4,4]_4]_0$. While somewhat simpler mathematically (because of the lower contraction order), we do not find their geometric interpretation to be as direct. The former term would strongly emphasize the $\pm\tilde{v}_4^0$ behavior that occurs for *P4I4* and *P4I5* (figures 8 and 9, respectively) as well as the trigonal distortion of *P4I3* and some of the *BCC-SC* distinction of *P4I5*. The second invariant would also capture the trigonal distortion of *P4I3* and some of the *BCC-SC* distinction of *P4I5*. However, as *P4I5* must be retained to fully describe the significance of the $\overline{\omega}_3$ factor, there is no apparent advantage is using these two invariants over the ones we have chosen. Additionally, we note the selection of *P4I3* as an invariant was motivated by the analogy to *P3I3,* where we have not found a simple replacement for the latter.

We also observe that any algebraic combination of a set of complete invariants would generate an additional set of complete invariants. For example, a simple linear combination of the base functions for *P3I2* (eq 28) and *P3I3* (eq 32) could be used to eliminate the dependence on $\overline{\omega}_2$ from one of the terms, and it would then only be retained in the complementary linear combination. While that might be advantageous for capturing the physical significance of the $\overline{\omega}_2$ factor, it would then force the modification of the *P3I2* term, which is currently used to identify the primary axes of the third order deformations. We generally conclude that the selection of a "best" set of complete and minimal invariants is subjective to some degree and may be application dependent. However, we think the current set is very well suited for the current application of MD simulations.

This work also generates an understanding of the significance of the "composite" vectors (as Lo and Don called them)[28], and then how to define the orientation functions. The original definition of terms like $N[4,4]_2$ is defining the relative combination weights of the addition of two rank 4 angular momentum vectors to form a rank 2 vector.[29] From a comparison of our graphical illustrations in Figures 2, 3 and 6, that addition of a rank *n* vector to itself in a manner to define a rank 2 vector identifies the best $C_2$ symmetry axes that define the original rank *n* vector. These also then define the HS axes for defining the orientation of that structure. The formation of composite vectors between two different initial vectors essentially forms a dot product between those two vectors. When applied to the rank 0 contraction between two rank 2 vectors, as shown in the Appendix, this defines the rotation angles between the HS frames of those two vectors. All the higher order contractions could be interpreted in a similar way. For example, the base function of P4I4 $\propto N[N[N[4,4]_2, 4]_2, N[4,4]_2]_0$ is dot product between $N[N[4,4]_2, 4]_2$ and $N[4,4]_2$. The latter identifies the best $C_2$ axes of the rank 4 vector, and the former 3-way contraction emphasizes the directionality and magnitude of the $\mu_4^0$ component.

The connection back to the original Cartesian tensor formulation from which this derivation began is also emphasized. As demonstrated by Coope[50-52] and Jerphagnon[30], any general $n \times n \times n$ tensor can be expressed in terms of the sums of the corresponding ST components. The various tensorial manipulations used to characterize complex material response (e.g., elasticity tensor, piezoelectric response)[78,79] could then be expanded out in the ST terms as a means of emphasizing underlying symmetry. The different trace contractions and outer product operations can be replaced by the analogous "composite" vector operations where the ranks of the products are reduced or raised, respectively. Computationally, all these manipulations are nested vector operations which are efficiently calculated on most machine architectures and are likely more efficient that the corresponding tensor operations. The extension to force or displacement vectors is not difficult and would create a basis set for characterizing those features as well. The connection to angular momentum expressions would also establish a framework for coupling to electronic spin.

## *5. Conclusions*

We developed a complete and symmetry-adapted set of descriptors, referred to as Strain Functional Descriptors (SFDs), for characterizing atomistic geometries. Using a Gaussian kernel, which has been shown to be the least-biased kernel for localizing information, discrete atomistic quantities are expressed as continuous functions. The SFDs are derived from $n^{\text{th}}$ order central moments (*i.e.*, gradients) of the continuum representation of the local number density using techniques from group theory. The SFDs characterize atomic environments in terms of shape, size, and orientation of local neighborhood and can identify presence/absence of symmetry elements. We show that the SFDs can be used to distinguish between different crystal symmetries, characterize phase transformations, and identify defect structures in finite temperature MD simulations. We also coupled SFDs to unsupervised machine learning techniques (classification algorithms) and demonstrated their ability to automatically identify complex dislocation networks, stacking faults, and vacancies in finite temperature MD simulation data. These classifiers can, in principle, be trained using small scale (hundreds of thousands of atoms) simulations, which can in turn be used for automated, *on-the-fly*, analysis of extreme-scale atomistic simulations.

*Appendix A: Relationship of orientation tensor products to Euler angles*

We start by defining an arbitrary rank 2 traceless tensor in its diagonal Cartesian form:

$$(\text{A.1})\quad \boldsymbol{M_{DT}} = M\begin{bmatrix} \chi_2^2 \sin\psi_M - \frac{1}{2}\chi_2^0 \cos\psi_M & 0 & 0 \\ 0 & -\chi_2^2 \sin\psi_M - \frac{1}{2}\chi_2^0 \cos\psi_M & 0 \\ 0 & 0 & \chi_2^0 \cos\psi_M \end{bmatrix}$$

This tensor has a magnitude $M$ and a skewness angle of $\psi_M$. The $\boldsymbol{\chi_l^m}$ vectors are similar to the $\boldsymbol{v_l^m}$ vectors but define the external axis frame. The $\boldsymbol{M_{DT}}$ tensor can then be rotated to an arbitrary frame using the Euler rotation matrix $\boldsymbol{E_r}$, using the zyz convention used by Edmonds (rotate about the initial z axis by α, then rotate β about the new y' axis, and then rotate γ about the new z" axis).[29]

$$(\text{A.2})\quad \boldsymbol{E_r} = \begin{bmatrix} \cos\beta\ \cos\alpha\ \cos\gamma - \sin\alpha\ \sin\gamma & \cos\beta\ \sin\alpha\ \cos\gamma + \cos\alpha\ \sin\gamma & -\sin\beta\ \cos\gamma \\ -\cos\beta\ \cos\alpha\ \sin\gamma - \sin\alpha\ \cos\gamma & -\cos\beta\ \sin\alpha\ \sin\gamma + \cos\alpha\ \cos\gamma & \sin\beta\ \sin\gamma \\ \sin\beta\ \cos\alpha & \sin\beta\ \sin\alpha & \cos\beta \end{bmatrix}$$

The arbitrarily oriented tensor is then defined as:

$$(\text{A.3})\quad \boldsymbol{M_2} = \boldsymbol{E_r M_{DT} E_r^T}$$

For simplicity, we then evaluate the contractions between this tensor and the two tensors that define the external axis frame: $\boldsymbol{\chi_2^0}$ and $\boldsymbol{\chi_2^2}$. These are in bold to emphasize that they are tensors, though their only non-zero components are $\chi_2^0$ and $\chi_2^2$, respectively. The norms of both tensors are 1. Those contractions are:

$$(\text{A.4})\quad \mathbf{N}[\boldsymbol{M_2}, \boldsymbol{\chi_2^0}]_{\mathbf{0}} = \frac{1}{4}\left((1 + 3\cos 2\beta)\cos\psi_M + 2\sqrt{3}\cos 2\alpha \sin^2\beta \sin\psi_M\right)$$

$$(\text{A.5})\quad \mathbf{N}[\boldsymbol{M_2}, \boldsymbol{\chi_2^2}]_{\mathbf{0}} = \frac{1}{4}\Big((\cos 2\alpha \cos 2\gamma\,(3 + \cos 2\beta) - 4\sin 2\alpha \sin 2\gamma \cos\beta)\sin\psi_M + 2\sqrt{3}\cos 2\gamma \sin^2\beta \cos\psi_M\Big)$$

$$(\text{A.6})\quad \mathbf{N}\left[\mathbf{N}[\boldsymbol{M_2}, \boldsymbol{\chi_2^0}]_{\mathbf{2}}, \boldsymbol{\chi_2^0}\right]_{\mathbf{0}} = \frac{1}{12\sqrt{2}}\left(\sqrt{3}(1 + 3\cos 2\beta)\cos\psi_M + 6\cos 2\alpha \sin^2\beta \sin\psi_M\right)$$

$$(\text{A.7})\quad \mathbf{N}\left[\mathbf{N}[\boldsymbol{M_2}, \boldsymbol{\chi_2^2}]_{\mathbf{2}}, \boldsymbol{\chi_2^2}\right]_{\mathbf{0}} = -\frac{1}{12\sqrt{2}}\left(\sqrt{3}(1 + 3\cos 2\beta)\cos\psi_M + 6\cos 2\alpha \sin^2\beta \sin\psi_M\right)$$

$$(\text{A.8})\quad \mathbf{N}\left[\mathbf{N}[\boldsymbol{M_2}, \boldsymbol{\chi_2^0}]_{\mathbf{2}}, \boldsymbol{M_2}\right]_{\mathbf{0}} = \frac{1}{12\sqrt{2}}\left(\sqrt{3}(1 + 3\cos 2\beta)\cos 2\psi_M - 6\cos 2\alpha \sin^2\beta \sin 2\psi_M\right)$$

(A.9) $\mathbf{N}\left[\mathbf{N}\left[\boldsymbol{M}_2,\chi_2^2\right]_2,\boldsymbol{M}_2\right]_0=\frac{1}{24}\Big(4\sqrt{6}\cos\beta\sin 2\alpha\sin 2\gamma\sin 2\psi_M+$
$\sqrt{2}\cos 2\gamma\left(6\cos 2\psi_M\sin^2\beta-\sqrt{3}\cos 2\alpha\,(3+\cos 2\beta)\sin 2\psi_M\right)\Big)$

We note two relationships between two pairs of these terms (A.4 and A.6, and A.6 and A.7):

(A.10) $\mathbf{N}\left[\boldsymbol{M}_2,\chi_2^0\right]_0=\sqrt{6}\,\mathbf{N}\left[\mathbf{N}\left[\boldsymbol{M}_2,\chi_2^0\right]_2,\chi_2^0\right]_0$

(A.11) $\mathbf{N}\left[\mathbf{N}\left[\boldsymbol{M}_2,\chi_2^0\right]_2,\chi_2^0\right]_0=-\mathbf{N}\left[\mathbf{N}\left[\boldsymbol{M}_2,\chi_2^2\right]_2,\chi_2^2\right]_0$

This leaves four independent relationships in A.4, A.5, A.8 and A.9. Equations A.4 and A.8 can both be solved for $\cos 2\alpha$, and by equating those two forms and then solving $\cos 2\beta$ produces the following relationship.

(A.12) $\cos 2\beta=\frac{4}{3}\frac{\sin 2\psi_M}{\sin 3\psi_M}\mathbf{N}\left[\boldsymbol{M}_2,\chi_2^0\right]_0+\frac{4\sqrt{6}}{3}\frac{\sin\psi_M}{\sin 3\psi_M}\mathbf{N}\left[\mathbf{N}\left[\boldsymbol{M}_2,\chi_2^0\right]_2,\boldsymbol{M}_2\right]_0-\frac{1}{3}$

This defines $\cos 2\beta$ in terms of known quantities. This can be substituted back into the previous relationships for $\cos 2\alpha$ to also define it in terms of those same quantities.

(A.13) $\cos 2\alpha=\frac{\sqrt{3}\left(\cos 2\psi_M\mathbf{N}\left[\boldsymbol{M}_2,\chi_2^0\right]_0-\sqrt{6}\cos\psi_M\mathbf{N}\left[\mathbf{N}\left[\boldsymbol{M}_2,\chi_2^0\right]_2,\boldsymbol{M}_2\right]_0\right)}{\sin 3\psi_M-\sin 2\psi_M\mathbf{N}\left[\boldsymbol{M}_2,\chi_2^0\right]_0-\sqrt{6}\sin\psi_M\mathbf{N}\left[\mathbf{N}\left[\boldsymbol{M}_2,\chi_2^0\right]_2,\boldsymbol{M}_2\right]_0}$

Equations A.5 and A.9 can be rearranged to isolate the terms containing $\sin 2\gamma$ and thereby eliminate them. The remaining expression contains only terms with $\cos 2\gamma$ which produces:

(A.14) $\cos 2\gamma=\frac{\sqrt{3}\left(\sin 2\psi_M\mathbf{N}\left[\boldsymbol{M}_2,\chi_2^2\right]_0+\sqrt{6}\sin\psi_M\mathbf{N}\left[\mathbf{N}\left[\boldsymbol{M}_2,\chi_2^2\right]_2,\boldsymbol{M}_2\right]_0\right)}{\sin 3\psi_M-\sin 2\psi_M\mathbf{N}\left[\boldsymbol{M}_2,\chi_2^0\right]_0-\sqrt{6}\sin\psi_M\mathbf{N}\left[\mathbf{N}\left[\boldsymbol{M}_2,\chi_2^0\right]_2,\boldsymbol{M}_2\right]_0}$

This defines the process for defining the rotation of an arbitrary rank 2 tensor into the diagonal HS form. We note that this process only allows α, β, and γ to be defined only over the range of 0-π. This lack of uniqueness arises because the ±directions in the axis directions are not distinguishable in the rank 2 framework (i.e. $(+z)^2=(-z)^2$). This same rotation process (defined as a rank *n* tensor) can be applied to an arbitrary rank *n* tensor to transform it to its pseudo-diagonal HS form.

Next, we consider defining the reference axis frame as a single tensor with some specified skewness angle $\psi_0$:

(A.15) $\boldsymbol{X}_2=\cos\psi_0\,\boldsymbol{\chi}_2^0+\sin\psi_0\,\boldsymbol{\chi}_2^2$

One can then express the three contractions between this and an arbitrary rank 2 tensor $\boldsymbol{M_2}$ in terms of eqs A.4-9. These can be reduced to:

(A.16) $\mathbf{N}[\boldsymbol{M_2}, \boldsymbol{X_2}]_{\mathbf{0}} = \cos\psi_0\, \mathbf{N}\left[\boldsymbol{M_2}, \boldsymbol{\chi_2^0}\right]_{\mathbf{0}} + \sin\psi_0\, \mathbf{N}\left[\boldsymbol{M_2}, \boldsymbol{\chi_2^2}\right]_{\mathbf{0}}$

(A.17) $\mathbf{N}[\mathbf{N}[\boldsymbol{M_2}, \boldsymbol{X_2}]_2, \boldsymbol{M_2}]_{\mathbf{0}} = \cos\psi_0\, \mathbf{N}\left[\mathbf{N}\left[\boldsymbol{M_2}, \boldsymbol{\chi_2^0}\right]_2, \boldsymbol{M_2}\right]_{\mathbf{0}} +$
$\sin\psi_0\, \mathbf{N}\left[\mathbf{N}\left[\boldsymbol{M_2}, \boldsymbol{\chi_2^2}\right]_2, \boldsymbol{M_2}\right]_{\mathbf{0}}$

(A.18) $\mathbf{N}[\mathbf{N}[\boldsymbol{M_2}, \boldsymbol{X_2}]_2, \boldsymbol{X_2}]_{\mathbf{0}} = \frac{1}{\sqrt{6}}\cos 2\psi_0\, \mathbf{N}\left[\boldsymbol{M_2}, \boldsymbol{\chi_2^0}\right]_{\mathbf{0}} - \frac{1}{\sqrt{6}}\sin 2\psi_0\, \mathbf{N}\left[\boldsymbol{M_2}, \boldsymbol{\chi_2^2}\right]_{\mathbf{0}}$

Additionally, the value of $\psi_0$ can be defined from

(A.19) $\cos 3\psi_0 = \sqrt{\frac{7}{2\sqrt{5}} \frac{\mathbf{N}[\mathbf{N}[\boldsymbol{X_2}, \boldsymbol{X_2}]_2, \boldsymbol{X_2}]_{\mathbf{0}}}{(\mathbf{N}[\boldsymbol{X_2}, \boldsymbol{X_2}]_{\mathbf{0}})^{3/2}}}$

The significance of the form $3\psi_0$ comes from eq A.1. Numerically, solutions of $\psi_0$, $\psi_0 + 2\pi/3$, and $\psi_0 + 4\pi/3$ are all equivalent. When such variations are substituted into A.1, the result is an interchange of the x, y and z axes. Consequently, a unique inversion of eqs A.16-19 to obtain the rotation angles α, β, and γ requires some additional information such as best alignment with the external axis frame, or best alignment between the two HS framework of two tensors. Such choices can be motivated by the particular application, but the intrinsic information is contained in 2 and 3-way contractions of the rank 2 tensors.

**Acknowledgements**
All the authors gratefully acknowledge the support of the U.S. Department of Energy through the LDRD Program at Los Alamos National Laboratory. The work was initialized under XWG7 and completed under XWH4, XX9A, and YW7T MATD. NM acknowledges the Center for Non-Linear Studies at LANL for support, and Kipton Barros and Nick Lubbers for fruitful discussions. EMK dedicates this work to Prof. Thomas J. Meyer (Univ. North Carolina at Chapel Hill, retired) for encouraging him in exploring this area of science.[56] This manuscript has been assigned LA-UR-26-28585.

**Declaration of generative AI and AI-assisted technologies in the manuscript preparation process**
During the preparation of this work, the author(s) used OpenAI GPT-6.1 Sol in order to visualize the strain gradient deformations in Figure 5. After using this tool/service, the author(s) reviewed and edited the content as needed and take(s) full responsibility for the content of the published article.

**Code availability statement**
The code for calculation of the descriptors is available at https://github.com/lanl/sfd. Mathematica notebooks can be obtained from EMK by reasonable request.

**References**

1 Lewis, A. C. *et al.* Quantitative Analysis and Feature Recognition in 3-D Microstructural Data Sets. *JOM*, 52-56 (2006).

2 Stukowski, A. Computational Analysis Methods in Atomistic Modeling of Crystals. *JOM* **66**, 399-407 (2013). https://doi.org/10.1007/s11837-013-0827-5

3 Honeycutt, J. D. & Andersen, H. C. Molecular Dynamics Study of Melting and Freezing of Small Lennard- Jones Clusters. *Journal of Physical Chemistry* **91**, 4950-4963 (1987).

4 Stukowski, A. Structure identification methods for atomistic simulations of crystalline materials. *Modelling and Simulation in Materials Science and Engineering* **20**, 045021 (2012). https://doi.org/10.1088/0965-0393/20/4/045021

5 Kelchner, C. L., Plimpton, S. J. & Hamilton, J. C. Dislocation nucleation and defect structure during surface indentation. *Phys. Rev. B* **58**, 11085-11088 (1998).

6 Steinhardt, P. J., Nelson, D., R. & Ronchetti, M. Bond-orientational order in liquids and glasses. *Physical Review B* **28**, 784-805 (1983).

7 Ackland, G. J. & Jones, A. P. Applications of local crystal structure measures in experiment and simulation. *Physical Review B* **73**, 054104 (2006). https://doi.org/10.1103/PhysRevB.73.054104

8 Maras, E., Trushin, O., Stukowski, A., Ala-Nissila, T. & Jónsson, H. Global transition path search for dislocation formation in Ge on Si(001). *Computer Physics Communications* **205**, 13-21 (2016). https://doi.org/10.1016/j.cpc.2016.04.001

9 Voronoi, G. Nouvelles applications des parametres continus a la theorie des formes quadratiques. II. *J. Reine Angew. Math.* **134**, 198-287 (1908).

10 Keys, A. S., Iacovella, C. R. & Glotzer, S. C. Characterizing complex particle morphologies through shape matching: Descriptors, applications, and algorithms. *Journal of Computational Physics* **230**, 6438-6463 (2011). https://doi.org/10.1016/j.jcp.2011.04.017

11 Larsen, P. M., Schmidt, S. & Schiøtz, J. Robust structural identification via polyhedral template matching. *Modelling and Simulation in Materials Science and Engineering* **24** (2016). https://doi.org/10.1088/0965-0393/24/5/055007

12 Stukowski, A., Bulatov, V. V. & Arsenlis, A. Automated identification and indexing of dislocations in crystal interfaces. *Modelling and Simulation in Materials Science and Engineering* **20** (2012). https://doi.org/10.1088/0965-0393/20/8/085007

13 Stukowski, A. A triangulation-based method to identify dislocations in atomistic models. *Journal of the Mechanics and Physics of Solids* **70**, 314-319 (2014). https://doi.org/10.1016/j.jmps.2014.06.009

14 Begau, C., Hua, J. & Hartmaier, A. A novel approach to study dislocation density tensors and lattice rotation patterns in atomistic simulations. *Journal of the Mechanics and Physics of Solids* **60**, 711-722 (2012). https://doi.org/10.1016/j.jmps.2011.12.005

15 Lazar, E. A., Han, J. & Srolovitz, D. J. Topological framework for local structure analysis in condensed matter. *Proc Natl Acad Sci U S A* **112**, E5769-5776 (2015). https://doi.org/10.1073/pnas.1505788112
16 Rosenbrock, C. W., Homer, E. R., Csányi, G. & Hart, G. L. W. Discovering the building blocks of atomic systems using machine learning: application to grain boundaries. *npj Computational Materials* **3**, 1-7 (2017). https://doi.org/10.1038/s41524-017-0027-x
17 Bartok, A. P., Payne, M. C., Kondor, R. & Csanyi, G. Gaussian approximation potentials: the accuracy of quantum mechanics, without the electrons. *Phys. Rev. Lett.* **104**, 136403 (2010). https://doi.org/10.1103/PhysRevLett.104.136403
18 Bartók, A. P., Kondor, R. & Csányi, G. On representing chemical environments. *Physical Review B* **87** (2013). https://doi.org/10.1103/PhysRevB.87.184115
19 Szlachta, W. J., Bartók, A. P. & Csányi, G. Accuracy and transferability of Gaussian approximation potential models for tungsten. *Physical Review B* **90** (2014). https://doi.org/10.1103/PhysRevB.90.104108
20 De, S., Bartok, A. P., Csanyi, G. & Ceriotti, M. Comparing molecules and solids across structural and alchemical space. *Phys Chem Chem Phys* **18**, 13754-13769 (2016). https://doi.org/10.1039/c6cp00415f
21 Deringer, V. L. & Csányi, G. Machine learning based interatomic potential for amorphous carbon. *Physical Review B* **95** (2017). https://doi.org/10.1103/PhysRevB.95.094203
22 Shapeev, A. V. Moment Tensor Potentials: A Class of Systematically Improvable Interatomic Potentials. *Multiscale Modeling & Simulation* **14**, 1153-1173 (2016). https://doi.org/10.1137/15m1054183
23 Drautz, R. Atomic cluster expansion for accurate and transferable interatomic potentials. *Physical Review B* **99**, 014104 (2019). https://doi.org/10.1103/PhysRevB.99.014104
24 Musil, F. *et al.* Physics-Inspired Structural Representations for Molecules and Materials. *Chemical Reviews* **121**, 9759-9815 (2021). https://doi.org/10.1021/acs.chemrev.1c00021
25 Hu, M.-K. Visual Pattern Recognition by Moment Invariants. *IRE Transactions on Information Theory* **8**, 179-187 (1962).
26 Sadjadi, F. A. & Hall, E. L. Three-Dimensional Moment Invariants. *IEEE Trans Pattern Anal Mach Intell* **PAMI-2**, 127-136 (1980).
27 Cyganski, D. & Orr, J. A. Applications of Tensor Theory to Object Recognition and Orientation Determination. *IEEE Trans. Patt. Analysis Machi. Intel.* **PAMI-7**, 662-673 (1985).
28 Lo, C.-H. & Don, H.-S. 3-D Moment Forms: Their Construction and Application to Object Identification and Positioning. *IEEE Transactions on Pattern Analysis and Machine Intelligence* **11**, 1053-1064 (1989). https://doi.org/10.1109/34.42836
29 Edmonds, A. R. Angular Momentum in Quantum Mechanics. *Princeton University Press, Princeton, NJ* **3rd ed.** (1974).

30 Jerphagnon, J., Chemla, D. & Bonneville, R. The description of the physical properties of condensed matter using irreducible tensors. *Advances in Physics* **27**, 609-650 (1978). https://doi.org/10.1080/00018737800101454
31 Ennis, D. B. & Kindlmann, G. Orthogonal tensor invariants and the analysis of diffusion tensor magnetic resonance images. *Magn Reson Med* **55**, 136-146 (2006). https://doi.org/10.1002/mrm.20741
32 Kindlmann, G., Ennis, D. B., Whitaker, R. T. & Westin, C. F. Diffusion tensor analysis with invariant gradients and rotation tangents. *IEEE Trans Med Imaging* **26**, 1483-1499 (2007). https://doi.org/10.1109/TMI.2007.907277
33 Trott, C. R., Hammond, S. D. & Thompson, A. P. SNAP: Strong Scaling High Fidelity Molecular Dynamics Simulations on Leadership-Class Computing Platforms. *Lecture Notes in Computer Science* **8488**, 19-34 (2014).
34 Wood, M. A. & Thompson, A. P. Extending the accuracy of the SNAP interatomic potential form. *J Chem Phys* **148**, 241721 (2018). https://doi.org/10.1063/1.5017641
35 Podryabinkin, E. V. & Shapeev, A. V. Active learning of linearly parametrized interatomic potentials. *Computational Materials Science* **140**, 171-180 (2017). https://doi.org/10.1016/j.commatsci.2017.08.031
36 Daw, M. S. & Baskes, M. I. Embedded-atom method: Derivation and application to impurities, surfaces, and other defects in metals. *Physical Review B* **29**, 6443-6453 (1984). https://doi.org/10.1103/PhysRevB.29.6443
37 Baskes, M. I. Modified embedded-atom potentials for cubic materials and impurities. *Physical Review B* **46**, 2727-2742 (1992). https://doi.org/10.1103/PhysRevB.46.2727
38 Baskes, M. I., Srinivasan, S. G., Valone, S. M. & Hoagland, R. G. Multistate modified embedded atom method. *Physical Review B* **75** (2007). https://doi.org/10.1103/PhysRevB.75.094113
39 Zimmerman, J. A., Bammann, D. J. & Gao, H. Deformation gradients for continuum mechanical analysis of atomistic simulations. *Int. J. Sol. Struct.* **46**, 238-253 (2009). https://doi.org/10.1016/j.ijsolstr.2008.08.036
40 Falk, M. L. & Langer, J. S. Dynamics of viscoplastic deformation in amorphous solids. *Physical Review E* **57**, 7192-7205 (1998).
41 Horstemeyer, M. F. & Baskes, M. I. Strain Tensors at the Atomic Scale. *Materials Research Society Symposium Proceedings* **578**, 15-20 (2000).
42 Shimizu, F., Ogata, S. & Li, J. Theory of Shear Banding in Metallic Glasses and Molecular Dynamics Calculations. *Materials Transactions* **48**, 2923-2927 (2007). https://doi.org/10.2320/matertrans.MJ200769
43 Gullett, P. M., Horstemeyer, M. F., Baskes, M. I. & Fang, H. A deformation gradient tensor and strain tensors for atomistic simulations. *Modelling and Simulation in Materials Science and Engineering* **16** (2008). https://doi.org/10.1088/0965-0393/16/1/015001
44 Lafourcade, P., Denoual, C. & Maillet, J.-B. Irreversible Deformation Mechanisms for 1,3,5-Triamino-2,4,6-Trinitrobenzene Single Crystal through Molecular Dynamics Simulations. *The Journal of Physical Chemistry C* **122**, 14954-14964 (2018). https://doi.org/10.1021/acs.jpcc.8b02983

45 Thompson, A. P., Swiler, L. P., Trott, C. R., Foiles, S. M. & Tucker, G. J. Spectral neighbor analysis method for automated generation of quantum-accurate interatomic potentials. *J.Comp. Phys.* **285**, 316-330 (2015). https://doi.org/10.1016/j.jcp.2014.12.018

46 Lindeberg, T. Generalized Gaussian Scale-Space Axiomatics Comprising Linear Scale-Space, Affine Scale-Space and Spatio-Temporal Scale-Space. *J. Math. Imag.Vis.* **40**, 36-81 (2010). https://doi.org/10.1007/s10851-010-0242-2

47 Arroyo, M. & Ortiz, M. Local maximum-entropy approximation schemes: a seamless bridge between finite elements and meshfree methods. *Int. J. Num. Meth. Eng.* **65**, 2167-2202 (2006). https://doi.org/10.1002/nme.1534

48 Cyron, C. J., Arroyo, M. & Ortiz, M. Smooth, second order, non-negative meshfree approximants selected by maximum entropy. *Int. J. Num. Meth. Eng.* **79**, 1605-1632 (2009). https://doi.org/10.1002/nme.2597

49 Grad, H. Note on N-Dimensional Hermite Polynomials. *Commun. Pure Appl. Math.* **2**, 325-330 (1949).

50 Coope, J. A. R., Snider, R. F. & McCourt, F. R. Irreducible Cartesian Tensors. *The Journal of Chemical Physics* **43**, 2269-2275 (1965). https://doi.org/10.1063/1.1697123

51 Coope, J. A. R. & Snider, R. F. Irreducible Cartesian Tensors. II. General Formulation. *Journal of Mathematical Physics* **11**, 1003-1017 (1970). https://doi.org/10.1063/1.1665190

52 Coope, J. A. R. Irreducible Cartesian Tensors. III. Clebsch-Gordan Reduction. *Journal of Mathematical Physics* **11**, 1591-1612 (1970). https://doi.org/10.1063/1.1665301

53 Levine, I. N., Busch, D. H. & Shull, H. *Quantum chemistry*. Vol. 6 (Pearson Prentice Hall Upper Saddle River, NJ, 2009).

54 Aubert, G. An alternative to Wigner d-matrices for rotating real spherical harmonics. *AIP Advances* **3** (2013). https://doi.org/10.1063/1.4811853

55 Cotton, F. A. *Chemical Applications of Group Theory*. (John Wiley & Sons, 1990).

56 Kober, E. M. & Meyer, T. J. Concerning the Absorption Spectra of the Ions [M(bpy)3]2+ (M = Fe, Ru, Os; bpy = 2,2'-Bipyridine). *Inorg. Chem.* **21**, 3967-3977 (1982).

57 Drautz, R. Atomic cluster expansion for accurate and transferable interatomic potentials. *Physical Review B* **99** (2019). https://doi.org/10.1103/PhysRevB.99.014104

58 Dusson, G. *et al.* Atomic cluster expansion: Completeness, efficiency and stability. *Journal of Computational Physics* **454**, 110946 (2022). https://doi.org/https://doi.org/10.1016/j.jcp.2022.110946

59 Goff, J. M., Sievers, C., Wood, M. A. & Thompson, A. P. Permutation-adapted complete and independent basis for atomic cluster expansion descriptors. *Journal of Computational Physics* (2024). https://doi.org/10.1016/j.jcp.2024.113073

60 Allen, A. E. A., Shinkle, E., Bujack, R. & Lubbers, N. Optimal invariant sets for atomistic machine learning. *npj Computational Materials* **12**, 75 (2026). https://doi.org/10.1038/s41524-025-01948-0

61 Luscher, D. J., McDowell, D. L. & Bronkhorst, C. A. A second gradient theoretical framework for hierarchical multiscale modeling of materials. *International Journal of Plasticity* **26**, 1248-1275 (2010). https://doi.org/10.1016/j.ijplas.2010.05.006

62 Admal, N. C., Marian, J. & Po, G. The atomistic representation of first strain-gradient elastic tensors. *Journal of the Mechanics and Physics of Solids* **99**, 93-115 (2017). https://doi.org/10.1016/j.jmps.2016.11.005

63 Plimpton, S. J. Fast Parallel Algorithms for Short–Range Molecular Dynamics. *J. Comput. Phys.* **117**, 1-19 (1995).

64 Stukowski, A. Visualization and analysis of atomistic simulation data with OVITO–the Open Visualization Tool. *Modell. Simul. Mater. Sci. Eng.* **18**, 015012 (2010). https://doi.org/10.1088/0965-0393/18/1/015012

65 Mishin, Y., Mehl, M. J., Papaconstantopoulos, D. A., Voter, A. F. & Kress, J. D. Structural stability and lattice defects in copper:Ab initio, tight-binding, and embedded-atom calculations. *Physical Review B* **63** (2001). https://doi.org/10.1103/PhysRevB.63.224106

66 Hennig, R. G., Lenosky, T. J., Trinkle, D. R., Rudin, S. P. & Wilkins, J. W. Classical potential describes martensitic phase transformations between theα,β, andωtitanium phases. *Physical Review B* **78** (2008). https://doi.org/10.1103/PhysRevB.78.054121

67 Stillinger, F. H. & Weber, T. A. Computer simulation of local order in condensed phases of silicon. *Phys. Rev. B* **31**, 5262 (1985).

68 Marinica, M.-C. *et al.* Interatomic potentials for modelling radiation defects and dislocations in tungsten. *Journal of Physics: Condensed Matter* **25**, 395502 (2013).

69 Ravelo, R., Germann, T., Guerrero, O., An, Q. & Holian, B. Shock-induced plasticity in tantalum single crystals: Interatomic potentials and large-scale molecular-dynamics simulations. *Physical Review B—Condensed Matter and Materials Physics* **88**, 134101 (2013).

70 Stukowski, A. & Arsenlis, A. On the elastic–plastic decomposition of crystal deformation at the atomic scale. *Modelling and Simulation in Materials Science and Engineering* **20** (2012). https://doi.org/10.1088/0965-0393/20/3/035012

71 Nye, J. F. Some Geometrical Relations in Dislocated Crystals. *Acta Metallurgica* **1**, 153-162 (1953).

72 Mishra, A., Suresh, S. A., Fensin, S. J., Mathew, N. & Kober, E. M. Learning from metastable symmetric-tilt grain boundaries using physics-based descriptors. *Physical Review Materials* **8**, 123605 (2024).

73 Tavenner, J. P., Gupta, A., Thompson, G. B., Kober, E. M. & Tucker, G. J. Learning grain boundary segregation behavior through fingerprinting complex atomic environments. *Communications Materials* **5**, 183 (2024).

74 Owens, C. B. *et al.* Feature engineering descriptors, transforms, and machine learning for grain boundaries and variable-sized atom clusters. *npj Computational Materials* **11**, 21 (2025).
75 Bishop, C. M. *Pattern Recognition and Machine Learning (Information Science and Statistics)*. (2006).
76 Pedregosa, F. *et al.* Scikit-learn: Machine Learning in Python. *J. Mach. Learn. Res.* **12**, 2825-2830 (2011).
77 Sander, J., Ester, M., Kriegel, H.-P. & Xu, X. Density-Based Clustering in Spatial Databases: The Algorithm GDBSCAN and Its Applications. *Data Mining and Knowledge Discovery* **2**, 169-194 (1998).
78 Lebensohn, R., Bringa, E. & Caro, A. A viscoplastic micromechanical model for the yield strength of nanocrystalline materials. *Acta Materialia* **55**, 261-271 (2007). https://doi.org/10.1016/j.actamat.2006.07.023
79 Lebensohn, R. A., Tomé, C. N. & CastaÑeda, P. P. Self-consistent modelling of the mechanical behaviour of viscoplastic polycrystals incorporating intragranular field fluctuations. *Philosophical Magazine* **87**, 4287-4322 (2007). https://doi.org/10.1080/14786430701432619

# Strain Functionals: A Complete and Symmetry-adapted Set of Descriptors to Characterize Atomistic Configurations

Edward M. Kober,*[,a] Jacob P. Tavenner,*[,2] Colin M. Adams,*[,3] Nithin Mathew*[,1,a]

*Group T-1, Theoretical Division, Los Alamos National Laboratory, Los Alamos, NM 87544
[1]Group XCP-5, X Computational Physics Division, Los Alamos National Laboratory, Los Alamos, NM 87544
[2]Computational Materials Science and Design Group, Baylor University, Waco, TX 76706
[3]Systems and Technology Research, Woburn, MA 01801

[a] Corresponding authors: emk@lanl.gov, mathewni@lanl.gov

## Supplemental Material

### ***S.1**: Transformation from tensor derivatives to traceless formulations*

The second derivative of the weight function with respect to the $\boldsymbol{r}$ vector, evaluated at $\boldsymbol{r} = \boldsymbol{r_a}$, generates a 3x3 matrix given below for the *x, y* and *z* components of the direction vector.

$$\sum \frac{\partial^2 w_b}{\partial r^2}\bigg|_{\boldsymbol{r_a}} = \sum \frac{\partial^2 exp\left(-\frac{|\mathbf{r}_b-\mathbf{r}|^2}{2\sigma^2}\right)}{\partial r^2}\Bigg|_{\boldsymbol{r_a}} = \frac{1}{\sigma^4}\sum \boldsymbol{D2}\ exp\left(-\frac{r_{ab}^2}{2\sigma^2}\right) \tag{S1.1}$$

$$\boldsymbol{D2} = \begin{bmatrix} x_{ab}^2-\sigma^2 & x_{ab}y_{ab} & x_{ab}z_{ab} \\ x_{ab}y_{ab} & y_{ab}^2-\sigma^2 & y_{ab}z_{ab} \\ x_{ab}z_{ab} & y_{ab}z_{ab} & z_{ab}^2-\sigma^2 \end{bmatrix} \tag{S1.2}$$

That matrix can then be separated into a portion that depends on the distance vector and one that depends only on σ. The former term can be simplify identified as the outer product of the $\boldsymbol{r_{ab}}$ vector with itself and the latter term is simply $\sigma^2$ times the 3x3 identify matrix, $\boldsymbol{I_2}$.

$$\boldsymbol{D2} = \begin{bmatrix} x_{ab}^2 & x_{ab}y_{ab} & x_{ab}z_{ab} \\ x_{ab}y_{ab} & y_{ab}^2 & y_{ab}z_{ab} \\ x_{ab}z_{ab} & y_{ab}z_{ab} & z_{ab}^2 \end{bmatrix} - \begin{bmatrix} \sigma^2 & 0 & 0 \\ 0 & \sigma^2 & 0 \\ 0 & 0 & \sigma^2 \end{bmatrix} = \boldsymbol{r_{ab}}\otimes\boldsymbol{r_{ab}} - \sigma^2\boldsymbol{I_2} \tag{S1.3}$$

The first matrix can then be made traceless and that trace added to the second matrix to result in an expression based on the traceless and diagonal (trace) components. The traceless component of a matrix/tensor $\boldsymbol{M}$ is denoted by the notation $[\mathbf{M}]_{\boldsymbol{T}}$. The diagonal term could be simply represented as a scalar, but the full expression is retained here to be consistent with the subsequent formulations for higher order tensors.

$$\boldsymbol{D2} = [\boldsymbol{r_{ab}}\otimes\boldsymbol{r_{ab}}]_{\boldsymbol{T}} + \left(\frac{r_{ab}^2}{3} - \sigma^2\right)\boldsymbol{I_2} \tag{S1.4}$$

The third derivative of the weight function generates the 3x3x3 tensor below.

$$\sum \frac{\partial^3 w_b}{\partial r^3}\bigg|_{r_a} = \sum \frac{\partial^3 exp\left(-\frac{|\mathrm{r}_b-\mathrm{r}|^2}{2\sigma^2}\right)}{\partial r^3}\Bigg|_{r_a} = \frac{1}{\sigma^6}\sum \boldsymbol{D3}\ exp\left(-\frac{r_{ab}^2}{2\sigma^2}\right) \qquad \text{(S1.5)}$$

$$\boldsymbol{D3} = \begin{bmatrix} \begin{bmatrix} 3\sigma^2 x_{ab} - x_{ab}^3 \\ \sigma^2 y_{ab} - x_{ab}^2 y_{ab} \\ \sigma^2 z_{ab} - x_{ab}^2 z_{ab} \end{bmatrix} & \begin{bmatrix} \sigma^2 y_{ab} - x_{ab}^2 y_{ab} \\ \sigma^2 x_{ab} - x_{ab} y_{ab}^2 \\ -x_{ab} y_{ab} z_{ab} \end{bmatrix} & \begin{bmatrix} \sigma^2 z_{ab} - x_{ab}^2 z_{ab} \\ -x_{ab} y_{ab} z_{ab} \\ \sigma^2 x_{ab} - x_{ab} z_{ab}^2 \end{bmatrix} \\ \begin{bmatrix} \sigma^2 y_{ab} - x_{ab}^2 y_{ab} \\ \sigma^2 x_{ab} - x_{ab} y_{ab}^2 \\ -x_{ab} y_{ab} z_{ab} \end{bmatrix} & \begin{bmatrix} \sigma^2 x_{ab} - x_{ab} y_{ab}^2 \\ 3\sigma^2 y_{ab} - y_{ab}^3 \\ \sigma^2 z_{ab} - y_{ab}^2 z_{ab} \end{bmatrix} & \begin{bmatrix} -x_{ab} y_{ab} z_{ab} \\ \sigma^2 z_{ab} - y_{ab}^2 z_{ab} \\ \sigma^2 y_{ab} - y_{ab} z_{ab}^2 \end{bmatrix} \\ \begin{bmatrix} \sigma^2 z_{ab} - x_{ab}^2 z_{ab} \\ -x_{ab} y_{ab} z_{ab} \\ \sigma^2 x_{ab} - x_{ab} z_{ab}^2 \end{bmatrix} & \begin{bmatrix} -x_{ab} y_{ab} z_{ab} \\ \sigma^2 z_{ab} - y_{ab}^2 z_{ab} \\ \sigma^2 y_{ab} - y_{ab} z_{ab}^2 \end{bmatrix} & \begin{bmatrix} \sigma^2 x_{ab} - x_{ab} z_{ab}^2 \\ \sigma^2 y_{ab} - y_{ab} z_{ab}^2 \\ 3\sigma^2 z_{ab} - z_{ab}^3 \end{bmatrix} \end{bmatrix} \qquad \text{(S1.6)}$$

As above the tensor can be separated into one part that does not depend on σ and one that does, as shown below.

$$\boldsymbol{D3} = -\begin{bmatrix} \begin{bmatrix} x_{ab}^3 \\ x_{ab}^2 y_{ab} \\ x_{ab}^2 z_{ab} \end{bmatrix} & \begin{bmatrix} x_{ab}^2 y_{ab} \\ x_{ab} y_{ab}^2 \\ x_{ab} y_{ab} z_{ab} \end{bmatrix} & \begin{bmatrix} x_{ab}^2 z_{ab} \\ x_{ab} y_{ab} z_{ab} \\ x_{ab} z_{ab}^2 \end{bmatrix} \\ \begin{bmatrix} x_{ab}^2 y_{ab} \\ x_{ab} y_{ab}^2 \\ x_{ab} y_{ab} z_{ab} \end{bmatrix} & \begin{bmatrix} x_{ab} y_{ab}^2 \\ y_{ab}^3 \\ y_{ab}^2 z_{ab} \end{bmatrix} & \begin{bmatrix} x_{ab} y_{ab} z_{ab} \\ y_{ab}^2 z_{ab} \\ y_{ab} z_{ab}^2 \end{bmatrix} \\ \begin{bmatrix} x_{ab}^2 z_{ab} \\ x_{ab} y_{ab} z_{ab} \\ x_{ab} z_{ab}^2 \end{bmatrix} & \begin{bmatrix} x_{ab} y_{ab} z_{ab} \\ y_{ab}^2 z_{ab} \\ y_{ab} z_{ab}^2 \end{bmatrix} & \begin{bmatrix} x_{ab} z_{ab}^2 \\ y_{ab} z_{ab}^2 \\ z_{ab}^3 \end{bmatrix} \end{bmatrix} + \sigma^2 \begin{bmatrix} \begin{bmatrix} 3x_{ab} \\ y_{ab} \\ z_{ab} \end{bmatrix} & \begin{bmatrix} y_{ab} \\ x_{ab} \\ 0 \end{bmatrix} & \begin{bmatrix} z_{ab} \\ 0 \\ x_{ab} \end{bmatrix} \\ \begin{bmatrix} y_{ab} \\ x_{ab} \\ 0 \end{bmatrix} & \begin{bmatrix} x_{ab} \\ 3y_{ab} \\ z_{ab} \end{bmatrix} & \begin{bmatrix} 0 \\ z_{ab} \\ y_{ab} \end{bmatrix} \\ \begin{bmatrix} z_{ab} \\ 0 \\ x_{ab} \end{bmatrix} & \begin{bmatrix} 0 \\ z_{ab} \\ y_{ab} \end{bmatrix} & \begin{bmatrix} x_{ab} \\ y_{ab} \\ 3z_{ab} \end{bmatrix} \end{bmatrix} \qquad \text{(S1.7)}$$

The first tensor is simply the three-fold outer product of $\boldsymbol{r_{ab}}$ with itself $[\boldsymbol{r_{ab}} \otimes \boldsymbol{r_{ab}} \otimes \boldsymbol{r_{ab}}]$, while the second has the symmetries of the trace of a symmetric rank 3 tensor. The trace of the symmetric rank 3 tensor ***D3*** is generally written as a simple vector as below, with three independent components.

$$\boldsymbol{T3} = Tr(\boldsymbol{r_{ab}} \otimes \boldsymbol{r_{ab}} \otimes \boldsymbol{r_{ab}}) = (r_{ab}^2 x_{ab}, r_{ab}^2 y_{ab}, r_{ab}^2 z_{ab}) \qquad \text{(S1.8)}$$

$$r_{ab}^2 = x_{ab}^2 + y_{ab}^2 + z_{ab}^2$$

In general, a rank 3 tensor will have three distinct traces that depend upon which axes the contraction occurs along. The average of those three traces is equal to the trace of the symmetric portion of the general rank 3 tensor.[30] Since ***D3*** is symmetric to begin with, these traces are all equivalent to ***T3***. That vector can be expanded back into the form of a symmetric rank 3 tensor by the formulation below, where the symmetric trace of a matrix/tensor ***M*** is designated by **[*M*]*S***.

$$[\boldsymbol{r_{ab}} \otimes \boldsymbol{r_{ab}} \otimes \boldsymbol{r_{ab}}]_{\boldsymbol{S}}(i,j,k) = \boldsymbol{T3}(i)\delta_{jk} + \boldsymbol{T3}(j)\delta_{ik} + \boldsymbol{T3}(k)\delta_{ij} \tag{S1.9}$$

The resulting tensor, given below, is seen to have the same symmetries as the second tensor in the expression for ***D3*** above and is simply related to it by the scalar $\left(\frac{\sigma}{r_{ab}}\right)^2$.

$$[\boldsymbol{r_{ab}} \otimes \boldsymbol{r_{ab}} \otimes \boldsymbol{r_{ab}}]_{\boldsymbol{S}} = \begin{bmatrix} \begin{bmatrix} 3r_{ab}^2 x_{ab} \\ r_{ab}^2 y_{ab} \\ r_{ab}^2 z_{ab} \end{bmatrix} & \begin{bmatrix} r_{ab}^2 y_{ab} \\ r_{ab}^2 x_{ab} \\ 0 \end{bmatrix} & \begin{bmatrix} r_{ab}^2 z_{ab} \\ 0 \\ r_{ab}^2 x_{ab} \end{bmatrix} \\ \begin{bmatrix} r_{ab}^2 y_{ab} \\ r_{ab}^2 x_{ab} \\ 0 \end{bmatrix} & \begin{bmatrix} r_{ab}^2 x_{ab} \\ 3r_{ab}^2 y_{ab} \\ r_{ab}^2 z_{ab} \end{bmatrix} & \begin{bmatrix} 0 \\ r_{ab}^2 z_{ab} \\ r_{ab}^2 y_{ab} \end{bmatrix} \\ \begin{bmatrix} r_{ab}^2 z_{ab} \\ 0 \\ r_{ab}^2 x_{ab} \end{bmatrix} & \begin{bmatrix} 0 \\ r_{ab}^2 z_{ab} \\ r_{ab}^2 y_{ab} \end{bmatrix} & \begin{bmatrix} r_{ab}^2 x_{ab} \\ r_{ab}^2 y_{ab} \\ 3r_{ab}^2 z_{ab} \end{bmatrix} \end{bmatrix} \tag{S1.10}$$

The traceless version of the original rank 3 tensor is then found by subtracting off the trace multiplied by (1/5), contrasted to the (1/3) factor used for the rank 2 tensors.

$$[\boldsymbol{r_{ab}} \otimes \boldsymbol{r_{ab}} \otimes \boldsymbol{r_{ab}}]_{\boldsymbol{T}} = [\boldsymbol{r_{ab}} \otimes \boldsymbol{r_{ab}} \otimes \boldsymbol{r_{ab}}] - \left(\frac{1}{5}\right)[\boldsymbol{r_{ab}} \otimes \boldsymbol{r_{ab}} \otimes \boldsymbol{r_{ab}}]_{\boldsymbol{S}} \tag{S1.11}$$

The expression for **D3** can then be rewritten in terms of a traceless tensor and a symmetric trace component.

$$\boldsymbol{D3} = -[\boldsymbol{r_{ab}} \otimes \boldsymbol{r_{ab}} \otimes \boldsymbol{r_{ab}}]_{\boldsymbol{T}} + \left(\frac{\sigma^2}{r_{ab}^2} - \frac{1}{5}\right)[\boldsymbol{r_{ab}} \otimes \boldsymbol{r_{ab}} \otimes \boldsymbol{r_{ab}}]_{\boldsymbol{S}} \tag{S1.12}$$

This has a very similar form to the expression for ***D2*** above given $[\boldsymbol{r_{ab}} \otimes \boldsymbol{r_{ab}}]_{\boldsymbol{S}} = r_{ab}^2 \boldsymbol{I_2}$
Similarly, the fourth derivative defines the expression

$$\sum \frac{\partial^4 w_b}{\partial r^4}\bigg|_{\boldsymbol{r_a}} = \sum \frac{\partial^4 exp\left(-\frac{|\mathrm{r}_b - \mathrm{r}|^2}{2\sigma^2}\right)}{\partial r^4}\Bigg|_{\boldsymbol{r_a}} = \frac{1}{\sigma^8}\sum \boldsymbol{D4}\; exp\left(-\frac{r_{ab}^2}{2\sigma^2}\right) \tag{S1.13}$$

The fourth rank tensor can be similarly separated into terms that depend upon different powers of $\sigma$ as below

$$\boldsymbol{D4} = \boldsymbol{r_{ab}} \otimes \boldsymbol{r_{ab}} \otimes \boldsymbol{r_{ab}} \otimes \boldsymbol{r_{ab}} - \frac{\sigma^2}{r_{ab}^2}[\boldsymbol{r_{ab}} \otimes \boldsymbol{r_{ab}} \otimes \boldsymbol{r_{ab}} \otimes \boldsymbol{r_{ab}}]_{\boldsymbol{S}} + \frac{\sigma^4}{2r_{ab}^4}[[\boldsymbol{r_{ab}} \otimes \boldsymbol{r_{ab}} \otimes \boldsymbol{r_{ab}} \otimes \boldsymbol{r_{ab}}]_{\boldsymbol{S}}]_{\boldsymbol{S}} \tag{S1.14}$$

Here, the first symmetric trace tensor is defined from the trace of the original tensor as below

$$\boldsymbol{T4} = Tr(\boldsymbol{r_{ab}} \otimes \boldsymbol{r_{ab}} \otimes \boldsymbol{r_{ab}} \otimes \boldsymbol{r_{ab}}) = r_{ab}^2 \begin{bmatrix} x_{ab}^2 & x_{ab}y_{ab} & x_{ab}z_{ab} \\ x_{ab}y_{ab} & y_{ab}^2 & y_{ab}z_{ab} \\ x_{ab}z_{ab} & y_{ab}z_{ab} & z_{ab}^2 \end{bmatrix} \quad \text{(S1.15)}$$

This can then be expanded back to rank 4 tensor as below

$$[\boldsymbol{r_{ab}} \otimes \boldsymbol{r_{ab}} \otimes \boldsymbol{r_{ab}} \otimes \boldsymbol{r_{ab}}]_{\boldsymbol{S}}(i,j,k,l) = \boldsymbol{T4}(i,j)\delta_{kl} + \boldsymbol{T4}(i,k)\delta_{jl} + \boldsymbol{T4}(i,l)\delta_{jk} + \boldsymbol{T4}(j,k)\delta_{il} + \boldsymbol{T4}(j,l)\delta_{ik} + \boldsymbol{T4}(k,l)\delta_{ij} \quad \text{(S1.16)}$$

Note that it has 6 independent quantities. The second symmetric trace tensor, which is a scalar, is defined from the trace of **T4** as below:

$$S4 = Tr\big(Tr(\boldsymbol{r_{ab}} \otimes \boldsymbol{r_{ab}} \otimes \boldsymbol{r_{ab}} \otimes \boldsymbol{r_{ab}})\big) = Tr\big(Tr(\boldsymbol{T4})\big) = r_{ab}^4 \quad \text{(S1.17)}$$

$$[\boldsymbol{r_{ab}} \otimes \boldsymbol{r_{ab}}]_{\boldsymbol{S}}(i,j) = S4\ \delta_{ij} = r_{ab}^4\ \delta_{ij} \quad \text{(S1.18)}$$

$$[[\boldsymbol{r_{ab}} \otimes \boldsymbol{r_{ab}} \otimes \boldsymbol{r_{ab}} \otimes \boldsymbol{r_{ab}}]_{\boldsymbol{S}}]_{\boldsymbol{S}}(i,j,k,l) = [\boldsymbol{r_{ab}} \otimes \boldsymbol{r_{ab}}]_{\boldsymbol{S}}(i,j)\delta_{kl} + [\boldsymbol{r_{ab}} \otimes \boldsymbol{r_{ab}}]_{\boldsymbol{S}}(i,k)\delta_{jl} + [\boldsymbol{r_{ab}} \otimes \boldsymbol{r_{ab}}]_{\boldsymbol{S}}(i,l)\delta_{jk} + [\boldsymbol{r_{ab}} \otimes \boldsymbol{r_{ab}}]_{\boldsymbol{S}}(j,k)\delta_{il} + [\boldsymbol{r_{ab}} \otimes \boldsymbol{r_{ab}}]_{\boldsymbol{S}}(j,l)\delta_{ik} + [\boldsymbol{r_{ab}} \otimes \boldsymbol{r_{ab}}]_{\boldsymbol{S}}(k,l)\delta_{ij} \quad \text{(S1.19)}$$

Note that this tensor has only 1 independent component, $r_{ab}^4$, which is simply related to the $r_{ab}^2$ term from the ***T4*** tensor. As above, the expression for ***D4*** can be further rewritten in terms of traceless tensors as below:

$$\boldsymbol{D4} = [\boldsymbol{r_{ab}} \otimes \boldsymbol{r_{ab}} \otimes \boldsymbol{r_{ab}} \otimes \boldsymbol{r_{ab}}]_{\boldsymbol{T}} - \left(\frac{\sigma^2}{r_{ab}^2} - \frac{1}{7}\right)[[\boldsymbol{r_{ab}} \otimes \boldsymbol{r_{ab}} \otimes \boldsymbol{r_{ab}} \otimes \boldsymbol{r_{ab}}]_{\boldsymbol{S}}]_{\boldsymbol{T}} + \left(\frac{\sigma^4}{2r_{ab}^4} - \frac{\sigma^2}{3r_{ab}^2} + \frac{1}{30}\right)[[\boldsymbol{r_{ab}} \otimes \boldsymbol{r_{ab}} \otimes \boldsymbol{r_{ab}} \otimes \boldsymbol{r_{ab}}]_{\boldsymbol{S}}]_{\boldsymbol{S}} \quad \text{(S1.20)}$$

*S.2 Description of strain tensors corresponding to l=2 deformation*

The deformations represented in Figure 1 for *l=2* can be directly mapped to Cartesian strains in mechanics literature as shown below. Note that the Cartesian *z* axis is designated as the primary axis consistent with the definitions of spherical harmonics. Off-diagonal elements can be generated/eliminated by rotation about the primary frame.

*l=2, m=0* (volume-preserving uniaxial strain state)

$$\begin{bmatrix} -\frac{1}{2}\varepsilon & 0 & 0 \\ 0 & -\frac{1}{2}\varepsilon & 0 \\ 0 & 0 & \varepsilon \end{bmatrix} \tag{S2.1}$$

*l=2, m=1* (shear strain)

$$\begin{bmatrix} 0 & 0 & 0 \\ 0 & -\varepsilon & 0 \\ 0 & 0 & \varepsilon \end{bmatrix}, \tag{S2.2}$$

$$\begin{bmatrix} -\varepsilon & 0 & 0 \\ 0 & 0 & 0 \\ 0 & 0 & \varepsilon \end{bmatrix} \tag{S2.3}$$

*l=2, m=2* (shear strain)

$$\begin{bmatrix} \varepsilon & 0 & 0 \\ 0 & -\varepsilon & 0 \\ 0 & 0 & 0 \end{bmatrix} \tag{S2.4}$$

*S.3 Third order moment relationships*

The second rank 3 contraction of the third order term, analogous to eq 19 of the second order term, is given in eq S3.1 and is zero by symmetry, as argued by Lo&Don. Further, it was found that the rank 3 contraction of a rank 3 ST with itself generates a zero vector, as shown in eq. S3.2. Consequently, none of these terms contain any additional information. This result further extends to all odd rank order contractions of odd rank ST.

$$\boldsymbol{N}[\boldsymbol{N}[\mathbf{3},\mathbf{3}]_3,\mathbf{3}]_0 = 0 \tag{S3.1}$$
$$\boldsymbol{N}[\mathbf{3},\mathbf{3}]_3 = \{0,0,0,0,0,0,0\} \tag{S3.2}$$

For completeness, the full expression for the norm of the rank 2 contraction of the third order term given in eq S3.3. This depends in a complicated fashion on differences the magnitudes and phase angles of the different components, where some patterns can be discerned. We note that all terms that involve the phase angles are all multiplied by $\mu_3^1$ such that they go to zero in the HS configuration giving eq 26 in the text.

$$\boldsymbol{N}[\boldsymbol{N}[\mathbf{3},\mathbf{3}]_2,\boldsymbol{N}[\mathbf{3},\mathbf{3}]_2]_0 = \frac{1}{420}(16\sqrt{5}(\mu_3^0)^4 + 32\sqrt{5}(\mu_3^0)^2(\mu_3^1)^2 + 21\sqrt{5}(\mu_3^1)^4 + 80\sqrt{5}(\mu_3^0)^2(\mu_3^2)^2 + 30\sqrt{5}(\mu_3^1)^2(\mu_3^2)^2 - 40\sqrt{5}(\mu_3^0)^2(\mu_3^3)^2 - 10\sqrt{5}(\mu_3^1)^2(\mu_3^3)^2 + 50\sqrt{5}(\mu_3^2)^2(\mu_3^3)^2 + 25\sqrt{5}(\mu_3^3)^4 - 40\sqrt{3}(\mu_3^0)(\mu_3^1)^2(\mu_3^2)\mathrm{Cos}[2(\omega_1 - \omega_2)] + 120\sqrt{5}(\mu_3^0)(\mu_3^1)(\mu_3^2)(\mu_3^3)\mathrm{Cos}[\omega_1 + 2\omega_2 - 3\omega_3] - 40\sqrt{3}(\mu_3^1)^3(\mu_3^3)\mathrm{Cos}[3(\omega_1 - \omega_3)] + 100\sqrt{3}(\mu_3^1)(\mu_3^2)^2(\mu_3^3)\mathrm{Cos}[\omega_1 - 4\omega_2 + 3\omega_3]) \quad \text{(S3.3)}$$

Performing similar higher order self-contractions of the rank 2 contraction generates terms that depend on the lower order contractions as shown in eq S3.4. The first expression is a nested series of three contractions with the $\boldsymbol{N}[\mathbf{3},\mathbf{3}]_2$ moiety, ending with the dot product. The second is dot product of the rank 2 contraction of that moiety with itself. These are equivalent and equal to a scaled norm of that quantity. Consequently, there is no further information in such contractions.

$$\boldsymbol{N}[\boldsymbol{N}[\boldsymbol{N}[\boldsymbol{N}[\mathbf{3},\mathbf{3}]_2,\boldsymbol{N}[\mathbf{3},\mathbf{3}]_2]_2,\boldsymbol{N}[\mathbf{3},\mathbf{3}]_2]_2,\boldsymbol{N}[\mathbf{3},\mathbf{3}]_2]_0 = \boldsymbol{N}[\boldsymbol{N}[\boldsymbol{N}[\mathbf{3},\mathbf{3}]_2,\boldsymbol{N}[\mathbf{3},\mathbf{3}]_2]_2,\boldsymbol{N}[\boldsymbol{N}[\mathbf{3},\mathbf{3}]_2,\boldsymbol{N}[\mathbf{3},\mathbf{3}]_2]_2]_0 = \frac{2}{7\sqrt{5}}(\boldsymbol{N}[\boldsymbol{N}[\mathbf{3},\mathbf{3}]_2,\boldsymbol{N}[\mathbf{3},\mathbf{3}]_2]_0)^2 \quad \text{(S3.4)}$$

The norms of the rank 4 and rank 6 contractions have fairly simple forms in the HS configuration, similar to that of the rank 2 contraction.

$$\boldsymbol{N}[\boldsymbol{N}[\mathbf{3},\mathbf{3}]_4,\boldsymbol{N}[\mathbf{3},\mathbf{3}]_4]_0 = \frac{1}{154}\{12(\mu_3^0)^4 - 24(\mu_3^0)^2((\mu_3^2)^2 - 4(\mu_3^3)^2) + 3(\mu_3^3)^2(2(\mu_3^2)^2 + (\mu_3^3)^2) + 28(\mu_3^2)^4\} \quad \text{(S3.5)}$$

$$\boldsymbol{N}[\boldsymbol{N}[\mathbf{3},\mathbf{3}]_6,\boldsymbol{N}[\mathbf{3},\mathbf{3}]_6]_0 = \frac{1}{924\sqrt{13}}\{400(\mu_3^0)^4 + 288(\mu_3^2)^4 + 926(\mu_3^2)^2(\mu_3^3)^2 + 463(\mu_3^3)^4 + 8(\mu_3^0)^2(142(\mu_3^2)^2 + 37(\mu_3^3)^2)\} \quad \text{(S3.6)}$$

In studying the behavior of these terms in the analysis of MD simulations, it was apparent that there was some direct relationship between this term and the norm of the rank 2 contraction (eq 26). Similar relationships were apparent also with the norm of the rank 6 contraction. Straightforward algebraic manipulations produced the following relationships in S3.7-9. From those relationships, it is apparent that only two of these four terms are independent. Further, any pair could be chosen as the independent metrics.

$$(\boldsymbol{N}[\mathbf{3},\mathbf{3}]_0)^2 = \frac{15}{7\sqrt{5}}\boldsymbol{N}[\boldsymbol{N}[\mathbf{3},\mathbf{3}]_2,\boldsymbol{N}[\mathbf{3},\mathbf{3}]_2]_0 + \frac{11}{14}\boldsymbol{N}[\boldsymbol{N}[\mathbf{3},\mathbf{3}]_4,\boldsymbol{N}[\mathbf{3},\mathbf{3}]_4]_0 \quad \text{(S3.7)}$$

$$(\boldsymbol{N}[\mathbf{3},\mathbf{3}]_0)^2 = -\frac{35}{24\sqrt{5}}\boldsymbol{N}[\boldsymbol{N}[\mathbf{3},\mathbf{3}]_2,\boldsymbol{N}[\mathbf{3},\mathbf{3}]_2]_0 + \frac{11\sqrt{13}}{24}\boldsymbol{N}[\boldsymbol{N}[\mathbf{3},\mathbf{3}]_6,\boldsymbol{N}[\mathbf{3},\mathbf{3}]_6]_0 \quad \text{(S3.8)}$$

$$(\boldsymbol{N}[\mathbf{3},\mathbf{3}]_0)^2 = \frac{7}{22}\boldsymbol{N}[\boldsymbol{N}[\mathbf{3},\mathbf{3}]_4,\boldsymbol{N}[\mathbf{3},\mathbf{3}]_4]_0 + \frac{3\sqrt{13}}{11}\boldsymbol{N}[\boldsymbol{N}[\mathbf{3},\mathbf{3}]_6,\boldsymbol{N}[\mathbf{3},\mathbf{3}]_6]_0 \quad \text{(S3.9)}$$

The full expression for the rank 4 skewness is given below. Note that all of the angular terms depend on the differences between different phase angles (and combinations thereof), and never upon a single phase angle. That is a consequence of the rotationally invariant form.

$$\boldsymbol{N}[\boldsymbol{N}[\boldsymbol{N}[\mathbf{3},\mathbf{3}]_4,\boldsymbol{N}[\mathbf{3},\mathbf{3}]_4]_4,\boldsymbol{N}[\mathbf{3},\mathbf{3}]_4]_0 = \frac{1}{11858\sqrt{13}}\{-12\sqrt{15}(\mu_3^0)(\mu_3^1)^2(\mu_3^2)(64(\mu_3^0)^2 + 119(\mu_3^1)^2 - 277(\mu_3^2)^2 - 497(\mu_3^3)^2)\mathrm{Cos}[2(\omega_1-\omega_2)] - 2970(\mu_3^1)^4(\mu_3^2)^2\mathrm{Cos}[4(\omega_1-\omega_2)] - 12240(\mu_3^0)^3(\mu_3^1)(\mu_3^2)(\mu_3^3)\mathrm{Cos}[\omega_1+2\omega_2-3\omega_3] - 14220(\mu_3^0)(\mu_3^1)^3(\mu_3^2)(\mu_3^3)\} \quad \text{(S3.10)}$$

To demonstrate the completeness of the chosen invariants, their forms where one of the magnitude terms is set to zero are examined. Note that this also zeroes out the coefficient for the $\cos(6\varpi_2)$ term. The results are shown below.

For $\mu_3^0 = 0$: (S3.11)

$$\boldsymbol{N}[\boldsymbol{N}[\mathbf{3},\mathbf{3}]_2,\boldsymbol{N}[\mathbf{3},\mathbf{3}]_2]_0 = \frac{1}{84\sqrt{5}}25(\mu_3^3)^2(2(\mu_3^2)^2+(\mu_3^3)^2)$$

$$\boldsymbol{N}[\boldsymbol{N}[\boldsymbol{N}[\mathbf{3},\mathbf{3}]_2,\boldsymbol{N}[\mathbf{3},\mathbf{3}]_2]_2,\boldsymbol{N}[\mathbf{3},\mathbf{3}]_2]_0 = \frac{125}{588\sqrt{30}}25(\mu_3^3)^4(3(\mu_3^2)^2+(\mu_3^3)^2)$$

$$\boldsymbol{N}[\boldsymbol{N}[\boldsymbol{N}[\mathbf{3},\mathbf{3}]_4,\boldsymbol{N}[\mathbf{3},\mathbf{3}]_4]_4,\boldsymbol{N}[\mathbf{3},\mathbf{3}]_4]_0 = \frac{1}{11858\sqrt{13}}\{7(\mu_3^2)^2 - 3(\mu_3^3)^2\}\{392(\mu_3^2)^4+144(\mu_3^2)^2(\mu_3^3)^2 + 27(\mu_3^3)^4\}$$

For $\mu_3^2 = 0$: (S3.12)

$$\boldsymbol{N}[\boldsymbol{N}[\mathbf{3},\mathbf{3}]_2,\boldsymbol{N}[\mathbf{3},\mathbf{3}]_2]_0 = \frac{1}{84\sqrt{5}}\{4(\mu_3^0)^2 - 5(\mu_3^3)^2\}^2$$

$$\boldsymbol{N}[\boldsymbol{N}[\boldsymbol{N}[\mathbf{3},\mathbf{3}]_2,\boldsymbol{N}[\mathbf{3},\mathbf{3}]_2]_2,\boldsymbol{N}[\mathbf{3},\mathbf{3}]_2]_0 = -\frac{1}{588\sqrt{30}}\{4(\mu_3^0)^2 - 5(\mu_3^3)^2\}^3$$

$$\boldsymbol{N}[\boldsymbol{N}[\boldsymbol{N}[\mathbf{3},\mathbf{3}]_4,\boldsymbol{N}[\mathbf{3},\mathbf{3}]_4]_4,\boldsymbol{N}[\mathbf{3},\mathbf{3}]_4]_0 = -\frac{81}{11858\sqrt{13}}\{2(\mu_3^0)^2 + (\mu_3^3)^2\}\{4(\mu_3^0)^4-94(\mu_3^0)^2(\mu_3^3)^2 + (\mu_3^3)^4\}$$

For $\mu_3^3 = 0$: (S3.13)

$$\boldsymbol{N}[\boldsymbol{N}[\mathbf{3},\mathbf{3}]_2,\boldsymbol{N}[\mathbf{3},\mathbf{3}]_2]_0 = \frac{16}{84\sqrt{5}}(\mu_3^0)^2\{(\mu_3^0)^2 + 5(\mu_3^2)^2\}$$

$$\boldsymbol{N}[\boldsymbol{N}[\boldsymbol{N}[\mathbf{3},\mathbf{3}]_2,\boldsymbol{N}[\mathbf{3},\mathbf{3}]_2]_2,\boldsymbol{N}[\mathbf{3},\mathbf{3}]_2]_0 = -\frac{16}{147\sqrt{30}}(\mu_3^0)^4 * \{(\mu_3^0)^2 - 15(\mu_3^2)^2\}$$

$$\boldsymbol{N}[\boldsymbol{N}[\boldsymbol{N}[\mathbf{3},\mathbf{3}]_4,\boldsymbol{N}[\mathbf{3},\mathbf{3}]_4]_4,\boldsymbol{N}[\mathbf{3},\mathbf{3}]_4]_0 = -\frac{82}{11858\sqrt{13}}\{9(\mu_3^0)^2 - 7(\mu_3^2)^2\}\{9(\mu_3^0)^4-30(\mu_3^0)^2(\mu_3^2)^2 + 49(\mu_3^2)^4\}$$

We note that these expressions contain terms that depend on the differences between the magnitudes of the remaining two terms. This shows that the three independent magnitude terms are uniquely defined by these expressions. Similarly, eqs 27 and 28 can be combined to generate two complementary terms where one does not have a dependence on the $\cos(6\varpi_2)$ term. The resulting polynomials in the magnitude terms also show that the three magnitude terms are then uniquely defined.

*S.4 Fourth order moment relationships*

The full expression for the norm of the rank 2 contraction of the rank 4 ST:

$$\boldsymbol{N}[\boldsymbol{N}[\mathbf{4},\mathbf{4}]_2,\boldsymbol{N}[\mathbf{4},\mathbf{4}]_2]_0 = \frac{1}{13860}(\sqrt{5}(400(\mu_4^0)^4 + 589(\mu_4^1)^4 + 64(\mu_4^2)^4 + 938(\mu_4^2)^2(\mu_4^3)^2 + 49(\mu_4^3)^4 - 112((\mu_4^2)^2 - 14(\mu_4^3)^2)(\mu_4^4)^2 + 784(\mu_4^4)^4 + (\mu_4^1)^2(758(\mu_4^2)^2 + 518(\mu_4^3)^2 - 952(\mu_4^4)^2) + 40(\mu_4^0)^2(20(\mu_4^1)^2 + 62(\mu_4^2)^2 - 7((\mu_4^3)^2 + 4(\mu_4^4)^2))) + 60(-42(\mu_4^0)(\mu_4^1)^2(\mu_4^2)\mathrm{Cos}[2(\theta_1 - \theta_2)] + 46\sqrt{7}(\mu_4^0)(\mu_4^1)(\mu_4^2)(\mu_4^3)\mathrm{Cos}[\theta_1 + 2\theta_2 - 3\theta_3] - 6\sqrt{35}(\mu_4^1)^3(\mu_4^3)\mathrm{Cos}[3(\theta_1 - \theta_3)] - 4\sqrt{35}(\mu_4^1)^2(\mu_4^2)(\mu_4^4)\mathrm{Cos}[2(\theta_1 + \theta_2 - 2\theta_4)] + 2\sqrt{7}(\mu_4^2)(\mu_4^4)(12(\mu_4^0)(\mu_4^2)\mathrm{Cos}[4(\theta_2 - \theta_4)] + 7\sqrt{5}(\mu_4^3)^2\mathrm{Cos}[2(\theta_2 - 3\theta_3 + 2\theta_4)]) + (\mu_4^1)(\mu_4^3)(9\sqrt{35}(\mu_4^2)^2\mathrm{Cos}[\theta_1 - 4\theta_2 + 3\theta_3] + 28(\mu_4^0)(\mu_4^4)\mathrm{Cos}[\theta_1 + 3\theta_3 - 4\theta_4] + 42\sqrt{5}(\mu_4^2)(\mu_4^4)\mathrm{Cos}[\theta_1 - 2\theta_2 - 3\theta_3 + 4\theta_4]))) \quad \text{(S4.1)}$$

The expression for the norm of the rank 2 contraction of the rank 4 ST in HS frame is given below

$$\begin{aligned}\boldsymbol{N}[\boldsymbol{N}[\mathbf{4},\mathbf{4}]_2,\boldsymbol{N}[\mathbf{4},\mathbf{4}]_2]_0 &= \frac{1}{13860}\{\sqrt{5}(400(\mu_4^0)^4 + 64(\mu_4^2)^4 + 14(\mu_4^2)^2\{(\mu_4^3)^2 - 8(\mu_4^4)^2\} \\ &+ 49\{(\mu_4^3)^4 + 32(\mu_4^3)^2(\mu_4^4)^2 + 16(\mu_4^4)^4\} \\ &+ 40(\mu_4^0)^2\{62(\mu_4^2)^2 - 7(\mu_4^3)^2 - 28(\mu_4^4)^2\}) \\ &+ 120\sqrt{7}\mu_4^2\mu_4^4\left(12\mu_4^0\mu_4^2\cos(4\varpi_2) + 7\sqrt{5}(\mu_4^3)^2\cos(2\{\varpi_2 - 3\varpi_3\})\right)\} \quad (S4.2)\end{aligned}$$

The expression for the skewness of the rank 2 contraction of the rank 4 ST in the HS frame is given below:

$$\boldsymbol{N}[\boldsymbol{N}[\boldsymbol{N}[\mathbf{4},\mathbf{4}]_2,\boldsymbol{N}[\mathbf{4},\mathbf{4}]_2]_2,\boldsymbol{N}[\mathbf{4},\mathbf{4}]_2]_0 = \frac{1}{58212\sqrt{770}}\Big\{\sqrt{7}\{8000(\mu_4^0)^6 + 512(\mu_4^2)^6 + 168(\mu_4^2)^4(67(\mu_4^3)^2 - 80(\mu_4^4)^2\} - 147(\mu_4^2)^2\{67(\mu_4^3)^4 + 812(\mu_4^3)^2(\mu_4^4)^2 - 320(\mu_4^4)^4\} - 343\{(\mu_4^3)^6 + 48(\mu_4^3)^4(\mu_4^4)^2 + 192(\mu_4^3)^2(\mu_4^4)^4 + 64(\mu_4^4)^6\} - 1200(\mu_4^0)^4\{100(\mu_4^2)^2 + 7(\mu_4^3)^2 + 28(\mu_4^4)^2\} - 60(\mu_4^0)^2\{800(\mu_4^2)^4 - 7(\mu_4^2)^2[167(\mu_4^3)^2 + 320(\mu_4^4)^2] - 49[(\mu_4^3)^4 + 20(\mu_4^3)^2(\mu_4^4)^2 + 16(\mu_4^4)^4]\} + 252(\mu_4^2)\Big[-12\sqrt{5}(\mu_4^0)(\mu_4^2)(\mu_4^4)[40(\mu_4^0)^2 + 16(\mu_4^2)^2 + 91(\mu_4^3)^2 - 56(\mu_4^4)^2]\mathrm{Cos}[4\varpi_2] - (\mu_4^3)^2\Big\{7(\mu_4^4)[-100(\mu_4^0)^2 + 35(\mu_4^2)^2 + 35(\mu_4^3)^2 + 224(\mu_4^4)^2]\mathrm{Cos}[2\{\varpi_2 - 3\varpi_3\}] + 9\sqrt{35}(\mu_4^0)\{25(\mu_4^2)^2\mathrm{Cos}[6\{\varpi_2 - 3\varpi_3\}] + 28(\mu_4^4)^2\mathrm{Cos}[2\{\varpi_2 + 3\varpi_3\}]\}\Big\}\Big]\Big\} \quad \text{(S4.3)}$$

The relationships between the $\boldsymbol{N}[\boldsymbol{N}[\mathbf{4},\mathbf{4}]_x, \boldsymbol{N}[\mathbf{4},\mathbf{4}]_x]_0$ for x = 2, 4, 6, 8 are given below

$$(\boldsymbol{N}[\mathbf{4},\mathbf{4}]_0)^2 = \frac{11\sqrt{5}}{49}\boldsymbol{N}[\boldsymbol{N}[\mathbf{4},\mathbf{4}]_2,\boldsymbol{N}[\mathbf{4},\mathbf{4}]_2]_0 + \frac{143}{98}\boldsymbol{N}[\boldsymbol{N}[\mathbf{4},\mathbf{4}]_4,\boldsymbol{N}[\mathbf{4},\mathbf{4}]_4]_0 \quad \text{(S4.4)}$$

$$(\boldsymbol{N}[\mathbf{4},\mathbf{4}]_0)^2 = \frac{77\sqrt{5}}{200}\boldsymbol{N}[\boldsymbol{N}[\mathbf{4},\mathbf{4}]_2,\boldsymbol{N}[\mathbf{4},\mathbf{4}]_2]_0 + \frac{11\sqrt{13}}{40}\boldsymbol{N}[\boldsymbol{N}[\mathbf{4},\mathbf{4}]_6,\boldsymbol{N}[\mathbf{4},\mathbf{4}]_6]_0 \quad \text{(S4.5)}$$

$$(\boldsymbol{N}[\mathbf{4},\mathbf{4}]_0)^2 = -\frac{28\sqrt{5}}{75}\boldsymbol{N}[\boldsymbol{N}[\mathbf{4},\mathbf{4}]_2,\boldsymbol{N}[\mathbf{4},\mathbf{4}]_2]_0 + \frac{13\sqrt{17}}{30}\boldsymbol{N}[\boldsymbol{N}[\mathbf{4},\mathbf{4}]_8,\boldsymbol{N}[\mathbf{4},\mathbf{4}]_8]_0 \quad \text{(S4.6)}$$

The full expression for the rank 4 skewness in HS coordinates is given below:

$$\boldsymbol{N}[\boldsymbol{N}[\boldsymbol{N}[\mathbf{4},\mathbf{4}]_4,\boldsymbol{N}[\mathbf{4},4]_4]_4,\boldsymbol{N}[\mathbf{4},\mathbf{4}]_4]_0 = \frac{1}{2004002}(17496(\mu_4^0)^6 - 79992(\mu_4^0)^4(\mu_4^2)^2 + 164804(\mu_4^0)^2(\mu_4^2)^4 - 28248(\mu_4^2)^6 - 394632(\mu_4^0)^4(\mu_4^3)^2 + 410508(\mu_4^0)^2(\mu_4^2)^2(\mu_4^3)^2 - 76104(\mu_4^2)^4(\mu_4^3)^2 + 460404(\mu_4^0)^2(\mu_4^3)^4 - 56889(\mu_4^2)^2(\mu_4^3)^4 - 27783(\mu_4^3)^6 + 139608(\mu_4^0)^4(\mu_4^4)^2 - 660352(\mu_4^0)^2(\mu_4^2)^2(\mu_4^4)^2 + 122346(\mu_4^2)^4(\mu_4^4)^2 - 966672(\mu_4^0)^2(\mu_4^3)^2(\mu_4^4)^2 - 12348(\mu_4^2)^2(\mu_4^3)^2(\mu_4^4)^2 - 37044(\mu_4^3)^4(\mu_4^4)^2 + 108584(\mu_4^0)^2(\mu_4^4)^4 - 116424(\mu_4^2)^2(\mu_4^4)^4 + 24696(\mu_4^3)^2(\mu_4^4)^4 + 8232(\mu_4^4)^6 + 24\sqrt{35}(\mu_4^0)(\mu_4^2)^2(\mu_4^4)(2818(\mu_4^0)^2 - 2937(\mu_4^2)^2 - 4837(\mu_4^3)^2 + 3738(\mu_4^4)^2)\mathrm{Cos}[4\bar{\omega}_2] + 198450(\mu_4^2)^4(\mu_4^4)^2\mathrm{Cos}[8\bar{\omega}_2] - 5040\sqrt{7}(\mu_4^0)^2(\mu_4^2)(\mu_4^3)^2(\mu_4^4)\mathrm{Cos}[2(\bar{\omega}_2 - 3\bar{\omega}_3)] + 26460\sqrt{7}(\mu_4^2)^3(\mu_4^3)^2(\mu_4^4)\mathrm{Cos}[2(\bar{\omega}_2 - 3\bar{\omega}_3)] + 26460\sqrt{7}(\mu_4^2)(\mu_4^3)^4(\mu_4^4)\mathrm{Cos}[2(\bar{\omega}_2 - 3\bar{\omega}_3)] + 70560\sqrt{7}(\mu_4^2)(\mu_4^3)^2(\mu_4^4)^3\mathrm{Cos}[2(\bar{\omega}_2 - 3\bar{\omega}_3)] + 52500\sqrt{5}(\mu_4^0)(\mu_4^2)^3(\mu_4^3)^2\mathrm{Cos}[6(\bar{\omega}_2 - \bar{\omega}_3)] + 58800\sqrt{5}(\mu_4^0)(\mu_4^2)(\mu_4^3)^2(\mu_4^4)^2\mathrm{Cos}[2(\bar{\omega}_2 + 3\bar{\omega}_3)]) \quad \text{(S4.7)}$$

The full expression for eq 43 in HS coordinates is given below:

$$\boldsymbol{N}[\boldsymbol{N}[\boldsymbol{N}[\mathbf{4},\mathbf{4}]_2,\mathbf{4}]_2,\boldsymbol{N}[\mathbf{4},\mathbf{4}]_2]_0 = \boldsymbol{N}[\boldsymbol{N}[\boldsymbol{N}[\mathbf{4},\mathbf{4}]_2,\boldsymbol{N}[\mathbf{4},\mathbf{4}]_2]_4,\mathbf{4}]_0 = \frac{1}{48510\sqrt{2}}\Big(\sqrt{35}\mu_4^0\Big(400(\mu_4^0)^4 - 416(\mu_4^2)^4 - 28(\mu_4^2)^2(89(\mu_4^3)^2 - 106(\mu_4^4)^2) + 49((\mu_4^3)^2 - 4(\mu_4^4)^2)^2 - 40(\mu_4^0)^2\big(13(\mu_4^2)^2 + 7((\mu_4^3)^2 + 4(\mu_4^4)^2)\big)\Big) + 280(\mu_4^2)^2\mu_4^4\big(41(\mu_4^0)^2 - 4(\mu_4^2)^2 + 21((\mu_4^3)^2 + (\mu_4^4)^2)\big)\mathrm{Cos}[4\bar{\omega}_2] - 9800\sqrt{5}\mu_4^0\mu_4^2(\mu_4^3)^2\mu_4^4\mathrm{Cos}[2(\bar{\omega}_2 - 3\bar{\omega}_3)] + 1750\sqrt{7}(\mu_4^2)^3(\mu_4^3)^2\mathrm{Cos}[6(\bar{\omega}_2 - \bar{\omega}_3)]\Big) \quad \text{(S4.8)}$$

The full expression for eq 45 is HS coordinates is given below:

$$\boldsymbol{N}[\boldsymbol{N}[\boldsymbol{N}[\mathbf{4},\mathbf{4}]_4,\mathbf{4}]_4,\boldsymbol{N}[[\mathbf{4},\mathbf{4}]_4,[\mathbf{4},\mathbf{4}]_4]_4]_0 = \quad \text{(S4.9)}$$

$$
\frac{1}{21042021\sqrt{286}}
\Big(472392\sqrt{7}(\mu_4^0)^7 - 848628\sqrt{7}(\mu_4^0)^5(\mu_4^2)^2 + 512800\sqrt{7}(\mu_4^0)^3(\mu_4^2)^4
- 1824548\sqrt{7}\mu_4^0(\mu_4^2)^6 + 632772\sqrt{7}(\mu_4^0)^5(\mu_4^3)^2 - 2319660\sqrt{7}(\mu_4^0)^3(\mu_4^2)^2(\mu_4^3)^2
- 3905734\sqrt{7}\mu_4^0(\mu_4^2)^4(\mu_4^3)^2 - 6906060\sqrt{7}(\mu_4^0)^3(\mu_4^3)^4
- 5595849\sqrt{7}\mu_4^0(\mu_4^2)^2(\mu_4^3)^4 - 3083913\sqrt{7}\mu_4^0(\mu_4^3)^6 + 4834872\sqrt{7}(\mu_4^0)^5(\mu_4^4)^2
- 3478440\sqrt{7}(\mu_4^0)^3(\mu_4^2)^2(\mu_4^4)^2 - 2585184\sqrt{7}\mu_4^0(\mu_4^2)^4(\mu_4^4)^2
- 6650280\sqrt{7}(\mu_4^0)^3(\mu_4^3)^2(\mu_4^4)^2 - 4228308\sqrt{7}\mu_4^0(\mu_4^2)^2(\mu_4^3)^2(\mu_4^4)^2
+ 1259496\sqrt{7}\mu_4^0(\mu_4^3)^4(\mu_4^4)^2 + 5972120\sqrt{7}(\mu_4^0)^3(\mu_4^4)^4
+ 2220876\sqrt{7}\mu_4^0(\mu_4^2)^2(\mu_4^4)^4 + 552916\sqrt{7}\mu_4^0(\mu_4^3)^2(\mu_4^4)^4 + 1105832\sqrt{7}\mu_4^0(\mu_4^4)^6
+ 21\sqrt{5}(\mu_4^2)^2\mu_4^4
\big(203980(\mu_4^0)^4 + 140268(\mu_4^2)^4
+ 8(\mu_4^0)^2(7519(\mu_4^2)^2 + 44954(\mu_4^3)^2 + 14819(\mu_4^4)^2)
+ 14(\mu_4^2)^2(12389(\mu_4^3)^2 + 25764(\mu_4^4)^2)
+ 49(2247(\mu_4^3)^4 + 5624(\mu_4^3)^2(\mu_4^4)^2 + 12(\mu_4^4)^4)\big)\mathrm{Cos}[4\bar{\omega}_2]
- 2933280\sqrt{7}\mu_4^0(\mu_4^2)^4(\mu_4^4)^2\mathrm{Cos}[8\bar{\omega}_2]
- 11077920(\mu_4^0)^3\mu_4^2(\mu_4^3)^2\mu_4^4\mathrm{Cos}[2(\bar{\omega}_2 - 3\bar{\omega}_3)]
+ 15321320\mu_4^0(\mu_4^2)^3(\mu_4^3)^2\mu_4^4\mathrm{Cos}[2(\bar{\omega}_2 - 3\bar{\omega}_3)]
+ 18398520\mu_4^0\mu_4^2(\mu_4^3)^4\mu_4^4\mathrm{Cos}[2(\bar{\omega}_2 - 3\bar{\omega}_3)]
- 15421280\mu_4^0\mu_4^2(\mu_4^3)^2(\mu_4^4)^3\mathrm{Cos}[2(\bar{\omega}_2 - 3\bar{\omega}_3)]
- 1665300\sqrt{35}(\mu_4^0)^2(\mu_4^2)^3(\mu_4^3)^2\mathrm{Cos}[6(\bar{\omega}_2 - \bar{\omega}_3)]
+ 14700\sqrt{35}(\mu_4^2)^5(\mu_4^3)^2\mathrm{Cos}[6(\bar{\omega}_2 - \bar{\omega}_3)]
+ 14700\sqrt{35}(\mu_4^2)^3(\mu_4^3)^4\mathrm{Cos}[6(\bar{\omega}_2 - \bar{\omega}_3)]
- 446880\sqrt{35}(\mu_4^2)^3(\mu_4^3)^2(\mu_4^4)^2\mathrm{Cos}[6(\bar{\omega}_2 - \bar{\omega}_3)]
- 720300\sqrt{5}(\mu_4^3)^4(\mu_4^4)^3\mathrm{Cos}[12\bar{\omega}_3]
- 1058400\sqrt{35}(\mu_4^0)^2\mu_4^2(\mu_4^3)^2(\mu_4^4)^2\mathrm{Cos}[2(\bar{\omega}_2 + 3\bar{\omega}_3)]
- 1519980\sqrt{35}(\mu_4^2)^3(\mu_4^3)^2(\mu_4^4)^2\mathrm{Cos}[2(\bar{\omega}_2 + 3\bar{\omega}_3)]
- 617400\sqrt{35}\mu_4^2(\mu_4^3)^4(\mu_4^4)^2\mathrm{Cos}[2(\bar{\omega}_2 + 3\bar{\omega}_3)]
+ 823200\sqrt{35}\mu_4^2(\mu_4^3)^2(\mu_4^4)^4\mathrm{Cos}[2(\bar{\omega}_2 + 3\bar{\omega}_3)]\Big)
$$